\documentclass[12pt,a4paper]{article}
\usepackage[utf8]{inputenc}
\usepackage[english]{babel}
\usepackage{amsmath}
\usepackage{subfigure}
\usepackage{cite}
\usepackage{amsfonts}
\usepackage{amssymb}
\usepackage{graphicx}
\newcommand{\be}{\begin{equation}}
\newcommand{\ee}{\end{equation}}
\newcommand{\bea}{\begin{eqnarray}}
\newcommand{\eea}{\end{eqnarray}}

\catcode`@=12
\def\la{\mathrel{\mathchoice {\vcenter{\offinterlineskip\halign{\hfil
$\displaystyle##$\hfil\cr<\cr\sim\cr}}}
{\vcenter{\offinterlineskip\halign{\hfil$\textstyle##$\hfil\cr<\cr\sim\cr}}}
{\vcenter{\offinterlineskip\halign{\hfil$\scriptstyle##$\hfil\cr<\cr\sim\cr}}}
{\vcenter{\offinterlineskip\halign{\hfil$\scriptscriptstyle##$\hfil\cr<\cr\sim
\cr}}}}}
\def\ga{\mathrel{\mathchoice {\vcenter{\offinterlineskip\halign{\hfil
$\displaystyle##$\hfil\cr>\cr\sim\cr}}}
{\vcenter{\offinterlineskip\halign{\hfil$\textstyle##$\hfil\cr>\cr\sim\cr}}}
{\vcenter{\offinterlineskip\halign{\hfil$\scriptstyle##$\hfil\cr>\cr\sim\cr}}}
{\vcenter{\offinterlineskip\halign{\hfil$\scriptscriptstyle##$\hfil\cr>\cr\sim
\cr}}}}}
\begin{document}
%\begin{flushright}
%SINP-APC-14/--
%\end{flushright}
\thispagestyle{empty}
\begin{center}
{\Large \bf
{Explaining Low Energy $\gamma$-ray Excess from the Galactic Centre using
a Two Component Dark Matter Model}}\\
\vspace{0.25cm}
\begin{center}
{\bf Anirban Biswas}\footnote{email: anirban.biswas@saha.ac.in}\\
\vspace{0.5cm}
{\it Astroparticle Physics and Cosmology Division,} \\
{\it Saha Institute of Nuclear Physics, Kolkata 700064, INDIA}\footnote{Present address:
Harish-Chandra Research Institute, Chhatnag Road, Jhusi, Allahabad, 211019, INDIA}
\end{center}
\vspace{1cm}
%%%%%%%%%%%%%%%%%%%%%%%%%%%%%%%%%%%%%%%%%%%%%%%%%%%%%%%%%%%%%%%%
{\bf ABSTRACT}
%%%%%%%%%%%%%%%%%%%%%%%%%%%%%%%%%%%%%%%%%%%%%%%%%%%%%%%%%%%%%%%%
\end{center}
%\begin{abstract}
Over the past few years, there has been a hint of the
$\gamma$-ray excess observed by the Fermi-LAT satellite
borne telescope from the regions surrounding the Galactic
Centre at an energy range $\sim 1$-$3$ GeV. The nature of this
excess $\gamma$-ray spectrum is found to be consistent with the
$\gamma$-ray emission expected from dark matter annihilation
at the Galactic Centre while disfavouring other known
astrophysical sources as the possible origin of this
phenomena. It is also reported that the spectrum and
morphology of this excess $\gamma$-rays can well be
explained by the dark matter particles having mass
in the range $30\sim 40$ GeV annihilating significantly into
${\rm b} \bar{\rm b}$ final state with an annihilation
cross section $\sigma {\rm v}\sim (1.4$ - $2.0)\times10^{-26}$
cm$^3/$s at the Galactic Centre.
In this work, we propose a two component dark matter model where
two different types of dark matter particles namely a complex
scalar and a Dirac fermion are considered. The stability
of both the dark sector particles are maintained by
virtue of an additional local U$(1)_{\rm X}$ gauge symmetry.
We find that our proposed scenario can provide a viable
explanation for this anomalous excess $\gamma$-rays
besides satisfying all the existing relevant theoretical as well as
experimental and observational bounds from LHC, PLANCK and LUX
collaborations. The allowed range of ``effective annihilation cross section"
of lighter dark matter particle for the ${\rm b} \bar{\rm b}$ annihilation channel thus
obtained is finally compared with the limits reported by the Fermi-LAT and
DES collaborations using data from various dwarf spheroidal galaxies.
    
%\end{abstract}
\vskip 1cm
%\qquad\quad Pacs: 95.35.+d, 98.80.Cq
%\vskip 1cm
%\quad\,\, Dark Matter, Beyond SM
\newpage
%%%%%%%%%%%%%%%%%%%%%%%%%%%%%%%%%%%%%%%%%%%%%%%%%%%%%%%%%%%%%%%%%%%%%%%%
\section{Introduction}
\label{intro}
%%%%%%%%%%%%%%%%%%%%%%%%%%%%%%%%%%%%%%%%%%%%%%%%%%%%%%%%%%%%%%%%%%%%%%%%
The existence of the dark matter (DM) in the Universe is
now an established fact by various astronomical measurements and observations
such as galaxy rotation curves, gravitational lensing of distant objects,
Bullet cluster etc. However, no information is still available
to us about the nature and the constituents of dark matter.
The most successful hypothesis until now is that the
dark matter of the Universe is composed of Weakly
Interacting Massive Particles or WIMPs. 
The particle nature of the dark matter can be explored mainly
in two ways. One of them is the process of direct detection
where the information about the mass of the dark matter
candidate along with its scattering cross section off the
detector nuclei can be obtained by measuring the
recoil energy of the latter as a result of
scattering of the dark matter particles with the nuclei.
The DM particles may also be trapped gravitationally within
the massive celestial objects like the Sun, Earth etc.
in case their escape velocities fall short of that required to
overcome the gravity at the central regions of these heavenly bodies.
Moreover, the centre of our Milky way galaxy is also
enriched with huge amount of dark matter. Annihilation of these trapped
dark matter particles can result in the production
of high energy neutrinos, positrons
(particle antiparticle pair in general), $\gamma$-rays etc.
Detection of such annihilation products will provide
valuable information about the constituents of the dark matter
in the Universe. This is known as the process of indirect detection.
More information about the properties of WIMPs and their detection
procedures (both direct and indirect) are discussed in Refs.
\cite{Jungman:1995df, Bertone:2004pz}.    
\paragraph{}
It has been claimed by several groups \cite{Goodenough:2009gk, Hooper:2010mq,
Boyarsky:2010dr, Hooper:2011ti, Abazajian:2012pn, Hooper:2013rwa, Abazajian:2014fta,
Daylan:2014rsa} in last few years after analysing the Fermi-LAT data \cite{fermidata} that
a hint of the $\gamma$-ray excess has been observed
by the Fermi-LAT satellite borne telescope \cite{Atwood:2009ez}
from the regions surrounding the Galactic
Centre (GC) at an energy range $\sim 1$-$3$ GeV. 
More recent analyses of Fermi-LAT data by Daylan et. al. \cite{Daylan:2014rsa}
have disfavoured the possibilities of its astrophysical origin and
strongly indicating that the spectrum of this anomalous excess
$\gamma$-rays is consistent with the emission expected from dark
matter annihilation at the Galactic Centre.
It is also reported in the same article \cite{Daylan:2014rsa} that
the observed $\gamma$-ray spectrum can be well explained by
a dark matter particle having mass in the range $\sim$ 30-40 GeV (or $\sim$ 7-10 GeV)
and annihilating significantly into
${\rm b }\bar{\rm b}$ (or $\tau^{+}$ ${\tau^{-}}$) final state with an
annihilation cross section $\sigma {\rm v}_{\rm b\bar{\rm b}}$ 
$\sim (1.4$ - $2.0)\times10^{-26}$ cm$^3/$s \footnote{annihilation
into $\tau^{+} {\tau^{-}}$ final state the required cross section is
${\sigma {\rm v}}_{\tau^{+} {\tau^{-}}}\sim 2.0 \times 10^{-27} {\rm cm}^3/{\rm s}$
(with local dark matter density = 0.4 GeV/cm$^3$) for a $\sim$ 10 GeV DM particle
\cite{Hooper:2013rwa}.}. Although, there are some
previous works \cite{Boucenna:2011hy, Alvares:2012qv, Alves:2014yha,
Berlin:2014tja, Agrawal:2014una, Izaguirre:2014vva, Cerdeno:2014cda,
Ipek:2014gua, Boehm:2014bia, Ko:2014gha, Abdullah:2014lla, Ghosh:2014pwa,
Martin:2014sxa, Wang:2014elb, Basak:2014sza, Detmold:2014qqa, 
Arina:2014yna, Okada:2014usa, Ghorbani:2014qpa, Banik:2014eda}
where different particle dark matter
models have been proposed to explain this low energy (GeV scale)
$\gamma$-ray excess from the neighbourhood regions of the Galactic
Centre but in most of these articles the authors have considered
single component dark matter model {\it i.e.} all the dark matter
present in the Universe are constituted by a single stable beyond
Standard Model particle. The larger dark matter mass
ranges which also give acceptable fits to
the Fermi-LAT data for the ${\rm b} \bar{\rm b}$
annihilation channel are discussed in Refs. \cite{Agrawal:2014oha, Calore:2014nla}.
A nonthermal decaying dark matter scenario explaining the
anomalous $\gamma$-ray excess from the GC is shown in Ref. \cite{Biswas:2015sva}. 
\paragraph{}
The Standard Model (SM) of particle physics does not contain
any stable particle which can play the role of DM. Therefore
it is generally assumed, in the existing literature, that the observed relic density
of the entire dark sector is contributed by a single beyond SM particle.
However, there are reasons to believe that the dark sector may also
possesses some diversity in its particle spectrum
like the visible sector of our Universe. One of the major
reasons is the similarity between the observed abundances
of both the dark and visible sector at the present epoch.   
Recently, various multicomponent dark matter models have been studied by
several groups \cite{Cao:2007fy, Profumo:2009tb, Feldman:2010wy, Aoki:2012ub,
Biswas:2013nn, Bhattacharya:2013hva, Bian:2013wna}.
In this present work, our endeavour is to explore
the possibility of having multicomponent DM scenario and
study whether it can provide a viable explanation of the
Fermi-LAT observed $\gamma$-ray excess from the regions
close the GC while simultaneously satisfying all the
existing theoretical, experimental as well as the observational
bounds.  
\paragraph{}
We propose a two component dark matter model where
the dark sector is composed of two different types of particles
namely, a complex scalar ($S$) and a Dirac fermion ($\psi$).
Our proposed model is an extension of the Standard Model
of particle physics where the scalar sector of the SM is
enlarged by the two SM gauge (${\rm SU}(3)_{\rm c} \times{\rm SU}(2)_{\rm L}
\times {\rm U}(1)_{\rm Y}$) singlet complex scalar fields $S$ and $\Phi_s$. 
The stability of these dark sector particles are ensured
by the application of an additional local ${\rm U(1)}_{\rm X}$
gauge symmetry under which only the two dark sector particles
$S$, $\psi$ and the complex scalar $\Phi_s$ transform nontrivially.
Therefore, the Lagrangian of this present two component dark matter
model remains invariant under the
${\rm SU}(2)_{\rm L} \times {\rm U}(1)_{\rm Y} \times {\rm U}(1)_{\rm X}$
gauge symmetry which breaks spontaneously to a residual ${\rm U(1)}_{\rm em}\times \mathbb{Z}_2$
symmetry when the complex scalars $\Phi$ (usual SM Higgs doublet)
and $\Phi_s$  acquire Vacuum Expectation Values (VEVs).
$S$ and $\psi$ are the only two fields in this model which
are odd under this residual $\mathbb{Z}_2$ symmetry.   
The effect of spontaneous breaking of local gauge symmetry is
manifested by the presence of five gauge bosons namely $W^\pm$, $Z$, $Z^\prime$ and $A$,
out of which one neutral gauge field ($A$) remains massless which is identified
as the ``photon" (mediator of electromagnetic interaction).
Thus, in the present scenario we have one
extra neutral gauge boson ($Z^\prime$) compared to the SM 
as we have considered a larger symmetry group here.
This neutral gauge boson ($Z^\prime$) is known as the
``dark photon" \cite{Lee:2013fda} due to its nature of interactions
with the electrically charged fermions of the Standard Model.  
Dark photon plays an important role in this proposed two component
dark matter model as it is the main interaction mediator through
which both the dark matter candidates interact mutually.
\paragraph{}
In the present two component dark matter model, the anomalous
$\gamma$-ray excess is produced by the hadronisation processes
of the ${\rm b}$ quarks, originated only from the self annihilation
($S S{^\dagger}\rightarrow {\rm b}\bar{\rm b}$) of the dark
matter candidate $S$ at the GC. Therefore as mentioned in Ref. \cite{Daylan:2014rsa},
the mass and the ${\rm b}\bar{\rm b}$ annihilation cross section
(actually annihilation cross section times relative velocity) 
of DM component $S$ need to be in the range 30 GeV$-$40 GeV and
$\sim(1.4-2.0)\times 10^{-26}$ cm$^3/$s respectively.
In order to satisfy the above condition on ${\sigma {\rm v}}_{{\rm b}\bar{\rm b}}$,
$S$ must annihilate significantly into ${{\rm b}\bar{\rm b}}$ final state which
is possible only when $M_{S}<M_{Z^{\prime}}$ such that $SS^{\dagger} \rightarrow Z^{\prime}Z^{\prime}$
annihilation mode is kinematically forbidden. 
On the other hand, since the dark matter component $\psi$ interacts
feebly with the SM particles (through tiny mixing between $Z$ and $Z^{\prime}$),
therefore its dominant interaction channels are
$\psi \bar{\psi} \rightarrow Z^{\prime} Z^{\prime},\,\,SS^{\dagger}$.
Both of these interaction modes become kinematically inaccessible
when $M_{\psi}<M_{Z^{\prime}},\,\,M_{S}$. However for $M_{\psi}>M_{S}$
atleast one of the interaction channels ($\psi \bar{\psi} \rightarrow SS^{\dagger}$)
is possible. Thus, throughout this present work we have assumed $M_{\psi}>M_{S}$ and 
consequently, the mass ($M_{\psi}$) of the dark matter component
$\psi$ is taken in the range $60\,\,{\rm GeV}$ to $150\,\,{\rm GeV}$.
We have found that our computed $\gamma$-ray fluxes from the Galactic Centre
region for $M_{S}=35\,\,{\rm GeV}$ and two different values of $M_{\psi}$,
namely $M_{\psi}=60\,\,{\rm GeV}$ and 80 GeV,
can explain the Fermi-LAT data well as long as the quantity called the ``effective
annihilation cross section" instead of the actual annihilation cross section
($\langle {\sigma {\rm v}}_{SS^\dagger\rightarrow {\rm b}\bar{\rm b}}\rangle$)
for the annihilation channel ${SS^\dagger\rightarrow {\rm b}\bar{\rm b}}$
lies in the range $(1.52-1.67)\times10^{-26}$ cm$^3$/s. Moreover, the $\gamma$-ray
fluxes obtained for $M_{\psi}\geq 100$ GeV become incompatible with the available
experimental data. The ``effective annihilation cross section"
($\langle {\sigma {\rm v}}_{SS^\dagger\rightarrow {\rm b}\bar{\rm b}}\rangle^\prime$)
of the DM component $S$ for the annihilation channel
${SS^\dagger\rightarrow {\rm b}\bar{\rm b}}$
is defined as the product of annihilation cross section
($\langle {\sigma {\rm v}}_{SS^\dagger\rightarrow {\rm b}\bar{\rm b}}\rangle$)
of the process $SS^\dagger\rightarrow {\rm b}\bar{\rm b}$ and the square of the
fractional contribution of the DM component $S$ to the total relic density.
\paragraph{} 
In our present two component dark matter model the lighter DM component
$S$ interacts with the visible sector (SM particles) mainly through the
exchange of scalar particles such as Higgs boson $h$ and
another neutral scalar boson $H$ \footnote
{for definition of $H$ see Eq. (\ref{hH-phiphis}) of Section \ref{model}.}.
There exist few single component dark matter models
\cite{Basak:2014sza, Banik:2014eda} in the literature
where the DM candidate also possesses similar type of
interactions with the SM particles. These dark matter
models are collectively known as the ``scalar portal" dark matter model.
It is shown in these articles that for the case of low mass dark matter
candidate (DM mass $M_{\rm DM}<M_h/2$ \footnote{$M_h$ is the mass of SM Higgs boson $h$,}),
the limit on the dark matter relic density from the PLANCK experiment \cite{Ade:2013zuv},
the bounds on spin independent scattering cross section of the DM particle
from the direct detection experiments like XENON 100
\cite{Aprile:2012nq}, LUX \cite{Akerib:2013tjd} and the limit on the
invisible branching ratio ($h \rightarrow {\rm DM} {\rm DM}$)
\cite{Belanger:2013kya} of the SM Higgs boson are satisfied
simultaneously only when the mass of the additional
neutral scalar field $H$ is two times the mass of the DM particle.
This is very fine tuning situation because there exists no symmetry in the theory
which can explain why the dark matter particle mass is exactly
half of the mass of the mediator particle.  
However, in the two component dark matter scenario we do required
such type of constraint on the mass of $H$. We have found that
in order to satisfy all the above mentioned experimental and observational
bounds, the mass of $H$ has to be in the range $\sim2M_{S}$ to $2M_{S}+\Delta M$,
where $\Delta M$ is $\sim 10$ GeV for $M_{S} = 30$ GeV to 40 GeV (see right panel
of Fig. \ref{plot4} in Section \ref{results}).   
\paragraph{}
This paper is organised as follows. In Section \ref{model}, we first propose
the present two component dark matter model and then study elaborately all
the relevant constraints that can be imposed on this proposed model
from the theoretical, experimental as well as the observational
bounds. Section \ref{boltz-eqn} describes
the coupled Boltzmann equations, required for computing the individual
relic densities of each dark matter candidate and hence
the overall relic density of the dark matter in the Universe.
The results that we have obtained by solving the coupled Boltzmann equations
numerically and using various experimental, observational and theoretical constraints
are discussed in Section \ref{results}. In Section \ref{1-3gev-gamma}
we calculate the $\gamma$-ray flux for this proposed two component dark matter scenario
and is compared with the available Fermi-LAT data. We have also compared the range of
allowed values of ``effective annihilation cross section" of $S$ for
 ${\rm b} \bar{\rm b}$ annihilation channel with the limits reported by
the Fermi-LAT and DES collaborations from their analyses using data
obtained from various dwarf spheroidal galaxies.  
Finally, in Section \ref{summary} we summarise our work. 
%%%%%%%%%%%%%%%%%%%%%%%%%%%%%%%%%%%%%%%%%%%%%%%%%%%%%%%%%%%%%%%%%%%%%%
\section{The Model}
\label{model}
%%%%%%%%%%%%%%%%%%%%%%%%%%%%%%%%%%%%%%%%%%%%%%%%%%%%%%%%%%%%%%%%%%%%%%
We propose a two component dark matter model where the dark sector is composed
of a complex scalar ($S$) and a Dirac fermion ($\psi$), both of which are singlets
under the Standard Model gauge group ${\rm SU(3)_{c}}\times{\rm SU}(2)_{\rm L}
\times {\rm U}(1)_{\rm Y}$.
Thus, in the scalar sector of the model we have the usual
Higgs doublet $\Phi$ and two complex SM gauge
singlets $S$, $\Phi_s$. Note that the present model is an extension of the Standard Model
of particle physics in all three sectors namely, gauge, fermionic as well as scalar sector.
The gauge symmetry group of the SM is enhanced by an additional
local $\rm{U}(1)_{X}$ symmetry under which all particles except
the SM particles (including the Higgs doublet $\Phi$) transform
nontrivially. Among the three complex scalars ($\Phi$, $\Phi_s$, $S$)
present in this model only two, namely $\Phi$ and $\Phi_s$ acquire VEVs.
Consequently, the local gauge symmetry
${\rm SU}(2)_{\rm L} \times {\rm U}(1)_{\rm Y}\times
{\rm U}(1)_{\rm X}$ is spontaneously broken which gives rise
to one extra massive neutral gauge boson ${\rm Z^{\prime}}$
in addition to four SM gauge bosons namely ${\rm W}^{\pm},\,\,{\rm Z}, \,\, A$.
After the spontaneous breaking of local ${\rm SU}(2)_{\rm L}
\times {\rm U}(1)_{\rm Y}\times {\rm U}(1)_{\rm X}$ symmetry the
Lagrangian of this model is left with a residual U(1)$_{\rm em}\times \mathbb{Z}_{2}$
symmetry under which only the dark sector particle $S$ and $\psi$
are odd. Now if we consider a renormalisable Lagrangian (terms upto mass dimension
four), the field $\psi$ becomes automatically stable as it is the only
fermionic field which transforms nontrivially under U(1)$_{\rm X}$. However,
by choosing proper U(1)$_{\rm X}$ charges ($Q_{\rm X}$) among the dark sector particles
($\psi$, $S$) and $\Phi_s$, $\psi$ can be made stable for the terms whose
mass dimension exceed four. This requires $Q_{\rm X}(\psi) \neq Q_{\rm X}(S)$
such that the stability violating terms of $\psi$ (decay terms)
arising from the higher dimensional effective operators (suppressed by
some high energy scale $\Lambda$) are either totally forbidden from
symmetry argument or suppressed by higher powers of $\Lambda$.
The assigned gauge charges and VEVs of
all the fields present in this model are given below in a
tabular form (see Table \ref{tab1}). Moreover, the other dark sector particle
$S$ becomes stable by application of the residual $\mathbb{Z}_2$ symmetry.    

\begin{table}[h!]
\begin{center}
\vskip 0.5cm
\begin{tabular} {|c|c|c|c|c|c|}
\hline
{\bf Field} & $\mathbf{{\rm SU(2)_{L}}}$ & $\mathbf{{\rm U(1)_{Y}}} $&
$\mathbf{{\rm U(1)_{X}}}$& {\bf VEV}\\
& {\bf charge} & {\bf charge} & {\bf charge}&\\
\hline
{$\Phi$} & 2 &$\frac{1}{2}$& 0& $v$\\
\hline
{$\Phi_s$} & 1 &$0$& $\frac{1}{2}$& ${v_s}$\\
\hline
{$S$} & 1 &$0$& $2$& $0$\\
\hline
{$\psi$} & 1 &$0$& $1$& $0$\\
\hline
{\rm $l_L$}& 2& -$\frac{1}{2}$& 0 & 0\\
\hline
{\rm $Q_L$}& 2& $\frac{1}{6}$& 0 & 0\\
\hline
{\rm $e_R$}& 1& -$1$& 0 & 0\\
\hline
{\rm $u_R$}& 1& $\frac{2}{3}$& 0 & 0\\
\hline
{\rm $d_R$}& 1& -$\frac{1}{3}$& 0 & 0\\
\hline
\end{tabular}
\end{center}
\caption{${\rm SU}(2)_{\rm L} \times {\rm U}(1)_{\rm Y} \times {\rm U}(1)_{\rm X}$
charges and VEVs of all fields (including SM fermions) involved in the
present model, where $l_L$, $Q_l$ are left handed lepton and quark doublet
while $e_R$, $u_R$ and $d_R$ represent right handed charge lepton, up type and
down type quark respectively.}
\label{tab1}
\end{table} 

The Lagrangian of the present model is then given by,
\begin{eqnarray}
\mathcal{L} \supset \mathcal{L}_{\rm gauge} + \mathcal{L}_{\rm fermion} +
\mathcal{L}_{\rm scalar} \,\,,
\label{lagrangian}
\end{eqnarray}
where $\mathcal{L}_{\rm gauge}$ is the Lagrangian of the gauge fields corresponding
to the gauge group U(1)$_{\rm Y}$ and U(1)$_{\rm X}$.
\begin{eqnarray}
\mathcal{L}_{\rm gauge} &=& -\frac{1}{4} \hat{B}_{\mu \nu} \hat{B}^{\mu \nu}
- \frac{1}{4} \hat{X}_{\mu \nu} \hat{X}^{\mu \nu} +
\frac{\chi}{2} \hat{X}_{\mu \nu} \hat{B}^{\mu \nu} \,,
\label{l-gauge}
\end{eqnarray}
with
\begin{eqnarray}
\hat{B}_{\mu \nu} &=& \partial_{\mu} \hat{B}_\nu - \partial_{\nu}
\hat{B}_{\mu} \,\,\,\, {\rm and} \,\,\, \hat{X}_{\mu \nu} = 
\partial_{\mu} \hat{X}_\nu - \partial_{\nu} \hat{X}_{\mu} \,\,.
\label{field-tensor}
\end{eqnarray}
%Here 
%$\hat{B}_{\mu \nu} = \partial_{\mu} \hat{B}_\nu -
%\partial_{\nu} \hat{B}_{\mu} \,\,\,\, {\rm and}
%\,\,\, \hat{X}_{\mu \nu} = \partial_{\mu} \hat{X}_\nu - \partial_{\nu} \hat{X}_{\mu} \,\,.$
In the above Eq. (\ref{field-tensor}) $B_{\mu}$ and $X_{\mu}$
represent gauge fields corresponding
to the unitary gauge groups U(1)$_{\rm Y}$ and U(1)$_{\rm X}$ respectively.
Hat notations on the gauge fields indicate that kinetic terms of $B_{\mu}$
and $X_{\mu}$ are not diagonal. The coefficient of kinetic
mixing term between the two U(1) gauge fields in Eq. (\ref{l-gauge})
is denoted by $\chi = \frac{\epsilon}{\cos \theta_{\rm w}}$  which is experimentally
constrained to be very small. Consider a GL(2, R) rotation from the basis $\hat{B}_{\mu},
\hat{X}_{\mu}$ $\rightarrow$ $B_{\mu}, X_{\mu}$ in such a way that with respect to 
new basis kinetic mixing term vanishes.
\begin{eqnarray}
\left(\begin{array}{c} B_{\mu} \\ X_{\mu}\end{array}\right) = \left(\begin{array}{cc}
1 &-\chi\\ 0 &\sqrt{1-\chi^2}\end{array}\right)\left(\begin{array}{c} \hat{B}_{\mu} \\
\hat{X}_{\mu}\end{array}\right)\,\,.
\end{eqnarray}
After such rotation, 
\begin{eqnarray} 
\hat{B}_{\mu} \simeq B_{\mu} + \chi X_{\mu} \,\,\,\,
{\rm and}\,\,\,
\hat{X}_{\mu} \simeq  X_{\mu} \,\, .
\label{hat-unhat-trans}
\end{eqnarray}
Since $\chi \ll$ 1 \cite{Gopalakrishna:2008dv, Davoudiasl:2012ag, Lee:2013fda},
we have ignored $\mathcal{O}(\chi^2)$ terms in the above
equation (Eq. (\ref{hat-unhat-trans})). 
Spontaneous breaking of ${\rm SU}(2)_{\rm L}\times {\rm U}(1)_{\rm Y}\times{\rm U(1)}_{\rm X}$
symmetry by VEVs of the neutral components
of the scalar doublet $\Phi$ (CP even part) and the
complex singlet scalar $\Phi_s$ respectively (see Table. \ref{tab1}),
results in a $3\times 3$ mass square mixing matrix between
the three neutral gauge bosons namely ${W_3}_{\mu}$, $B_{\mu}$, $X_{\mu}$. 
\begin{eqnarray}
\mathcal{M}^2_{\rm gauge} = \frac{v^2}{4}\left(\begin{array}{ccc}
g^2 &-g g^\prime & -\chi g g^\prime  \\
-g g^\prime & {g^\prime}^2 & \chi {g^\prime}^2\\
-\chi g g^\prime  & \chi {g^\prime}^2  & g^2_{\rm X} \left(\frac{v_s}{v}\right)^2
\end{array}\right)\,\, .
\label{gauge-mixing}
\end{eqnarray}
After diagonalising this mass square mixing matrix by an orthogonal
matrix $O(\theta_{\rm NB}, \theta_{\rm W})$ we obtain three physical neutral gauge
fields which are denoted by $Z_{\mu}$, $A_{\mu}$ and ${Z^\prime}_{\mu}$. The
eigenstates of the matrix $\mathcal{M}^2_{\rm gauge}$ namely,
$Z_{\mu}$, $A_{\mu}$ and ${Z^\prime}_{\mu}$ are
linearly related to ${W_3}_{\mu}$, $B_{\mu}$, $X_{\mu}$ by an
orthogonal transformation which is given by,
\begin{eqnarray}
\hspace{2cm}\left(\begin{array}{c} Z_{\mu} \\A_{\mu} \\{Z^\prime}_{\mu}\end{array}\right)
&=& O(\theta_{\rm NB}, \theta_{\rm W})
\left(\begin{array}{c} W^3_{\mu}\\ B_{\mu}\\ X_{\mu}\end{array}\right)\,\, ,  
\end{eqnarray}
with
\begin{eqnarray}
O(\theta_{\rm NB}, \theta_{\rm W}) &=&
\left(\begin{array}{ccc}
\cos\theta_{\rm NB} \cos\theta_{\rm W} &-\cos\theta_{\rm NB}
\sin\theta_{\rm W}& -\sin\theta_{\rm NB}\\
\sin\theta_{\rm W}&\cos\theta_{\rm W} & 0\\
\sin\theta_{\rm NB} \cos\theta_{\rm W}&-\sin\theta_{\rm NB}
\sin\theta_{\rm W}&\cos\theta_{\rm NB}
\end{array}\right)\,\, ,
\label{u-matrix}
\end{eqnarray} 
where $\theta_{\rm W}$ and $\theta_{\rm NB}$ are the usual weak mixing angle and the
mixing angle between two neutral gauge bosons $Z$ and $Z^\prime$ respectively.
The expressions of $\theta_{\rm W}$ and $\theta_{\rm NB}$ are given by,
\begin{eqnarray}
\theta_{\rm W} = \tan^{-1}\left(\frac{g^\prime}{g}\right)\ , \ \ \ \ \ 
\theta_{\rm NB} = \frac{1}{2}\,\tan^{-1}\left(\frac{2 \epsilon \tan\theta_{\rm W}}
{1 - \frac{g^2_{\rm X}}{{g^2 + {g^\prime}^2}} \frac{v^2_s}{v^2}}\right) \,\, .
\label{gauge-mix-angle}
\end{eqnarray}
Among the three physical neutral gauge bosons one remains
massless which is identified as the `photon'. The masses
of other two neutral bosons namely $Z$ and $Z^\prime$ are given by,
\begin{eqnarray}
M_{Z} &=& \sqrt{\frac{g^2_{\rm Z} v^2 + g^2_{\rm X} v^2_s}{8} + \frac{1}{8}
\sqrt{(g^2_{\rm Z} v^2 - g^2_{\rm X} v^2_s)^2 +
4 (g^\prime g_{\rm Z}v^2 \chi)^2} }\ ,\nonumber \\
M_{Z^\prime} &=& \sqrt{\frac{g^2_{\rm Z} v^2 + g^2_{\rm X} v^2_s}{8} - \frac{1}{8}
\sqrt{(g^2_{\rm Z} v^2 - g^2_{\rm X} v^2_s)^2 + 4 (g^\prime g_{\rm Z}v^2 \chi)^2} }\ ,
\label{Z-Z'mass}
\end{eqnarray}
where
\begin{eqnarray}
g_{\rm Z} &=& \sqrt{g^2 + {g^\prime}^2} \ , \nonumber
\end{eqnarray}
using the condition $\chi \ll 1$ as mentioned before, Eq. (\ref{Z-Z'mass}) reduces to
\begin{eqnarray}
M^2_{Z} &\simeq & \frac{g^2_{\rm Z} v^2}{4}\ , \nonumber \\
M^2_{Z^\prime} &\simeq & \frac{g^2_{X} v^2_s}{4} \ .
\label{Z-Z'approx-mass}
\end{eqnarray}
In Eq. (\ref{lagrangian}), $\mathcal{L}_{\rm fermion}$ refers to the Lagrangian of
the singlet Dirac fermion $\psi$, which is given by,
\begin{eqnarray}
\mathcal{L}_{\rm fermion} = \bar{\psi}(i {{D\!\!\!\!\slash}_\psi} - M_{\psi})\psi \ ,
\end{eqnarray}
where the covariant derivative ${{D\!\!\!\!\slash}_\psi}$ of the field $\psi$ is defined as, 
\begin{eqnarray}
{{D\!\!\!\!\slash}_\psi} \psi &=& \gamma^{\mu} D_{\mu} \psi \, , \nonumber \\
&=& \gamma^\mu \left(\partial_{\mu} + i g_{\rm X}X_{\mu}\right)\psi \ . 
\end{eqnarray} 

The scalar sector Lagrangian $\mathcal{L}_{\rm scalar}$ (in Eq. (\ref{lagrangian})) of
the present model has the following form
\begin{eqnarray}
\mathcal{L}_{\rm scalar} &=& ({D_{\phi}}_{\mu} \Phi)^\dagger ({D_{\phi}}^{\mu} \Phi) + 
({D_{\phi_s}}_{\mu} \Phi_s)^\dagger ({D_{\phi_s}}^{\mu} \Phi_s) +
({D_{s}}_{\mu} S)^\dagger ({D_{s}}^{\mu} S) \nonumber \\
&& -~V(\Phi, \Phi_s, S)\,\,\,,
\end{eqnarray}
with
\begin{eqnarray}
V(\Phi, \Phi_s, S) &=& \mu^2 (\Phi^\dagger\Phi) + \lambda (\Phi^\dagger\Phi)^2 +
\mu^2_s (\Phi_s^\dagger\Phi_s) + \lambda_s (\Phi_s^\dagger\Phi_s)^2
+ \rho^2 (S^\dagger S) \nonumber \\ && + \kappa (S^\dagger S)^2 
 + \lambda_1 (\Phi^\dagger\Phi)(\Phi_s^\dagger\Phi_s) +
\lambda_2 (\Phi^\dagger\Phi)(S^\dagger S) \nonumber \\
&&+ \lambda_3 (\Phi_s^\dagger \Phi_s)(S^\dagger S) \,\, ,
\label{scalar-potential}
\end{eqnarray}
where 
\begin{eqnarray}
{D_{\phi}}_{\mu}\Phi &=& \left(\partial_{\mu} + i \frac{g}{2}
\sigma^a {W_a}_{\mu } + i \frac{g^\prime}{2} (B_{\mu} +
\chi X_{\mu})\right)\Phi \ , \nonumber \\
{D_{\phi_s}}_{\mu} \Phi_s &=& \left(\partial_{\mu} +
i \frac{g_{\rm X}}{2} X_{\mu} \right)\Phi_s \ , \nonumber \\
{D_{s}}_{\mu} S &=& \left(\partial_{\mu} +
i 2\,g_{\rm X}\,X_{\mu} \right)S \ , 
\end{eqnarray}
are the covariant derivatives of the scalar doublet $\Phi$ and
two complex scalar singlets $\Phi_s$, $S$ respectively.
Similar to the gauge sector, the scalar sector also exhibits mixing between
two real scalars namely $\phi^0$ (neutral CP even part of the doublet $\Phi$)
and $\phi^0_s$ (real part of the complex scalar $\Phi_s$) after spontaneous breaking
of gauge symmetry (${\rm SU}(2)_{\rm L}\times {\rm U}(1)_{\rm Y}\times{\rm U}(1)_{\rm X}$). 
The mass square mixing matrix between these two real scalars are given by,
\begin{eqnarray}
\mathcal{M}^2_{\rm scalar} = \left(\begin{array}{cc}
2\lambda v^2 ~~&~~ \lambda_1 v_s v \\
~~&~~\\
\lambda_1 v_s v ~~&~~ 2 \lambda_s v^2_s
\end{array}\right) \,\,.
\end{eqnarray}
Diagonalising $\mathcal{M}^2_{\rm scalar}$ by an orthogonal matrix $O(\alpha)$, we obtain
two real physical scalars namely $h$ and $H$. The old basis states ($\phi^0,~\phi^0_s$)
and the eigenstates ($h,~H$) of the matrix $\mathcal{M}^2_{\rm scalar}$ are
linearly related by the orthogonal matrix $O(\alpha)$ which is given by,
\begin{eqnarray}
\left(\begin{array}{c} h \\ H\end{array}\right)
=\left(\begin{array}{cc}\cos\alpha ~-\sin\alpha
\\ \sin\alpha ~~~~\cos\alpha \end{array}\right)
\left(\begin{array}{c} \phi^0 \\ \phi^0_s\end{array}\right) \,\, .
\label{hH-phiphis}
\end{eqnarray}
The mixing angle $\alpha$ and the masses of the physical real scalars $h$ and $H$ are 
\begin{eqnarray}
\alpha &=& \frac{1}{2}~\tan^{-1}\left(\frac{\frac{\lambda_1}{\lambda_s}\frac{v}{v_s}}
{1 - \frac{\lambda}{\lambda_s}\frac{v^2}{v^2_s}}\right) \,\, ,\\
M_{h} &=& \sqrt{\lambda v^2 + \lambda_s v^2_s + 
\sqrt{(\lambda v^2 - \lambda_s v^2_s)^2 + (\lambda_1 v v_s)^2} }\ ,\nonumber \\
M_{H} &=& \sqrt{\lambda v^2 + \lambda_s v^2_s - 
\sqrt{(\lambda v^2 - \lambda_s v^2_s)^2 + (\lambda_1 v v_s)^2} } \,\ .
\end{eqnarray}
Between these two real scalars, $h$ plays the role of SM Higgs boson.
The mass term of the scalar field $S$ which does not mix with 
other components of the scalar sector is given by,
\begin{eqnarray}
M_{S} = \sqrt{\rho^2 + \frac{\lambda_2 v^2}{2} + \frac{\lambda_3 v^2_s}{2}}.
\end{eqnarray}

Both the fermionic field $\psi$ and complex scalar field $S$ remain
decoupled from the visible sector of the model. Thus, they can be 
viable components of dark matter. From the above discussion it is
evident that the present scenario involves 10 unknown parameters
namely, masses of two dark matter components namely $M_S$ and $M_{\psi}$,
mass of one neutral scalar ($M_{H}$)\footnote{as we identify one of
the neutral scalars namely $h$ with the SM Higgs boson, therefore
its mass is fixed at $\sim 125.5$ GeV \cite{Aad:2012tfa, Chatrchyan:2012ufa}.},
mass of extra neutral gauge boson ($M_{Z^\prime}$), the gauge
coupling $g_{\rm X}$ corresponding to the gauge group U(1)$_{\rm X}$, neutral scalars
mixing angle ($\alpha$), coefficient of kinetic mixing term ($\epsilon$),
quartic self coupling ($\kappa$)\footnote{Through out this work we have kept
$\kappa$ fixed at 0.1} of $S$ and other two quartic couplings
$\lambda_2$, $\lambda_3$ of $\Phi$ between $S$ and $\Phi_s$
respectively. The allowed ranges of these parameters will
be restricted by imposing both experimental, observational
as well as theoretical bounds mentioned below.  
\begin{itemize}
\item {\bf Vacuum Stability}\,- In order to obtain a stable vacuum, the scalar 
potential $V(\Phi, \Phi_s, S)$ (Eq. (\ref{scalar-potential})) of the
present model must be bounded from below. This will be maintained
if the following conditions are satisfied,
\begin{eqnarray}
&&\lambda \geq 0, \lambda_s \geq 0, \kappa \geq 0,\nonumber \\
&&\lambda_1 \geq - 2\sqrt{\lambda\,\lambda_s},\nonumber \\
&&\lambda_2 \geq - 2\sqrt{\lambda\,\kappa},\nonumber \\
&&\lambda_3 \geq - 2\sqrt{\lambda_s\,\kappa},\nonumber \\
&&\sqrt{\lambda_1+2\sqrt{\lambda\,\lambda_s}}\sqrt{\lambda_2+2\sqrt{\lambda\,\kappa}}
\sqrt{\lambda_3+2\sqrt{\lambda_s\,\kappa}} \nonumber \\ 
&&+ 2\,\sqrt{\lambda \lambda_s \kappa} + \lambda_1 \sqrt{\kappa}
+ \lambda_2 \sqrt{\lambda_s} + \lambda_3 \sqrt{\lambda} \geq 0 \,\,\,\,.
\end{eqnarray}
\item {\bf Zero VEV of ${S}$}\,- In the present scenario we assume that one among
the three scalars, namely $S$ does not possess any vacuum expectation value.
The VEV of other two scalars are $v$ and $v_s$ respectively (see Table \ref{tab1}). Hence the
ground state of the model is ($v,\,v_s,\,0$) which requires
\begin{eqnarray}
\mu^2 < 0, ~~~~~\mu^2_s < 0~~~~~~{\rm and}~~~~~\rho^2 > 0\,\,.
\end{eqnarray}
 
\item {\bf PLANCK Limit}\,- The total relic density ($\Omega_{\rm T} h^2$) of the dark matter
components must lie within the range \cite{Ade:2013zuv} specified by
the PLANCK experiment. The PLANCK limit for the relic density of
the dark matter in the Universe is
\begin{eqnarray}
~~~~~~~~~~~~~~~~~~0.1172 < \Omega_{\rm DM} h^2 < 0.1226~~~{\rm at}~~68\%~~{\rm C.L.}
\label{planck-limit}
\end{eqnarray}
\item {\bf Limits from Dark Matter Direct Detection Experiments}\,-
In our two component dark matter model
both the dark matter components namely $S$ and $\psi$ can interact
with detector nuclei placed at various underground laboratories.
The component $\psi$ scatters off the detector nuclei only via
exchange of $Z$ and $Z^{\prime}$ bosons, while for the other
component $S$, the dominant contribution comes mainly through
the exchange of scalar particles such as SM like Higgs boson ($h$)
and $H$. Fig. \ref{direct-detec} shows the Feynman diagrams
for the scattering of both the dark matter components with
the detector nucleon ($N$).
\begin{figure}[h!]
\centering
\includegraphics[height=4cm,width=6cm]{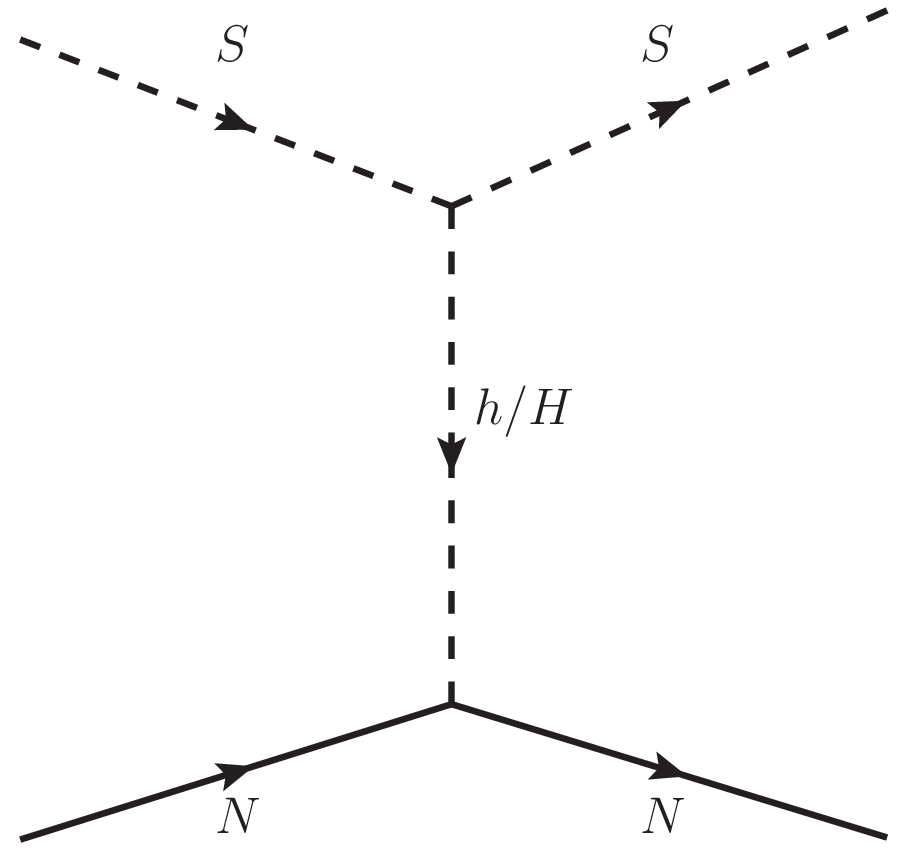} ~~~~~~~~~~~~~~~~~
\includegraphics[height=4cm,width=6cm]{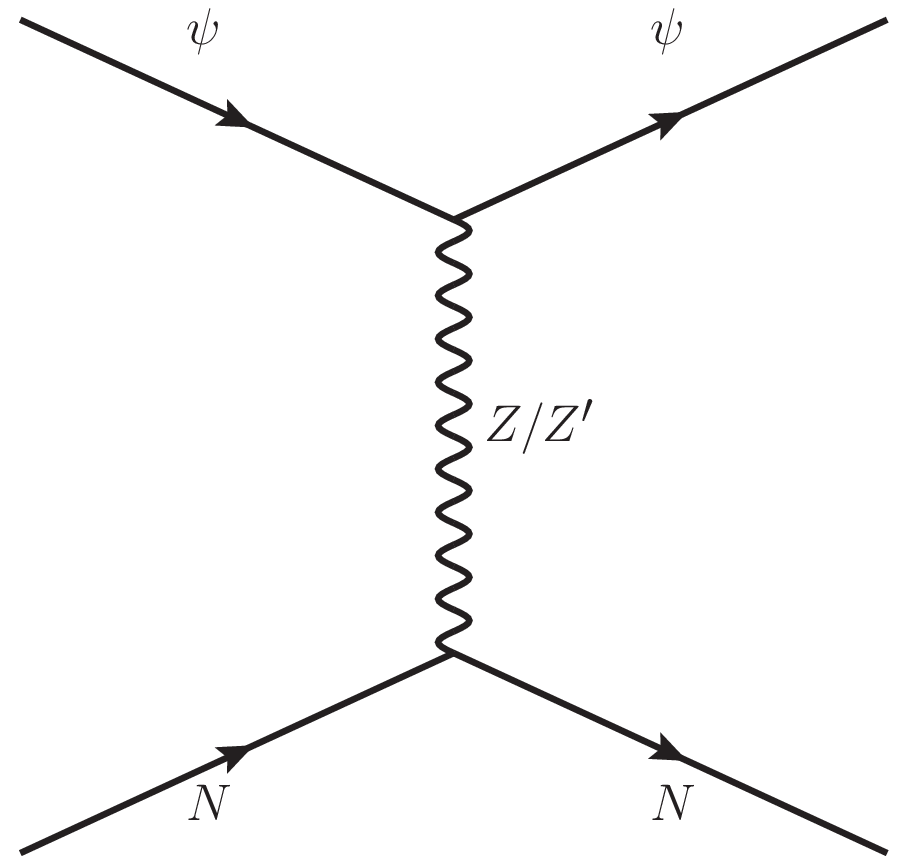}
\caption{Feynman diagrams for the elastic scattering between both the
dark matter candidates $S$, $\psi$ and the nucleon $N$ of the detector material.}
\label{direct-detec}
\end{figure}
The spin independent scattering cross section between the dark matter component $\psi$
and nucleon $N$ is given by \cite{Jia:2013lza},
\begin{eqnarray}
\sigma^{\psi N \rightarrow \psi N}_{\rm SI} \simeq \frac{\sqrt{2}G_{\rm F}
M^2_{Z}~g_{\rm X}^2~Q_{\rm X}(\psi)^2 \sin^2\theta_{\rm NB} \cos^2\theta_{\rm NB}
~\mu_{\psi N}^2~F^z_1}{16\pi}\left(\frac{1}{M^2_{Z}}-\frac{1}
{M^2_{Z^\prime}} \right)^2 \,,\nonumber \\
\label{psiN-scat} 
\end{eqnarray}
where   
\begin{eqnarray*}
\mu_{\psi N} = \frac{M_{\psi}M_N}{(M_{\psi} + M_N)} \ 
\end{eqnarray*}
is the reduced mass between $\psi$ and $N$, $F^z_1 = -0.5$ is the form factor for neutron
and $Q_{\rm X}(\psi) = {1}$ is the U(1)$_{\rm X}$ charge of $\psi$.
The expression of spin independent scattering cross section for the process 
$S N \rightarrow S N$ is given by, 
\begin{eqnarray}
\sigma^{SN\rightarrow SN}_{\rm SI} \simeq
\frac{\mu^2_{SN}}{4\pi}\left(\frac{M_N f}{M_S v}\right)^2
\left(\frac{g_{SSh} \cos \alpha}{M^2_h} + \frac{g_{SSH} \sin \alpha}{M^2_H}\right)^2 \ ,
\label{SN-scat}
\end{eqnarray}
with
\begin{equation*}
\mu_{S N} = \frac{M_{S}M_N}{(M_{S} + M_N)} \ , 
\end{equation*}
where
\begin{eqnarray}
g_{SSh} &=& -\left(\lambda_2 v \cos \alpha - \lambda_3 v_s \sin \alpha \right)\ , \nonumber \\
g_{SSH} &=& -\left(\lambda_2 v \sin \alpha + \lambda_3 v_s \cos \alpha \right)\ 
\label{gssh}
\end{eqnarray}
are the couplings for the vertex $SSh$ and $SSH$ respectively. $f$ is the relevant form
factor. Throughout this work, we have adopted the value of $f=0.3$ \cite{Mambrini:2011ik,
Giedt:2009mr, Barbieri:2006dq}.

The Dark sector in our model is composed of two kinds of particles,
one of which is scalar (however, it has different antiparticle\footnote{particle and
antiparticle possess different U(1)$_{\rm X}$ charges.} (complex scalar))
while the other component is fermionic (Dirac fermion) in nature.
Their masses as well as the contributions to the total relic density of
the dark matter are different in general. Consequently the number densities
$n_{S}$, $n_{\psi}$ for these two components at the present epoch are also different.
Therefore when we compare the spin independent scattering cross sections
computed for both the dark matter components ($S$ and $\psi$)
within the framework of the present two component DM model 
with the available experimental data (exclusion plots from various ongoing dark
matter direct detection experiments such as LUX \cite{Akerib:2013tjd},
XENON-100 \cite{Aprile:2012nq} etc.),
one has to keep in mind that the exclusion plots are computed with the assumption
that all the dark matter present in the Universe are same in nature {\it i.e.}
interaction rates with the detector nuclei are same for all the dark matter particles.
However, this assumption is certainly not true in our case as we have two component
dark matter scenario. Hence we need to rescale both
$\sigma^{\psi N\rightarrow \psi N }_{\rm SI}$ and
$\sigma^{S N\rightarrow S N }_{\rm SI}$ (Eqs. (\ref{psiN-scat})-(\ref{SN-scat}))
by appropriate factors consistent with the present consideration
that we have two types of dark matter in the Universe.
We define the spin independent ``effective scattering cross section"
between the detector nucleon $N$ and the dark matter component $i$ as,
\begin{eqnarray}
{{\sigma^\prime}^{\,i N\rightarrow i N }_{\rm SI}} = \frac{n_i}{n_i + n_j}
{\sigma^{i N\rightarrow i N }_{\rm SI}} \, , 
\label{effective-scatter}
\end{eqnarray}
with $i, j = S, \psi$ and $i\neq j$, $n_i$ is the number
density of the dark mater component $i$ at the present epoch.
For a viable two component dark matter model
it is desirable that 
\begin{eqnarray}
{{\sigma^\prime}^{\,i N\rightarrow i N }_{\rm SI}} < \sigma^{Exp}_{\rm SI} (M_i)\ .
\end{eqnarray}
In the above $M_i$ is the mass of the $i^{th}$ type dark matter and
$\sigma^{Exp}_{\rm SI} (M_i)$ is the experimental upper bound for the spin
independent scattering cross section between dark matter particle of mass $M_i$
and the nucleon $N$. In this work we have used results only from LUX experiment
for constraining the relevant parameter space of this model
since LUX imposes strongest limits (exclusion plot)
in $\sigma_{\rm SI}-M_{\rm DM}$ plane until now.
\item {\bf Precision electroweak observable}\, - We have already shown earlier,
in this section, that the dark photon $Z^{\prime}$ possesses a nonzero
mixing with the SM gauge boson $Z$ due to the presence of a kinetic mixing term
in the Lagrangian (Eq. (\ref{l-gauge})) between the hypercharge gauge boson
$\hat{B}_{\mu}$ and the U(1)$_{\rm X}$ gauge boson $\hat{X}_{\mu}$. As a
result of this nonzero $Z-Z^{\prime}$ mixing, the values of electroweak precision
observables, namely the mass ($M_{Z}$) and the decay width ($\Gamma_{Z}$)
of $Z$ boson, the rho parameter ($\rho$) as well as the electroweak
oblique parameters ($S$, $T$, $U$ parameters or Peskin Takeuchi parameters
\cite{Peskin:1991sw}), get shifted from their SM predictions. Therefore,
the experimentally measured values of these parameters constrain the allowed
ranges of kinetic mixing parameter $\epsilon$ and the mixing angle $\theta_{\rm NB}$. 
Thus while computing the allowed range for the $Z-Z^{\prime}$ mixing angle
$\theta_{\rm NB}$ we have considered the following constraints,
\begin{itemize}
\item[•] The deviation of the physical $Z$ boson mass from its SM value
should remain within the range given by \cite{Agashe:2014kda}, \cite{Hook:2010tw}
\begin{eqnarray}
\frac{M_{Z}-M_{Z_0}}{M_{Z_0}}\leq 2.4\times 10^{-5} \,\,,
\end{eqnarray}  
where $M_{Z_0}$ is the mass of $Z$ boson predicted from the SM of particle physics. 
\item[•] The nonzero mixing angle $\theta_{\rm NB}$ between $Z$
and $Z^\prime$ allows the physical $Z$ boson to decay into the final
state comprised of particle and anti-particle of dark matter candidate $S$.
This additional decay mode is known as the non-Standard invisible decay
channel of $Z$ \footnote{In this present model SM $Z$ boson can also decay into a
final state containing $\psi$ and $\bar{\psi}$ through its nonzero mixing
with $Z^{\prime}$. This decay channel is also another non-standard
decay mode of the SM $Z$ boson. However, this decay mode is kinemetically 
forbidden for the entire adopted range of $M_{\psi}$.}.
Therefore, the total decay width
of $Z$ boson is
\begin{eqnarray} 
\Gamma^{\rm tot}_{Z} =\sum_{f}
\Gamma_{Z\rightarrow f\bar{f}}+\Gamma_{Z\rightarrow S S^\dagger}\,\,,
\end{eqnarray}
where $f$ is any SM model fermion except $t$ quark.
In order to satisfy the precisely measured decay width of Z boson,
$\Gamma_Z = 2.4952 \pm 0.0023$ GeV \cite{Agashe:2014kda}, as reported by the
LEP \cite{ALEPH:2005ab}, the non-Standard invisible
decay width ($\Gamma_{Z\rightarrow S S^\dagger}$)
of $Z$ boson should obey the following limit
\begin{eqnarray}
\Gamma_{Z\rightarrow S S^\dagger} \leq 2.3\,\,\,{\rm MeV}\,\,.
\end{eqnarray}  
The expressions of $\Gamma_{Z\rightarrow f\bar{f}}$ and
$\Gamma_{Z\rightarrow S S^\dagger}$ are given by
\begin{eqnarray}
\Gamma_{Z\rightarrow f\bar{f}} &=& \frac{n_c\,g^2 M_{Z}}{48\pi \cos^2 \theta_{\rm W}}
(a^2_f + b^2_f)\bigg(1+\frac{2(a^2_f-2 b^2_f)}{(a^2_f+b^2_f)}
\frac{M^2_f}{M^2_{Z}}\bigg) \sqrt{1-\frac{4 M^2_f}{M^2_Z}} \,\,, 
\label{zff}\\
\Gamma_{Z\rightarrow S S^\dagger} &=&
\frac{g^2_{\rm X}Q_{\rm X}(S)^2\sin^2\theta_{\rm NB} M_Z}
{48\pi}\Bigg(1-\frac{4 M^2_S}{M^2_Z}\Bigg)^{3/2},
\label{z-decay-width}
\end{eqnarray}
with
\begin{eqnarray}
a_f &=& t_3 a - 2 Q_f (a-\cos\theta_{\rm NB}\cos^2\theta_{\rm W})\,\,,\\
b_f &=& t_3 a\,\,,
\end{eqnarray}
and
\begin{eqnarray}
a &=& \cos \theta_{\rm NB}+\epsilon \tan\theta_{\rm W}\sin\theta_{\rm NB}\,\,.
\label{a}
\end{eqnarray}
In the above Eq. (\ref{z-decay-width}) $M_f$, $Q_f$ and $t_3$ are the mass,
electrical charge and isospin quantum number of the SM fermion $f$ while $Q_{\rm X}(S) = 2$
is the U(1)$_{\rm X}$ charge of the dark matter particle $S$ (Table \ref{tab1}) while
$n_c=1\,(3)$ is the colour charge of the SM lepton (quark).  
\item[•] $Z-Z^{\prime}$ mixing shifts the value of the tree level $\rho$ parameter
from its Standard Model value which is equal to one. For small value of
mixing angle ($\theta_{\rm NB}$) between $Z$ and $Z^\prime$ the deviation
of $\rho$ parameter from its SM value is given by \cite{Ko:2013zsa}
\begin{eqnarray}
\Delta \rho \simeq \epsilon \tan \theta_{\rm W}\frac{M^2_{Z_{0}}}{M^2_{Z}}
\theta_{\rm NB}\,\,.
\end{eqnarray}
Thus, in order to keep the value of $\rho$ parameter within the range
as specified in Refs. \cite{Agashe:2014kda, ALEPH:2005ab},
the allowed ranges of $\epsilon$ as well as $\theta_{\rm NB}$ will be constrained.
   
\item[•] There will be some extra contributions to the Electroweak oblique
parameters \cite{Peskin:1991sw}, mainly in $S$ and $T$ parameter, which result in from the
nonzero mixing between the SM gauge boson $Z$ and dark gauge boson $Z^{\prime}$.
The $Z-Z^{\prime}$ contributions to the $S$ and $T$ parameters for the lowest
order in $\theta_{\rm NB}$ are given in Ref. \cite{Babu:1997st}.
We have varied the relevant parameters, namely $\epsilon$, $M_{Z^{\prime}}$,
in such a way that the electroweak oblique parameters always lie within
the range as given in Refs. \cite{Baak:2014ora} 
\begin{eqnarray}
S = 0.05 \pm 0.11\,,\,\,\, T = 0.09 \pm 0.13\,,\,\,\, U = 0.01 \pm 0.11 .
\end{eqnarray}
\end{itemize}
\item {\bf Constraint from LHC results}
\begin{itemize}
\item[•] {\bf Dilepton Production}\,-
Moreover, there exists strong upper bound on the mixing angle
$\theta_{\rm NB}$ between $Z$ and $Z^{\prime}$ from ATLAS
\cite{Aad:2014cka} and CMS \cite{CMS:2013qca}
collaborations for the processes like $pp\rightarrow Z^{\prime} X \rightarrow f
\bar{f} X$ ($f=e$, $\mu$) since the fermiophobic $Z^{\prime}$
couples to the SM leptons only through the mixing angle
$\theta_{\rm NB}$. In Ref. \cite{Cline:2014dwa} the upper limits on the kinetic
mixing parameter $\chi\,\,\left(\equiv \frac{\epsilon}{\cos \theta_{\rm W}}\right)$
are given with the mass of $Z^{\prime}$ varies in the range
166 GeV$\la M_{Z^{\prime}} \la 3$ TeV. These limits are obtained
from the upper bounds on ``signal cross section times branching fraction"
$\sigma\,{\rm BR}$ for the above mentioned processes as reported by the
ATLAS collaboration \cite{Aad:2014cka}. For lighter $Z^{\prime}$, which
we are considering in this work, the upper bounds on $\chi$,
obtained from electroweak precision data (EWPD) constraints \cite{Hook:2010tw},
are also given in Ref. \cite{Cline:2014dwa}. From the Fig. 6 of Ref. \cite{Cline:2014dwa}
it is evident that the upper bound on kinetic mixing parameter $\chi$,
using EWPD as constraints, varies from $\sim 10^{-1.55}$ to $\sim 10^{-1.83}$ for
$10\,{\rm GeV}\leq M_{Z^\prime}\leq 80$ GeV.
The corresponding upper bound on the mixing angle
$\theta_{\rm NB}$ (computed using Eq. (\ref{gauge-mix-angle}))
varies in the range $\sim$ 0.01 rad to $\sim$ 0.03 rad.
\item[•] {\bf Signal strength of the SM like Higgs boson $h$} -
The signal strength ratio of Higgs boson for a particular
decay channel ($h\rightarrow X \bar{X}$) is given by,
\begin{eqnarray}
R_{X \bar{X}}~~ = \frac{\sigma}{\sigma_{\rm SM}}
\frac{{\rm BR}(h \rightarrow X \bar{X})}
{~~~~{\rm BR}(h \rightarrow X \bar{X})^{\rm SM}}\ ,
\label{signal-strenght}
\end{eqnarray}
where $X$ is any SM fermion / gauge boson.
In Eq. (\ref{signal-strenght}) $\sigma$ and BR$(h \rightarrow X \bar{X})$ are
the Higgs production cross section and the branching ratio of the decay channel
$h \rightarrow X \bar{X}$ for this present model respectively. Similar quantities
for the SM are denoted by $\sigma_{\rm SM}$ and
BR$(h \rightarrow X \bar{X})^{\rm SM}$ respectively.
In order to satisfy LHC results \cite{CMS:yva}
we take $R_{X \bar{X}} \geq 0.8$ for any SM particle $X$.
\item[•] {\bf Invisible decay width of Higgs boson ($\Gamma_{\rm inv}$)}\,-
In this present two component dark matter model, the SM like
Higgs boson $h$ can decay into final states that consist of 
$S$, $S^{\dagger}$, if the kinematical condition $M_h>2M_S$ is satisfied.
Moreover if $M_h>2M_H$, $h$ can also decay into
a pair of $H$ which subsequently decays into $S$ and $S^{\dagger}$.
Therefore, the actual invisible decay width of the SM like Higgs boson $h$ is given by
\begin{eqnarray}
\Gamma^{\rm inv}_{h} = \Gamma_{h\rightarrow S S^{\dagger}} +
\Gamma_{h\rightarrow H H}\times({\rm BR}(H\rightarrow S S^{\dagger}))^2 \,\,,
\end{eqnarray} 
with
\begin{eqnarray}
\Gamma_{h\rightarrow S S^{\dagger}} &=&
\frac{g^2_{SSh}}{16 \pi M_h}\sqrt{1- \frac{4 M^2_S}{M^2_h}} \,\,, \\
\Gamma_{h\rightarrow HH} &=&
\frac{g^2_{HHh}}{8 \pi M_h}\sqrt{1- \frac{4 M^2_H}{M^2_h}}\,\,
\end{eqnarray}
and
\begin{eqnarray}
g_{HHh} &=& -\frac{\lambda_1}{2}\,
\Bigg(v\,\cos^3 \alpha-v_s\,\sin^3 \alpha - 2\,v_s
\cos^2 \alpha\,\sin \alpha - 2\,v\sin^2 \alpha\,\cos \alpha\Bigg)
\nonumber\\&&
+3\,\lambda_s\,v_s\,\cos^2 \alpha\,\sin \alpha
-3\lambda\,v \cos \alpha\,\sin^2 \alpha  \,\,,
\end{eqnarray}
is the coupling strength for the vertex involving two $H$ and one $h$
while ${\rm BR}(H\rightarrow S S^{\dagger})$ is the branching fraction
of $H$ decaying into $SS^{\dagger}$ final state.
Throughout the work, we assume that the total invisible branching ratio
(BR$_{\rm inv}$) of the SM like Higgs boson ($h$) is less than
20\% \cite{Belanger:2013kya} of its total decay width ($\Gamma_h$).
\end{itemize} 
\end{itemize}
%%%%%%%%%%%%%%%%%%%%%%%%%%%%%%%%%%%%%%%%%%%%%%%%%%%%%%%%%%%%%%%%%%%%%%%%%%%%%%%%%%%
\section{Solution of Coupled Boltzmann Equations of two Dark Matter Components
$\psi$ and $S$}
\label{boltz-eqn}
\begin{figure}[h!]
\centering
\includegraphics[height=4cm,width=5.5cm]{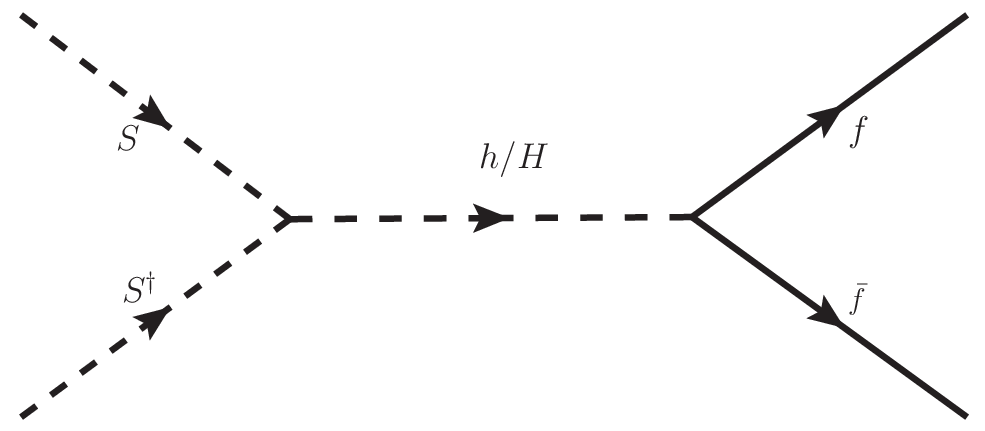}
\hspace{0.5cm}
\includegraphics[height=4.0cm,width=4cm]{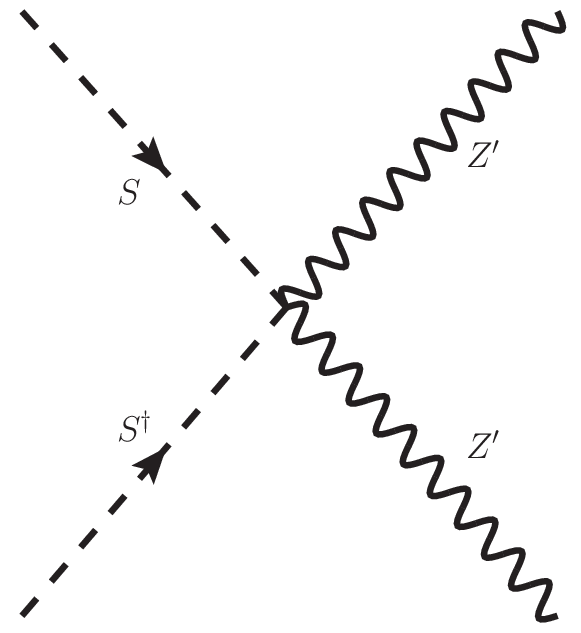}
\hspace{0.5cm}
\includegraphics[height=4.0cm,width=5.5cm]{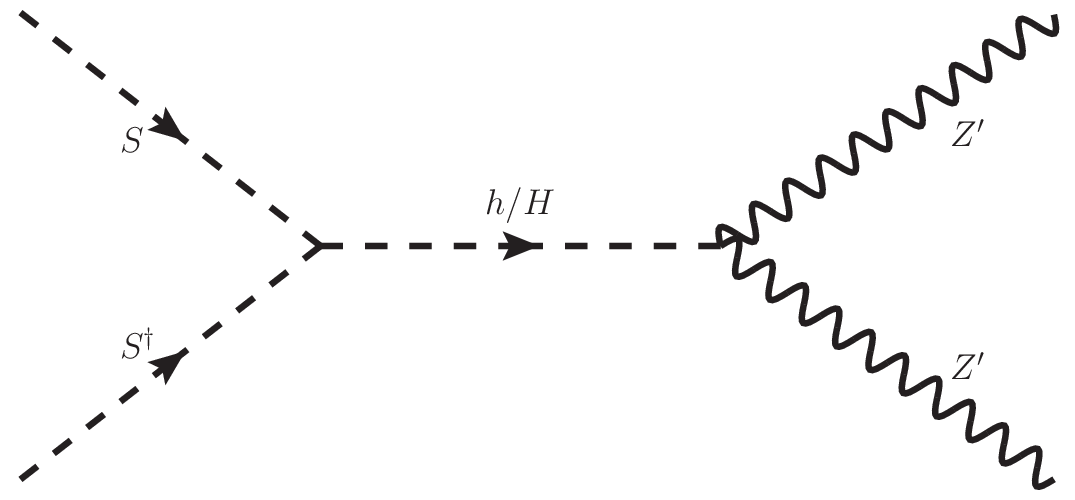}
\includegraphics[height=4.0cm,width=5.5cm]{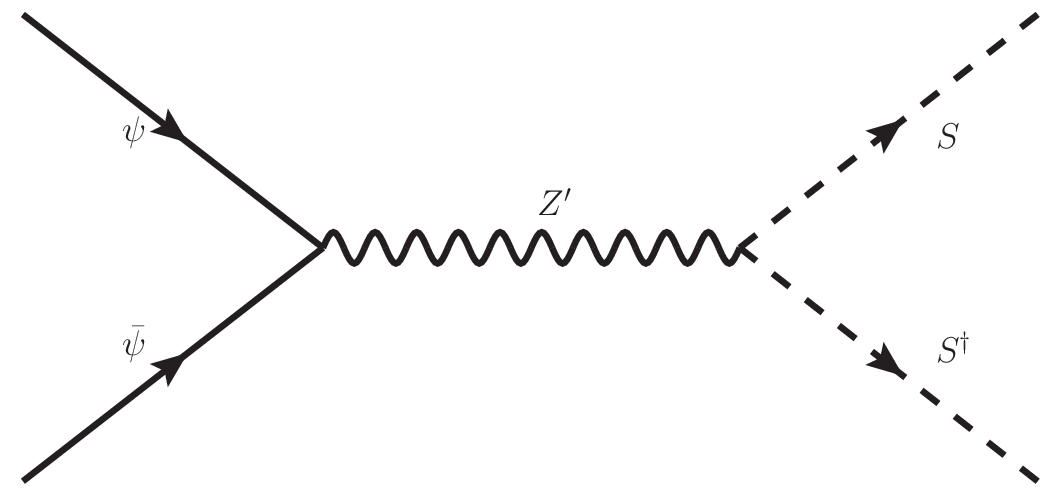}
\includegraphics[height=4.0cm,width=5.5cm]{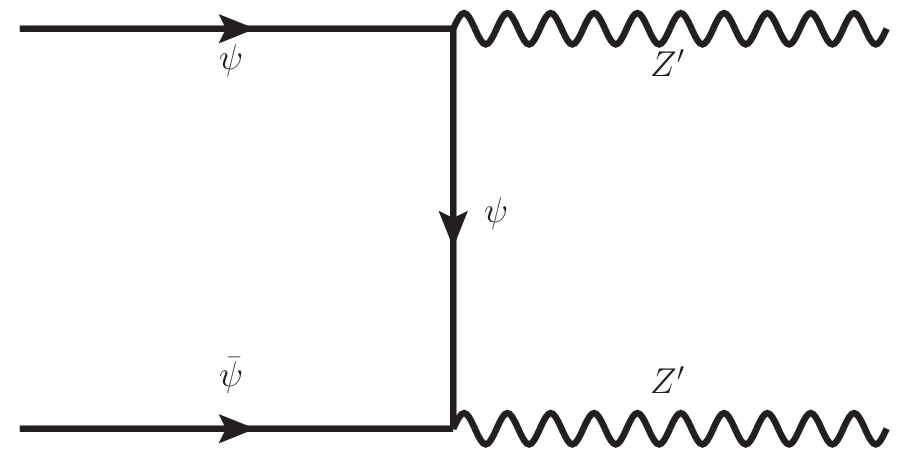}
\includegraphics[height=4.0cm,width=5.5cm]{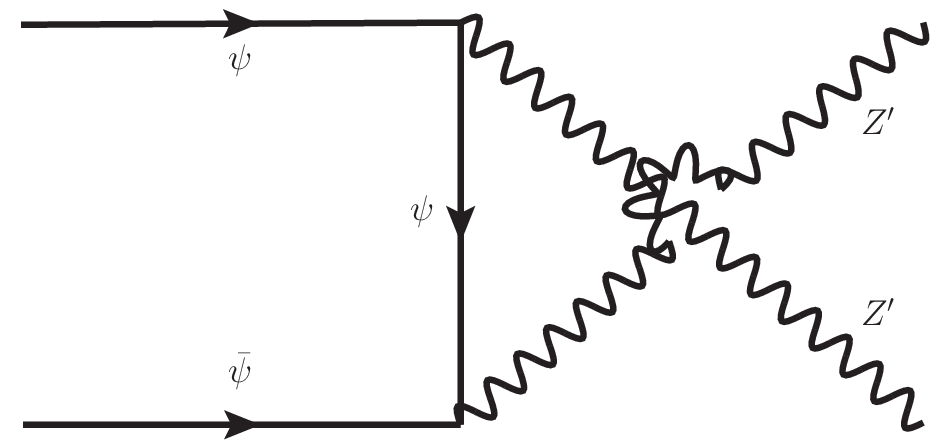}
\caption{Feynman diagrams for the dominant self annihilation processes of
both the dark matter components $S$, $\psi$.}
\label{anni-dia}
\end{figure}

%%%%%%%%%%%%%%%%%%%%%%%%%%%%%%%%%%%%%%%%%%%%%%%%%%%%%%%%%%%%%%%%%%%%%%%%%%%%%%%%%%%%
In the present model, the dark sector has two different types of
particles namely, a Dirac fermion $\psi$ and a complex scalar $S$. Therefore, the total
relic density of the dark matter in the Universe must have contributions
from both of these dark sector particles. In order to compute
the total relic density as well as individual relic densities of
each dark matter candidate, it is essential to solve two coupled Boltzmann
equations \cite{Biswas:2013nn, Belanger:2011ww} which describe the evolution of
number densities of both the dark matter candidates. The equations are given by,
\begin{eqnarray}
\frac{dn_{\psi}}{dt} + 3n_{\psi}{H} &=& 
-\langle{\sigma {\rm{v}}}_{\psi \bar{\psi} \rightarrow {X \bar{X}}}\rangle 
\left(n_{\psi}^2 -(n_{\psi}^{\rm eq})^2\right)
- {\langle{\sigma {\rm{v}}}_{\psi \bar{\psi}
\rightarrow S S^\dagger}\rangle} \left(n_{\psi}^2 - 
\frac{(n_{\psi}^{\rm eq})^2}{(n_{S}^{\rm eq})^2}n_{S}^2\right) \,\, ,
%\nonumber\, , 
%
%\\
\label{boltz-eq1} \\
%\end{eqnarray}
%\begin{eqnarray}
\frac{dn_{S}}{dt} + 3n_{S}{H} &=& -\langle{\sigma {\rm{v}}}
_{S S^\dagger \rightarrow {X \bar{X}}}\rangle \left(n_{S}^2 -
(n_{S}^{\rm eq})^2\right) 
+ {\langle{\sigma {\rm{v}}}_{\psi \bar{\psi} \rightarrow S S^\dagger}\rangle} 
\left(n_{\psi}^2 - \frac{(n_{\psi}^{\rm eq})^2}{(n_{S}^{\rm eq})^2}n_{S}^2\right)\ , 
%\nonumber \\18 running,
\label{boltz-eq2} 
\end{eqnarray}
where $n_{\psi}$, $n_S$ denote the number densities of $\psi$ and $S$ while
their equilibrium values are denoted by $n_{\psi}^{\rm eq}$, $n_S^{\rm eq}$ respectively, $H$
is the Hubble's constant. ${\sigma {\rm{v}}}_{SS^\dagger \rightarrow X \bar{X}}$
%\footnote{$~i^c$ represents the antiparticle of dark matter component $i$,
%{\it i.e.} $i^c = S^{\dagger}(\bar{\psi})$ for the dark matter candidate $S(\psi)$.}  
describes the self annihilation cross section of the dark matter component $S$
into the SM particles as well as other non SM particles such as $Z^{\prime}$,
$H$\footnote{$~X$ represents any particle in
the present model, expect $\psi$, satisfying the
kinematic condition $s > 4 M^2_{X}$. Where $s$ being the centre of mass energy
and $M_{X}$ is the mass of $X$. We will see later that the annihilation channel
$SS^{\dagger}\rightarrow H H$ is kinametically not possible within the allowed ranges
of $M_H$ for a particular value of $M_S$ (right panel of Fig. \ref{plot4}).}
while the same quantity for the other DM component ($\psi$) is denoted by
${\sigma {\rm{v}}}_{\psi \bar{\psi} \rightarrow {X \bar{X}}}$\footnote{For the
DM component $\psi$, ${\rm X} \neq h$, $H$ as the Lagrangian of the present model does
not contain any of these vertices namely $Zhh$, $Z^{\prime}hh$, $ZHH$ and $Z^{\prime}HH$.}. 
The interaction between the two components of dark matter
is described by the annihilation cross section
${\sigma {\rm{v}}}_{\psi \bar{\psi} \rightarrow S S^\dagger}$ where
the heavier dark matter component (say $\psi$) annihilates to produce other
component $S$. New gauge boson $Z^\prime$ is the main exchange
particle for this interaction. The Feynman diagrams for
all the processes (self annihilations of both the dark matter components and
dark matter conversion from heavier to lighter) which are relevant for
the evolution of number densities of $S$ and $\psi$
are given in Fig. \ref{anni-dia}.
The annihilation cross sections for the processes shown in
Fig. \ref{anni-dia} are given by,
\begin{eqnarray}
\sigma_{\psi \bar{\psi}\rightarrow Z^{\prime} Z^{\prime}} &=&
\frac{(Q_{\rm X}(\psi)\,g_{\rm X}\,\cos \theta_{\rm NB})^4}{8 \pi s(s-4M^2_{\psi})}
\Bigg[-\frac{s \sqrt{1-\frac{4 M_{\psi }^2}{s}} \sqrt{1-\frac{4 M^2_{Z^\prime}}{s}} \left(s
M_{\psi }^2+4 M_{\psi }^4+2 M^4_{Z^\prime}\right)}
{M_{\psi }^2 \left(s-4M^2_{Z^\prime}\right)+M^4_{Z^\prime}} + \nonumber \\
&&\hspace{-2cm}\frac{\left(4 M_{\psi }^2 \left(s-2 M^2_{Z^\prime}\right)-8 M_{\psi }^4+
4 M^4_{Z^\prime}+s^2\right)}{s-2M^2_{Z^{\prime}}}
%\Bigg\{\log \frac{2M^2_{Z^{\prime}}-s + \sqrt{\left(1-\frac{4M^2_{\psi}}{s}\right)
%\left(1-\frac{4M^2_{Z^{\prime}}}{s}\right)}}
%{2M^2_{Z^{\prime}}-s -\sqrt{\left(1-\frac{4M^2_{\psi}}{s}\right)
%\left(1-\frac{4M^2_{Z^{\prime}}}{s}\right)}} \nonumber \\
%&&+
\ln \frac{2M^2_{Z^{\prime}}-s\left(1+\sqrt{\left(1-\frac{4M^2_{\psi}}{s}\right)
\left(1-\frac{4M^2_{Z^{\prime}}}{s}\right)} \right)}
{2M^2_{Z^{\prime}}-s\left(1-\sqrt{\left(1-\frac{4M^2_{\psi}}{s}\right)
\left(1-\frac{4M^2_{Z^{\prime}}}{s}\right)}\right)}
%\Bigg\}
\Bigg]\,\,, \nonumber\\
\label{psipsiz1z1}
\end{eqnarray} 
\begin{eqnarray}
\sigma_{SS^\dagger\rightarrow f \bar{f}} &=& n_c \frac{m^2_f}{8\pi s}
\left(s - 4m^2_f\right)
\sqrt{\frac{s-{4 m^2_f}}{s-{4M^2_S}}} 
\left[
\frac{\left(\frac{g_{SSh}}{v}\right)^2 \cos^2 \alpha}{\left(s-M^2_h\right)^2
+ \left(\Gamma_h M_h\right)^2}
+\frac{\left(\frac{g_{SSH}}{v}\right)^2
\sin^2 \alpha}{\left(s-M^2_H\right)^2 + \left(\Gamma_H M_H\right)^2}
\right.\nonumber\\&&\left.
+\frac{\left(\frac{g_{SSh}\,g_{SSH}}{v^2}\right)
\sin 2\alpha \{\left(s-M^2_h\right)\left(s-M^2_H\right)
+ \Gamma_h \Gamma_H M_h M_H\}}{\{\left(s-M^2_h\right)^2 +
\left(\Gamma_h M_h\right)^2\}\{\left(s-M^2_H\right)^2 +
\left(\Gamma_H M_H\right)^2\}}
%\right.\nonumber\\&&\left.
\right] \,\, ,
\label{sigmaff}
\end{eqnarray}
\begin{eqnarray}
\sigma_{S S^{\dagger} \rightarrow Z^{\prime} Z^{\prime}} &=&
\frac{1}{16 \pi s} \sqrt{\frac{s-4 M^2_{Z^{\prime}}}
{s-4 M^2_{\psi}}} \left(|\mathfrak{g}|^2\frac{s^2}{M^4_{Z^{\prime}}} \right)
\Bigg(1-\frac{4 M^2_{Z^{\prime}}}{s} +\frac{12 M^4_{Z^\prime}}{s^4}\Bigg)\,\,,
\label{ssz1z1}
\end{eqnarray}
where
\begin{eqnarray}
\mathfrak{g} &=& g_{\rm X}\,Q_{\rm X}(S)\,\cos \theta_{\rm NB} - \frac{g_{SSH}\,g_{HZ^{\prime}Z^{\prime}}\,
[(s-M^2_H)-i \Gamma_H M_H]} {(s-M^2_H)^2+(\Gamma_H M_H)^2} \nonumber \\
&&-\frac{g_{SSh}\,g_{hZ^{\prime}Z^{\prime}}\,
[(s-M^2_h)-i \Gamma_h M_h]} {(s-M^2_h)^2+(\Gamma_h M_h)^2}\,\,,
\label{g}
\end{eqnarray}
with
\begin{eqnarray}
g_{hZ^{\prime}Z^{\prime}} &=& \frac{g^2_{\rm Z}\, v}{4} \Bigg\{b^2\,\cos \alpha 
- \left(\frac{g_{\rm X}\,Q_{\rm X}(\Phi_s)}{g_{\rm Z}\,Q_{\rm Y}(\Phi)}\right)^2
\frac{v_s}{v}\sin \alpha \cos^2 \theta_{\rm NB}\Bigg\}\,\,,\nonumber\\
g_{HZ^{\prime}Z^{\prime}} &=& \frac{g^2_{\rm Z}\, v}{4} \Bigg\{b^2\,\sin \alpha
+ \left(\frac{g_{\rm X}\,Q_{\rm X}(\Phi_s)}{g_{\rm Z}\,Q_{\rm Y}(\Phi)}\right)^2
\frac{v_s}{v}\cos \alpha \cos^2 \theta_{\rm NB}\Bigg\}\,\,,
\label{ghzz}
\end{eqnarray}
and 
\begin{eqnarray}
b &=& \sin \theta_{\rm NB} - \epsilon \tan \theta_{\rm W} \cos \theta_{\rm NB}\,\,. 
\end{eqnarray}
Further,
\begin{eqnarray}
\sigma_{\psi \bar{\psi} \rightarrow S {S}^\dagger} &=& 
\frac{\left(g^2_{\rm X}Q_{\rm X}(\psi)Q_{\rm X}(S)
\cos^2 \theta_{\rm NB}\right)^2}{48\pi}
\frac{\left(s-4M^2_S\right)}{\left(s-M^2_{Z^\prime}\right)^2}
\sqrt{\frac{s-{4 M^2_S}}{s-{4M^2_{\psi}}}} \left(1 + \frac{2M^2_{\psi}}{s}\right) \, ,
\nonumber \\
\label{sigma-psi}
\end{eqnarray}
where $\Gamma_h$, $\Gamma_H$ (Eqs. (\ref{sigmaff}, \ref{g}))
are the decay widths corresponding to the physical scalar fields $h$, $H$.
In the above, $Q_{\rm X}(\psi) = 1$, $Q_{\rm X}(S)=2$ and
$Q_{\rm X}(\Phi_s)=\frac{1}{2}$ are the U(1)$_{\rm X}$ gauge charges
of $\psi$, $S$ and $\Phi_s$ respectively while $Q_{\rm Y} (\Phi) = \frac{1}{2}$
denotes the $U(1)_{\rm Y}$ charge of $\Phi$ (see Table \ref{tab1}). The
quantity $g_{hZ^{\prime} Z^{\prime}}$ ($g_{HZ^{\prime} Z^{\prime}}$) in
Eq. (\ref{ghzz}) represents the coupling between scalar field
$h$ ($H$) and two $Z^{\prime}$'s\footnote{The expressions of the coupling terms
$g_{hZZ}$, $g_{HZZ}$ can be readily obtained from Eq. (\ref{ghzz})
by replacing the terms $b$ and $\cos \theta_{\rm NB}$ with the quantities
$a$ (Eq. (\ref{a})) and $\sin \theta_{\rm NB}$ respectively.} 
Moreover, the expressions of the coupling terms $g_{SSH}$ and $g_{SSh}$
appearing in Eqs. (\ref{sigmaff}, \ref{g}) are given in Eq. (\ref{gssh}). 
In Eqs. (\ref{boltz-eq1}, \ref{boltz-eq2})
the symbol $\langle...\rangle$ implies the quantity within
the curly bracket is thermally averaged. A general expression
of thermally averaged annihilation cross section for the
annihilation channel ${A A^\dagger \rightarrow B \bar{B}}$ is
given by \cite{Gondolo:1990dk},
\begin{eqnarray}
\left\langle \sigma {\rm v}_{A A^\dagger \rightarrow  B \bar{B}}\right\rangle 
&=& \frac{1}{8 M_A^4 T K_2^2\left(\frac{M_A}{T}\right)}
\int_{4M_A^2}^\infty \,\frac{1}{2}\sigma_{A A^\dagger \rightarrow B \bar{B}}\,
(s-4M_A^2)\,\sqrt{s}\,K_1\left(\frac{\sqrt{s}}{T}\right)\,ds \,\, , \nonumber \\
\label{thermal-ave}
\end{eqnarray}
where $K_1,~K_2$ are the modified Bessel functions of order 1, 2 and
$T$ is the temperature of the Universe. The Extra $\frac{1}{2}$ factor
before $\sigma_{A A^\dagger \rightarrow B \bar{B}}$
arises due to the fact that the initial state particles of the annihilation
channel ($A A^\dagger \rightarrow  B \bar{B}$) are not identical
(see Refs. \cite{Gondolo:1990dk, Srednicki:1988ce} for more discussions).
More specifically, no extra $\frac{1}{2}$ factor would be needed
if the particle and its antiparticle involved in the initial state
of an annihilation process are identical in nature
({\it e.g.} Majorana fermion or real scalar field). 
As mentioned earlier, the quantity 
$\langle{\sigma {\rm{v}}}_{\psi \bar{\psi} \rightarrow {X \bar{X}}}\rangle$
appearing in Eq. (\ref{boltz-eq1}) is the thermally averaged total annihilation
cross section of $\psi$ due to all possible final state particles including
both the SM as well as non SM particles. However, $\psi$ can interact
with the visible world (SM particles) only through the exchange of 
$Z$ and $Z^\prime$ bosons. Therefore, resulting annihilation cross section of the
channel $\psi \bar{\psi} \rightarrow Y \bar{Y}$ ($Y$ is any SM particle
except the Higgs boson $h$ and $Z$ boson\footnote{For $ZZ$ final state
$\sigma {\rm v}_{\psi \bar{\psi} \rightarrow Z Z}$ is proportional to
$\sin^4 \theta_{\rm NB}$.}) is proportional to $\sin^2 \theta_{\rm NB}$,
sine of the mixing angle between the two neutral gauge bosons. Now for
small values of $\theta_{\rm NB}$, which is required to satisfy
experimental constraints \cite{Agashe:2014kda}, one can use the approximation $\sin \theta_{\rm NB}
\simeq \theta_{\rm NB}$. Thus, under this circumstance the term
$\sin \theta_{\rm NB} \propto \epsilon$, the kinetic mixing parameter
between U(1)$_{\rm Y}$ and U(1)$_{\rm X}$ gauge fields (see Eq. (\ref{gauge-mix-angle})
for the expression of $\theta_{\rm NB}$).  
Hence, $\langle{{\sigma {\rm v}}_{\psi \bar{\psi}\rightarrow X \bar{X}}}\rangle \sim$
$\langle{{\sigma {\rm v}}_{\psi \bar{\psi}\rightarrow Z^{\prime} Z^{\prime}}}\rangle$ as
the value of $\chi=\frac{\epsilon}{\cos \theta_{\rm W}}$\footnote{In
this work we adopt $\epsilon \sim 10^{-3}$ which
satisfy all the constraints mentioned in the previous section.}
is severely constrained to be very small \cite{Gopalakrishna:2008dv,
Davoudiasl:2012ag, Lee:2013fda, Agashe:2014kda}, which implies
that the probability of producing two $Z^{\prime}$ in the
final state is much more than that for the SM particles from the self annihilation
of the dark matter candidate $\psi$. Hence, we can neglect the contributions
arising from the annihilation channels $\psi \bar{\psi} \rightarrow Y \bar{Y}$ in Eq. (\ref{boltz-eq1}). 
\paragraph{}
Let us introduce two dimensionless variables namely $Y_{i} = \frac{n_i}{\rm s}$
and $x_i = \frac{M_i}{T}$ for $i$ = $\psi$, $S$. $Y_i$ and $M_i$ are
the comoving number density and mass of the dark matter component $i$.
The entropy density of the Universe is denoted by $\rm s$.
In terms of these two dimensionless variables Eq. (\ref{boltz-eq1})
and Eq. (\ref{boltz-eq2}) take the following forms,
\begin{eqnarray}
&&\frac{dY_{\psi}}{dx_{\psi}} =
-\left(\frac{45G}{\pi}\right)^{-\frac{1}{2}}
\frac{M_{\psi}}{x_{\psi}^2}{\sqrt{g_\star}}
\left({\langle{\sigma {\rm{v}}}_
{\psi \bar{\psi}\rightarrow {Z^{\prime} Z^{\prime}}}\rangle}
\left(Y_{\psi}^2-(Y_{\psi}^{\rm eq})^2\right) + 
{\langle{\sigma {\rm{v}}}_{\psi\bar{\psi}\rightarrow SS^\dagger}\rangle}\left(Y_{\psi}^2 - 
\frac{(Y_{\psi}^{\rm eq})^2}{(Y_{S}^{\rm eq})^2}Y_{S}^2\right)
\right)
\,\, , \nonumber \\  
\label{boltz-eq-y2}\\
%\end{eqnarray}
%\begin{eqnarray}
&&\frac{dY_{S}}{dx_{S}} =-\left(\frac{45G}{\pi}\right)^{-\frac{1}{2}}
\frac{M_{S}}{x_{S}^2}\sqrt{g_\star}\left({\langle{\sigma {\rm{v}}}_
{SS^\dagger\rightarrow {X \bar{X}}}\rangle}
\left(Y_{S}^2-(Y_{S}^{\rm eq})^2\right) 
 - {\langle{\sigma {\rm{v}}}_{\psi\bar{\psi}\rightarrow SS^\dagger}\rangle} \left(Y_{\psi}^2 - 
\frac{(Y_{\psi}^{\rm eq})^2}{(Y_{S}^{\rm eq})^2}Y_{S}^2\right)\right)
\,\, , \nonumber\\ 
\label{boltz-eq-y1}
\end{eqnarray} 
where $G$ is the Gravitational constant.
The quantity $g_{\star}$ is defined as \cite{Gondolo:1990dk},
\begin{eqnarray}
\sqrt{g_\star} = \frac{h_{{\rm{eff}}}(T)}{\sqrt{g_{{\rm{eff}}}(T)}}
\left(1 + \frac{1}{3}\frac{d\,{\rm ln}(h_{{\rm{eff}}}(T))}{d\,{\rm ln}(T)}\right) \,\, ,
\label{gstar}
\end{eqnarray}
with $g_{{\rm{eff}}}(T)$ and $h_{{\rm{eff}}}(T)$ are the effective
degrees of freedom corresponding to the energy and entropy densities
of the Universe. They are related to energy and entropy densities
through the relations
%\begin{eqnarray}
$\rho = g_{{\rm{eff}}}(T)\frac{\pi^2}{30}T^4, \, 
{\rm s} = h_{{\rm{eff}}}(T)\frac{2\pi^2}{45}T^3$.

We have solved Eqs. (\ref{boltz-eq-y2}, \ref{boltz-eq-y1})
numerically to obtain the values of comoving number
densities ($Y_{\psi}$, $Y_{S}$) for both the dark matter components
$\psi$ and $S$ respectively at the present temperature $T_0$ of
the Universe. Using these values of $Y_i$ ($i = \psi,\,\, S$) at
$T_0$ the relic density of each dark matter component can then
be computed using the following equation \cite{Edsjo:1997bg, Biswas:2011td}
\begin{eqnarray}
\Omega_i h^2 = 2.755\times 10^8 \left(\frac{M_i}{\rm GeV}\right) Y_{i}(T_0)\,\,.
\label{omega_ind}
\end{eqnarray}
The total relic density ($\Omega_{\rm T}h^2$) of the dark matter is simply the sum of the
relic densities of each dark matter component which can be written as \cite{Biswas:2013nn},
\begin{eqnarray}
\Omega_{\rm T} h^2 = \Omega_{\psi} h^2 + \Omega_{S} h^2\,\,. 
\label{omega_tot}
\end{eqnarray}
%%%%%%%%%%%%%%%%%%%%%%%%%%%%%%%%%%%%%%%%%%%%%%%%%%%%%%%%%%%%%%%%%%%%%%%%%%  
\section{Results}
\label{results}
\begin{figure}[h!]
\includegraphics[height=9cm,width=7.2cm,angle=-90]{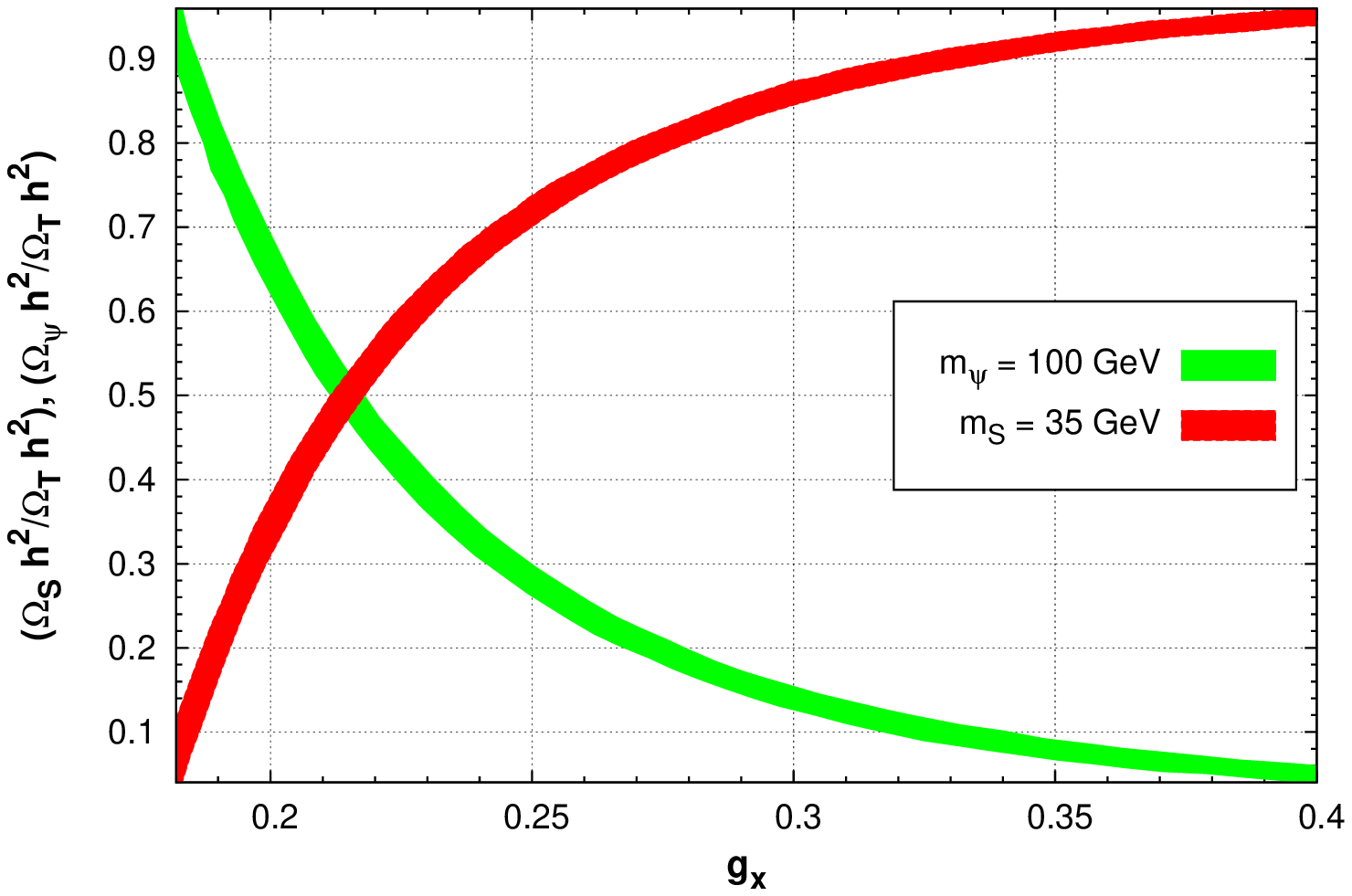}
\includegraphics[height=9cm,width=7cm,angle=-90]{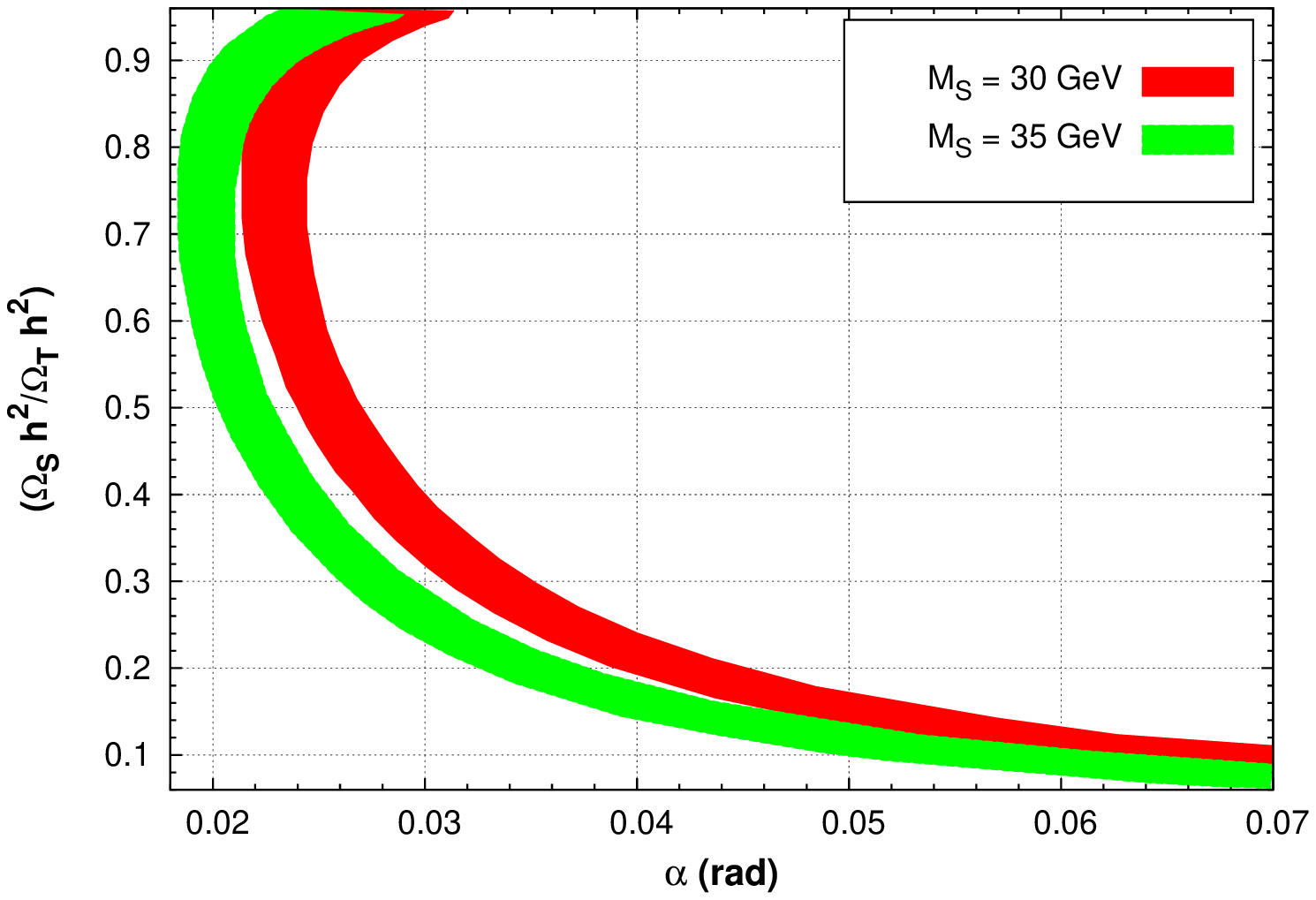}
\caption{Left panel : Variations of $\frac{\Omega_{S}h^2}{\Omega_{\rm T} h^2}$, 
$\frac{\Omega_{\psi}h^2}{\Omega_{\rm T} h^2}$ with the gauge coupling $g_{\rm X}$
for $M_{S}=35$ GeV and $M_{\psi} = 100$ GeV.
Right Panel : Variations of $\frac{\Omega_{S}h^2}{\Omega_{\rm T} h^2}$
with the mixing angle $\alpha$ between the neutral scalars $h$, $H$ for
$M_{S} = 30$ GeV, $M_{\psi} = 100$ GeV and
$M_{S} = 35$ GeV, $M_{\psi} = 100$ GeV respectively.}
\label{plot1}
\end{figure}

%%%%%%%%%%%%%%%%%%%%%%%%%%%%%%%%%%%%%%%%%%%%%%%%%%%%%%%%%%%%%%%%%%%%%%%%%%%
In this section we describe the effects of the model
parameters namely $M_S$, $M_{\psi}$, $M_H$, $M_{Z^\prime}$,
$\alpha$, $\lambda_2$, $\lambda_3$ on the relic
densities of both the dark matter components. In this present
two component dark matter scenario the role of heavier dark matter
candidate is played by the component $\psi$. Hence, throughout the work
we assume $M_{\psi}>M_{S}$ and the value of $M_{S}$ is taken in the range
$30 \sim 40$ GeV as this would be required to explain the observed 
$\gamma$-ray excess, from the regions close to the GC, at an energy range $1\sim3$ GeV
by the annihilation of the dark matter component $S$.
The ranges of the model parameters adopted in this work are given below,
\begin{eqnarray}
\begin{array}{cccccc}
1.0\times 10^{-3}&\le& \lambda_2, \lambda_3&\le& 1.0\times 10^{-2}\,\,,\\ 
1.0\times 10^{-2}&\le&\alpha& < & 1.4\times 10^{-1}\,\, ,\\
30\text{ GeV}&\le & M_{Z^\prime} & \le& 75\text{ GeV}\,\, ,\\
40\text{ GeV}&\le & M_{H} & \le& 100\text{ GeV}\,\, ,\\
60\text{ GeV}&\le & M_{\psi} & \le& 150\text{ GeV}\,\, ,\\
30\text{ GeV}&\le & M_{S} & \le& 40\text{ GeV}\,\, .
%-(500 \text{ GeV})^2&\le&\kappa_2&\le&(500 \text{ GeV})^2\\
%-1000\text{ GeV}&\le& \kappa_3&\le& 1000\text{ GeV},
\label{para-ranges}
\end{array}
\end{eqnarray}
From now on, whenever we use any
specific values of the model parameters, we mention
it explicitly with the other parameters scanned over their entire
considered range given in Eq. (\ref{para-ranges}).
Moreover in this work, we have considered $\epsilon\sim10^{-3}$ 
which is equivalent to the kinetic mixing parameter $\chi\sim 1.13\times 10^{-3}$.
The corresponding $Z-Z^{\prime}$ mixing angle lies in the
range $\sim (0.6-1.65)\times 10^{-3}$ rad
for the variation of the entire considered range of $M_{Z^{\prime}}$
mentioned above. We have checked that both the values
of kinetic mixing parameter and mixing angle $\theta_{\rm NB}$
satisfy all the relevant constraints listed in Section \ref{model}.
In order to study the variations of the relic densities
of each of the dark matter components, namely $S$ and $\psi$,
with the model parameters mentioned above,
we define a ratio $\frac{\Omega_i h^2}{\Omega_{\rm T}h^2}$
which represents the fractional contribution of the dark matter
component $i$ ($i = S, \psi$) to the overall dark matter
relic density ($\Omega_{\rm T} h^2$). As mentioned earlier, for the computations
of individual relic densities ($\Omega_{i} h^2$) of both the
dark matter candidates we need to solve two coupled Boltzmann equations 
(Eqs. (\ref{boltz-eq-y2}), (\ref{boltz-eq-y1})) numerically
%which are given in the previous section (Section \ref{boltz-eqn})
using Eqs. (\ref{psipsiz1z1}-\ref{thermal-ave}) and Eq. (\ref{gstar}).
\paragraph{}
In left panel of Fig. \ref{plot1} we plot the variations of the fractional
contributions of both the dark matter components $\psi$ and $S$ with
gauge coupling $g_{\rm X}$ of U(1)$_{\rm X}$ gauge group for
$M_{S}=35$ GeV, $M_{\psi}=100$ GeV, $M_{Z^\prime} = 50$ GeV and $M_{H}=75$ GeV. 
%The neutral scalars mixing angles $\alpha$ and the two quartic couplings
%$\lambda_2$, $\lambda_3$ are scanned over their
%entire considered range ($0.01\leq\alpha\leq 0.1$,
%$0.001\leq\lambda_2,\lambda_3\leq0.01$).
From the left panel of Fig. \ref{plot1}
it is seen that the fractional contribution $\frac{\Omega_\psi h^2}{\Omega_{\rm T} h^2}$
($\frac{\Omega_S h^2}{\Omega_{\rm T} h^2}$)
of the dark matter component $\psi$ ($S$) decreases (increases)
with gauge coupling $g_{\rm X}$. This
nature can be explained from the fact that the annihilation
cross section of $\psi$ into $Z^{\prime} Z^{\prime}$ final state
is directly proportional to $g^4_{\rm X}$ (see Eq. (\ref{psipsiz1z1})).
Therefore, as the gauge coupling $g_{\rm X}$ increases the annihilation
cross section of $\psi$ for the channel $\psi\bar{\psi}
\rightarrow Z^{\prime} Z^{\prime}$ increases which in turn
decreases the relic density of the dark matter component $\psi$.
On the other hand, the coupling term between the two coupled
Boltzmann equations (Eq. (\ref{boltz-eq-y1}, \ref{boltz-eq-y2})) is
proportional to the annihilation cross section
${\sigma {\rm v}}_{\psi\bar{\psi}\rightarrow SS^\dagger}$ which
is also directly proportional to $g^4_{\rm X}$ (see Eq. (\ref{sigma-psi})).
Now, any increment in $g_{\rm X}$ will naturally increase the quantity
${\sigma {\rm v}}_{\psi\bar{\psi}\rightarrow SS^\dagger}$ and hence
the coupling strength between the two coupled Boltzmann equations
gets amplified which further reduces the contribution of the heavier component
$\psi$ to the total DM relic density. Physically it indicates more
and more $S$ particles production in the final state from
the pair annihilation of the heavier dark matter component $\psi$.
Consequently, the individual relic density of the lighter DM component $S$
as well as its the fractional contribution to the total DM relic density
increases. The variations of the fractional relic density contribution of
$S$ with the neutral scalars mixing angle $\alpha$ are shown in the right
panel of Fig. \ref{plot1}. In this plot (right panel of Fig. \ref{plot1})
the red coloured region describes the allowed zone in the
$\frac{\Omega_{S}h^2}{\Omega_{\rm T} h^2}$ -Vs-$\alpha$ plane
which satisfies all the constraints listed in Section \ref{model} 
for $M_{S} = 30$ GeV, $M_H$= 65 GeV while the green coloured 
region is for the case when $M_S=35$ GeV, $M_H=75$ GeV. Both of these plots
are computed for heavier dark matter mass $M_{\psi}=$ 100 GeV and $M_{Z^\prime} = 50$ GeV.
From the right panel of Fig. \ref{plot1} it is evident
that in order to satisfy all the relevant constraints
listed in Section \ref{model}, the maximum allowed
value of the neutral scalars mixing angle $\alpha$ is $\sim$ 0.07 rad. 
%From the right panel of Fig. \ref{plot1} we see that the
%ratio $\frac{\Omega_{S}h^2}{\Omega_{\rm T} h^2}$ decreases
%as $\alpha$ increases.  
\begin{figure}[h!]
\includegraphics[height=9cm,width=7cm,angle=-90]{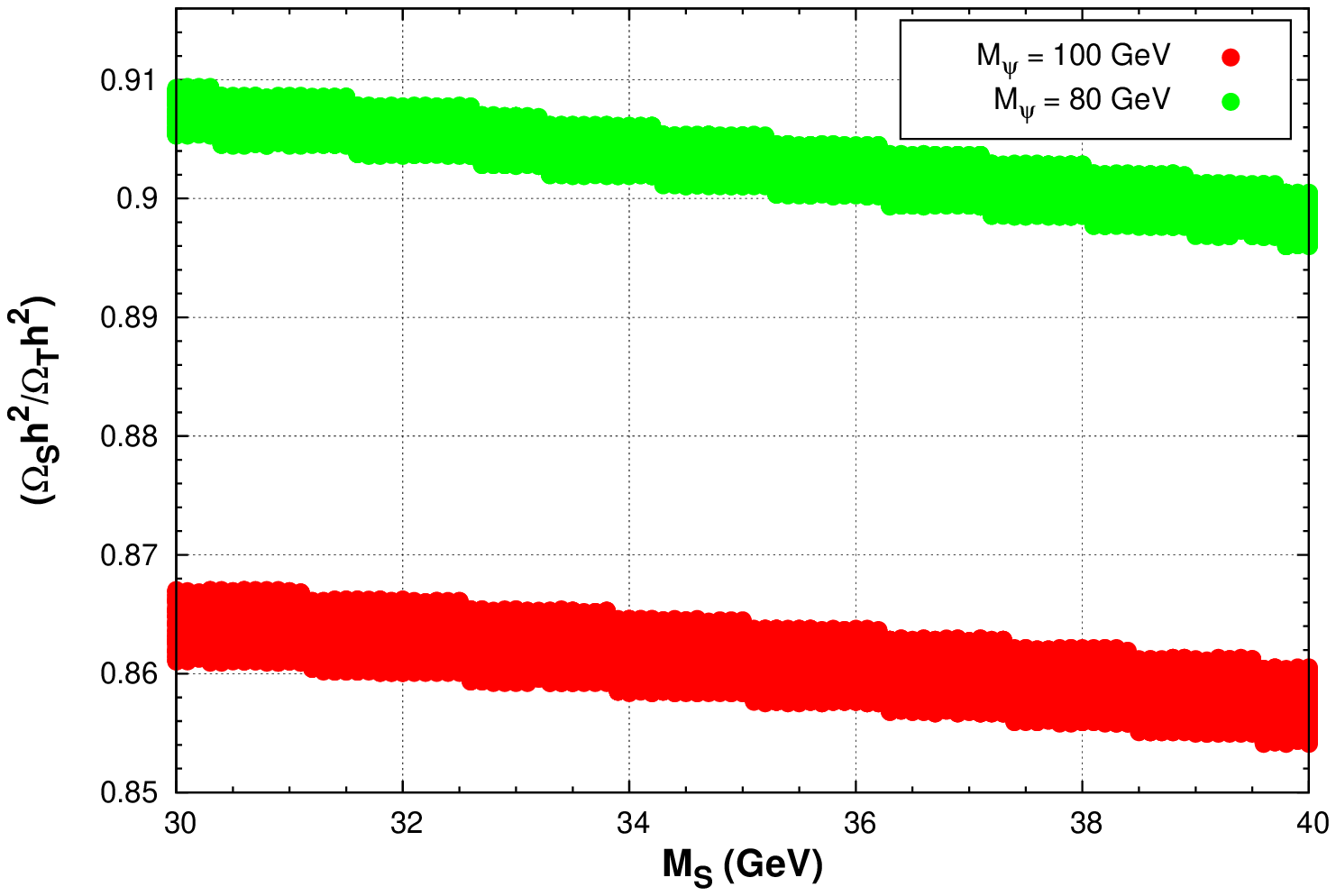}
\includegraphics[height=9cm,width=7cm,angle=-90]{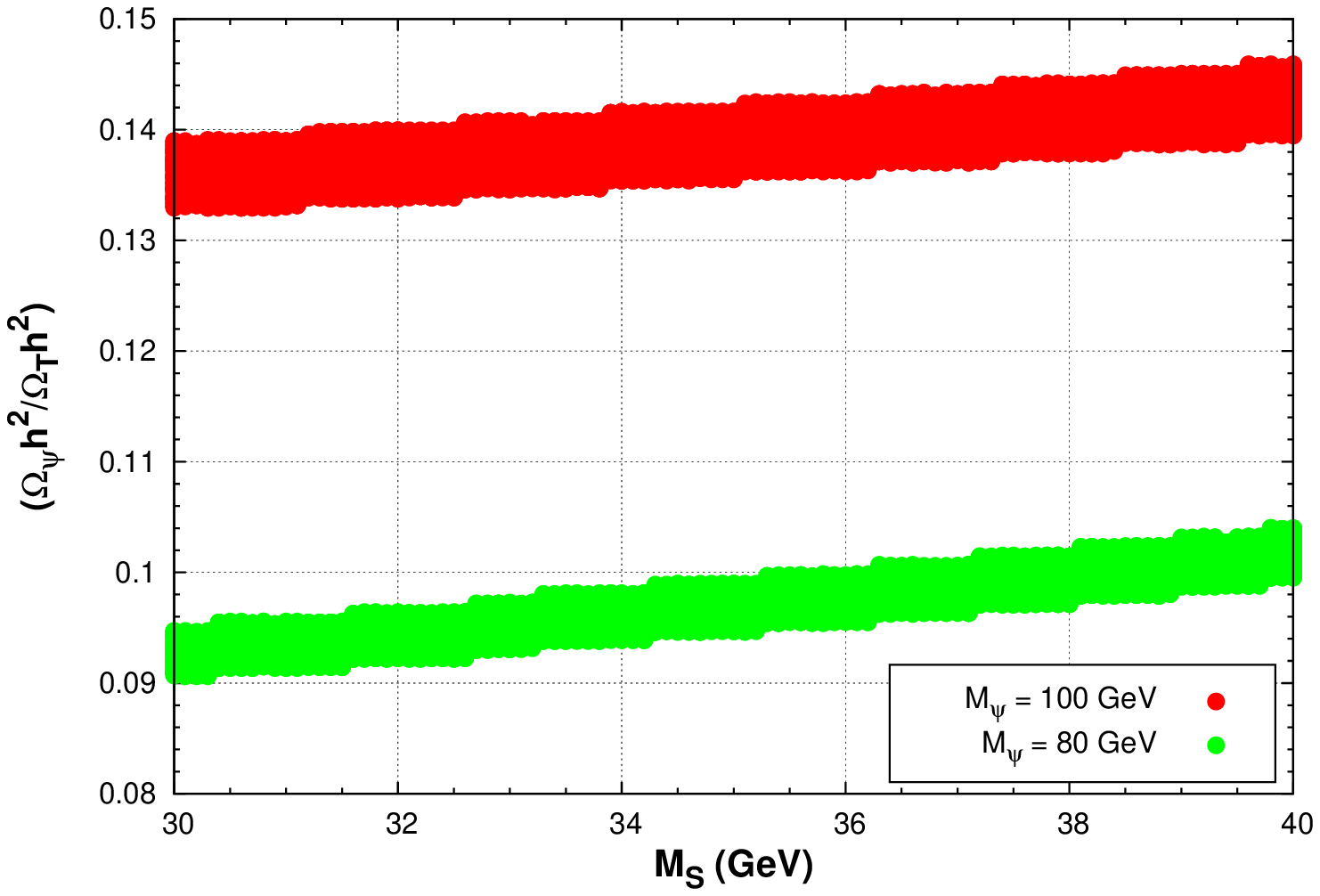}
\caption{Variations of $\frac{\Omega_{S}h^2}{\Omega_{\rm T} h^2}$ (left panel)
and $\frac{\Omega_{\psi}h^2}{\Omega_{\rm T} h^2}$ (right panel) with the mass of $S$
for $M_{\psi} = 80$, 100 GeV respectively.}
\label{plot2}
\end{figure}
\paragraph{}
We also calculate the variations of the ratios $\frac{\Omega_{S}h^2}{\Omega_{\rm T} h^2}$,
$\frac{\Omega_{\psi}h^2}{\Omega_{\rm T} h^2}$ with the mass of the
lighter dark matter component $S$ for two different values of $M_{\psi}$ and
the results are plotted in both the panels of Fig.\ref{plot2}. In the
left panel of Fig. \ref{plot2} we show the variations of the fractional
contribution of $S$ to the total dark matter relic density with $M_S$
for $M_{\psi} = 80$ GeV (green coloured band) and 100 GeV (red coloured band).
The right panel of the same figure (Fig. \ref{plot2}) shows how the
fractional contribution of the heavier dark matter component $\psi$
to the total relic density varies with the mass of lighter dark matter
component $S$. Similar to the plots in the left panel of Fig. \ref{plot2},
in the right panel too, the green and red coloured band represent
allowed zones for $M_{\psi}=80$ GeV and $100$ GeV respectively.
From the left panel of Fig. \ref{plot2} it appears that the
contribution of $S$ to the overall DM relic density
decreases as its mass increases. A possible reason could be the
increase of annihilation cross section (${\sigma {\rm v}}_{SS^\dagger\rightarrow f \bar{f}}$)
of the dark matter component $S$ into the final
states comprised of fermion and antifermion pair
($SS^\dagger\rightarrow f \bar{f}$, $f$ being any SM fermion except top quark)
with $M_S$ (see Eq. (\ref{sigmaff})). Therefore the relic density of $S$ decreases. This
reduction in $\Omega_{S} h^2$ must be compensated by
an increment towards the relic density of $\psi$ such that
the total relic density of both the dark matter candidates always
lies within the range provided by the PLANCK experiment. Also,
the annihilation cross section of $\psi$ for the channel
$\psi \bar{\psi} \rightarrow S S^\dagger$ decreases with $M_{S}$
(see Eq. (\ref{sigma-psi})) which further increases the
individual density of $\psi$ and hence, its the contribution 
to the overall density. This feature
is revealed in the plots of the right panel of Fig. \ref{plot2}.  
\begin{figure}[h!]
\includegraphics[height=9cm,width=7cm,angle=-90]{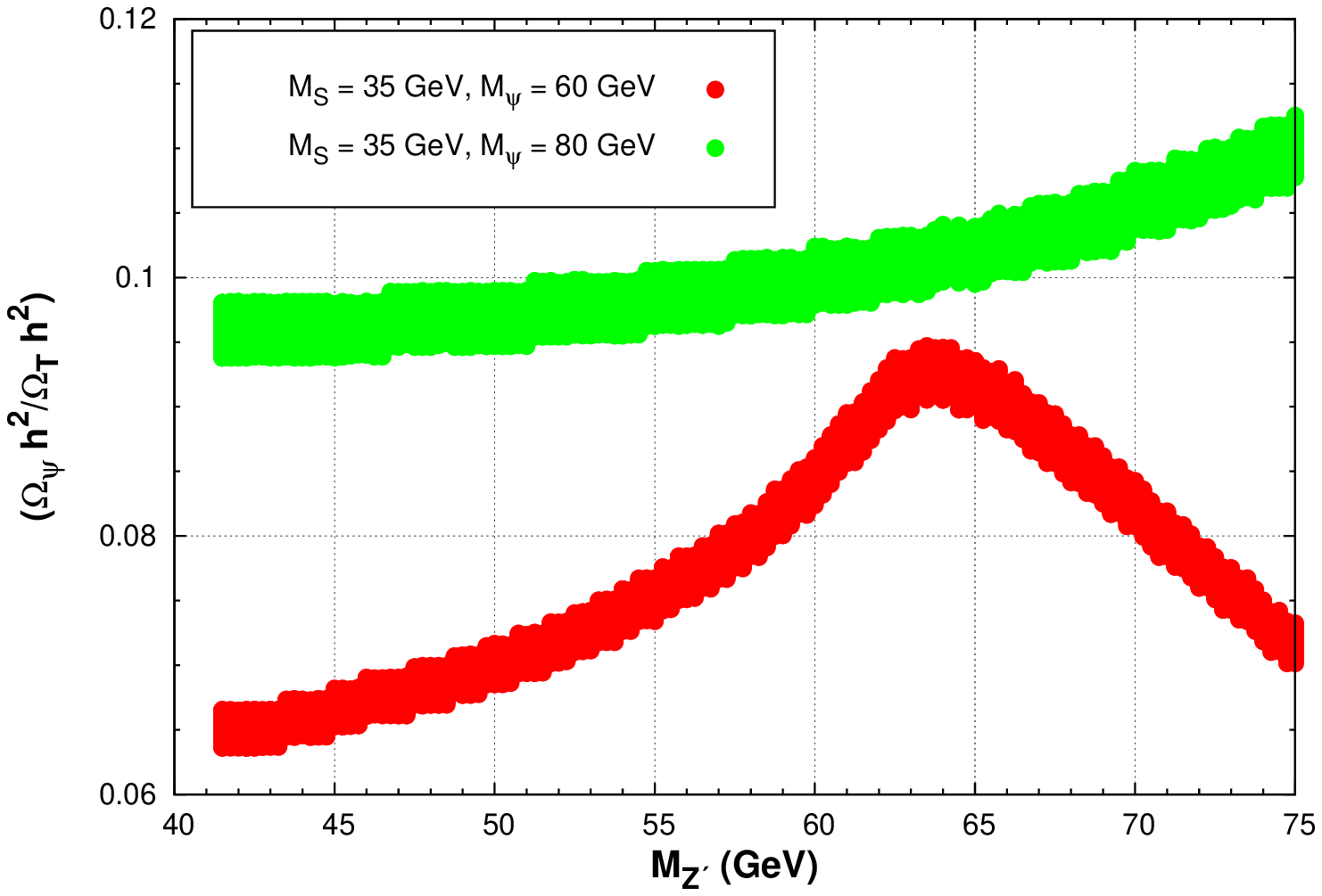}
\includegraphics[height=9cm,width=7cm,angle=-90]{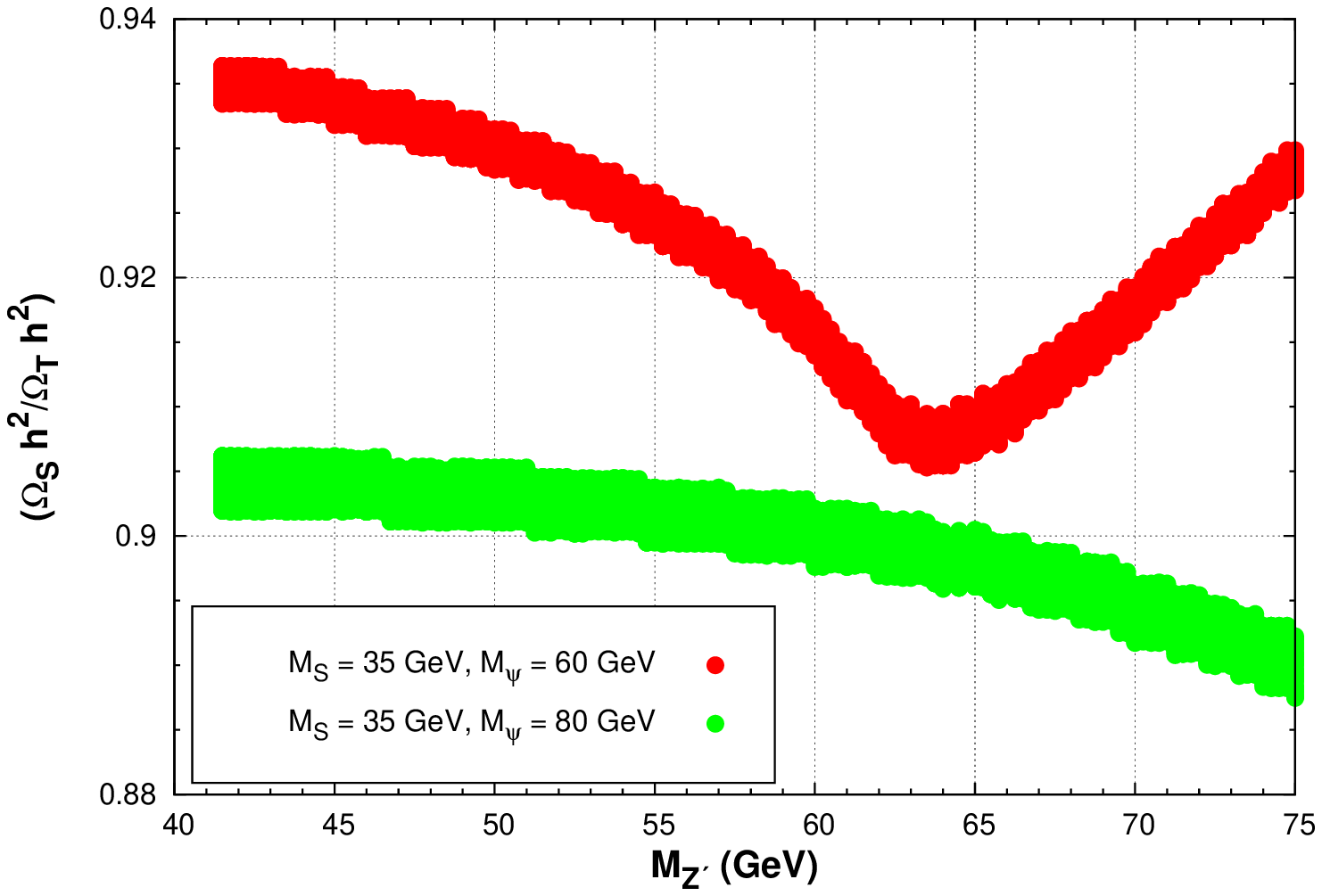}
\caption{Variations of $\frac{\Omega_{\psi}h^2}{\Omega_{\rm T} h^2}$ (left panel)
and $\frac{\Omega_{S}h^2}{\Omega_{\rm T} h^2}$ (right panel) with the mass $(M_{Z^\prime})$
of $Z^\prime$ for two different values of $M_{\psi} = 60$ GeV, 80 GeV while $M_{S}$ remains
fixed at 35 GeV.}
\label{plot3}
\end{figure}
\paragraph{}
It has been discussed earlier that the dark matter component $\psi$
can interact with both the visible (SM particles) and invisible
(other dark matter component $S$) world mainly through the exchange of gauge boson $Z^\prime$.
Therefore, the number density of the heavier dark matter component $\psi$ 
at present epoch (the relic density of $\psi$) should depend on the
mass of $Z^\prime$. The two main annihilation channels (see Fig. \ref{anni-dia})
of $\psi$ behave exactly the opposite way with respect to the variation of $M_{Z^\prime}$
and $\psi \bar{\psi} \rightarrow Z^{\prime} Z^{\prime}$ being its dominant annihilation
mode. While the annihilation cross section for the channel
$\psi \bar{\psi} \rightarrow Z^{\prime} Z^{\prime}$ (Eq. (\ref{psipsiz1z1}))
decreases with $M_{Z^{\prime}}$, the same quantity (Eq. (\ref{sigma-psi})) for 
the interaction mode in which heavier dark matter components
annihilate to produce lighter dark matter components ($\psi \bar{\psi}\rightarrow
SS^\dagger$) increases with the mass of $Z^\prime$ boson as long as
$M_{Z^\prime} < \sqrt{s}$ \footnote{which is the case we are considering.}
condition holds. Hence, the present day abundance of
the dark matter component $\psi$ is guided by the combined effects
of both the above mentioned annihilation processes.
In left panel of Fig. \ref{plot3} we show the variation of
fractional relic density of $\psi$ with
the mass of $Z^{\prime}$. The red coloured band is for $M_{\psi} =60$ GeV
while the variation of $\frac{\Omega_{\psi} h^2}{\Omega_{\rm T} h^2}$
for $M_{\psi}=80$ GeV is represented by green coloured band. From this
figure it is seen that for $M_{\psi}=60$ GeV, the quantity
$\frac{\Omega_{\psi} h^2}{\Omega_{\rm T} h^2}$ initially increases
with $M_{Z^{\prime}}$ (when $M_{Z^{\prime}} \la 63$ GeV) and thereafter
it decreases as $M_{Z^{\prime}}$ goes beyond 63 GeV. A possible reason
of this nature could be when $s \geq 4M^2_{Z^{\prime}}$ 
$\psi \bar{\psi} \rightarrow Z^{\prime} Z^{\prime}$ is the dominant annihilation
mode of $\psi$, therefore the fractional relic density of $\psi$
enhances with the increase of $M_{Z^{\prime}}$. However for $s < 4M^2_{Z^{\prime}}$
the DM component $\psi$ annihilates only through the channel
$\psi \bar{\psi} \rightarrow S S^{\dagger}$ as the annihilation
process $\psi \bar{\psi} \rightarrow Z^{\prime} Z^{\prime}$
is now kinematically forbidden. Consequently, with respect to
the variation of $M_{Z^{\prime}}$ the ratio
$\frac{\Omega_{\psi} h^2}{\Omega_{\rm T} h^2}$ changes
in the opposite direction as compared to the former case.
On the other hand for $M_{\psi}=80$ GeV and above, $s$ is always
greater than  $4M^2_{Z^{\prime}}$,
therefore the interaction mode $\psi \bar{\psi} \rightarrow Z^{\prime} Z^{\prime}$
is always kinametically accessible.
Hence for the entire considered range of $M_{Z^{\prime}}$,
the fractional contribution of $\psi$ towards the total DM relic density
increases with the mass of $Z^{\prime}$.
The change in the mass of the $Z^\prime$ boson has no significant
effect on the interactions of the lighter dark matter component $S$
with the SM fermions as those occur mainly through the exchange of neutral scalars
namely $h$ and $H$. However, the changes occur in $\Omega_{\psi} h^2$
due to the variation of $M_{Z^{\prime}}$ should be compensated by an
equal and opposite changes in the relic density of the DM component $S$
such that the total relic density should always satisfies the range
predicted by the satellite borne experiment PLANCK. This is shown
in the right panel of Fig. \ref{plot3} where the variation
of fractional relic density of $S$ with $M_{Z^{\prime}}$ are plotted.
Both the plots in Fig. \ref{plot3} have been computed for a fixed value
of $M_S = 35$ GeV.     
\begin{figure}[h!]
\includegraphics[height=9cm,width=7cm,angle=-90]{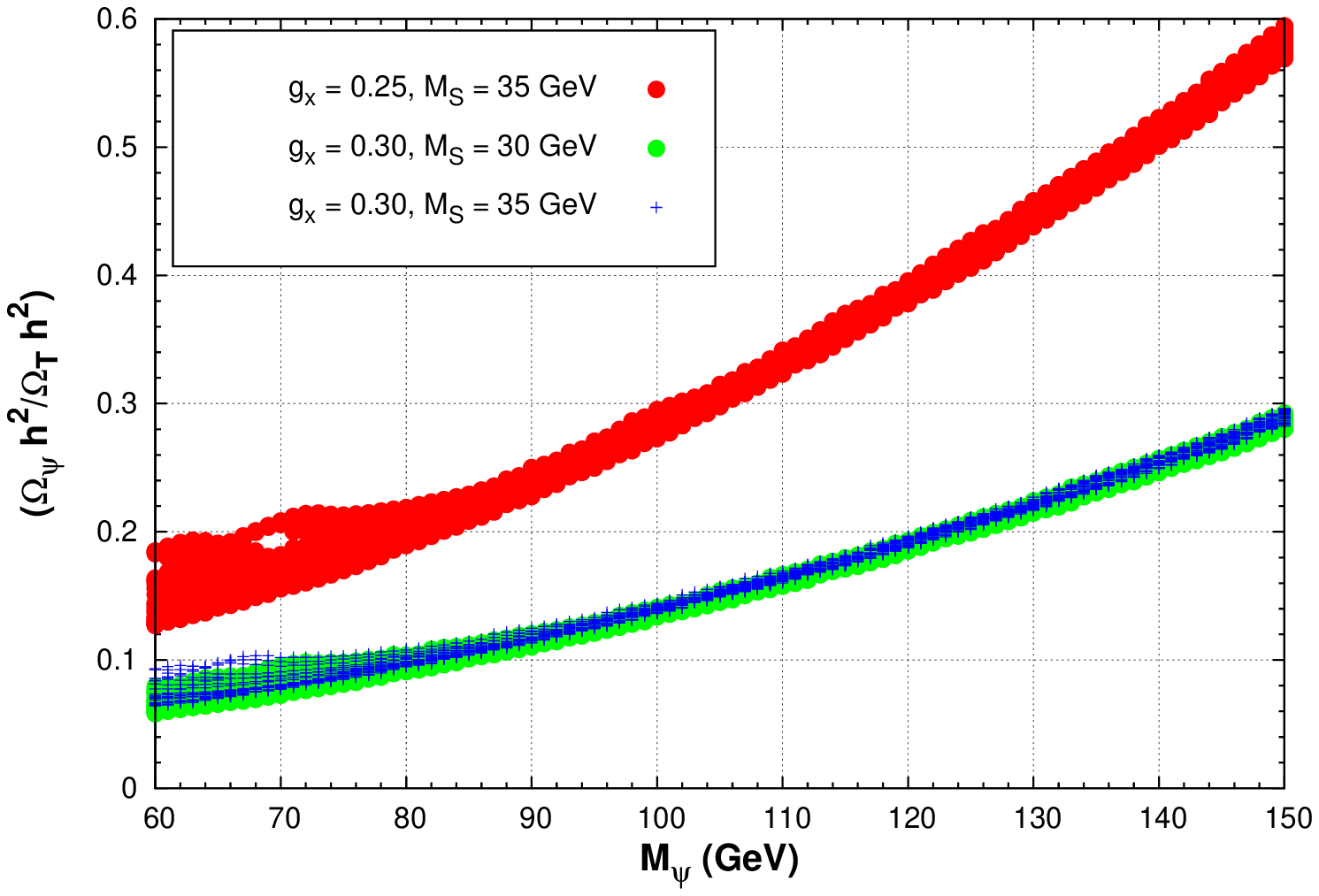}
\includegraphics[height=9cm,width=7cm,angle=-90]{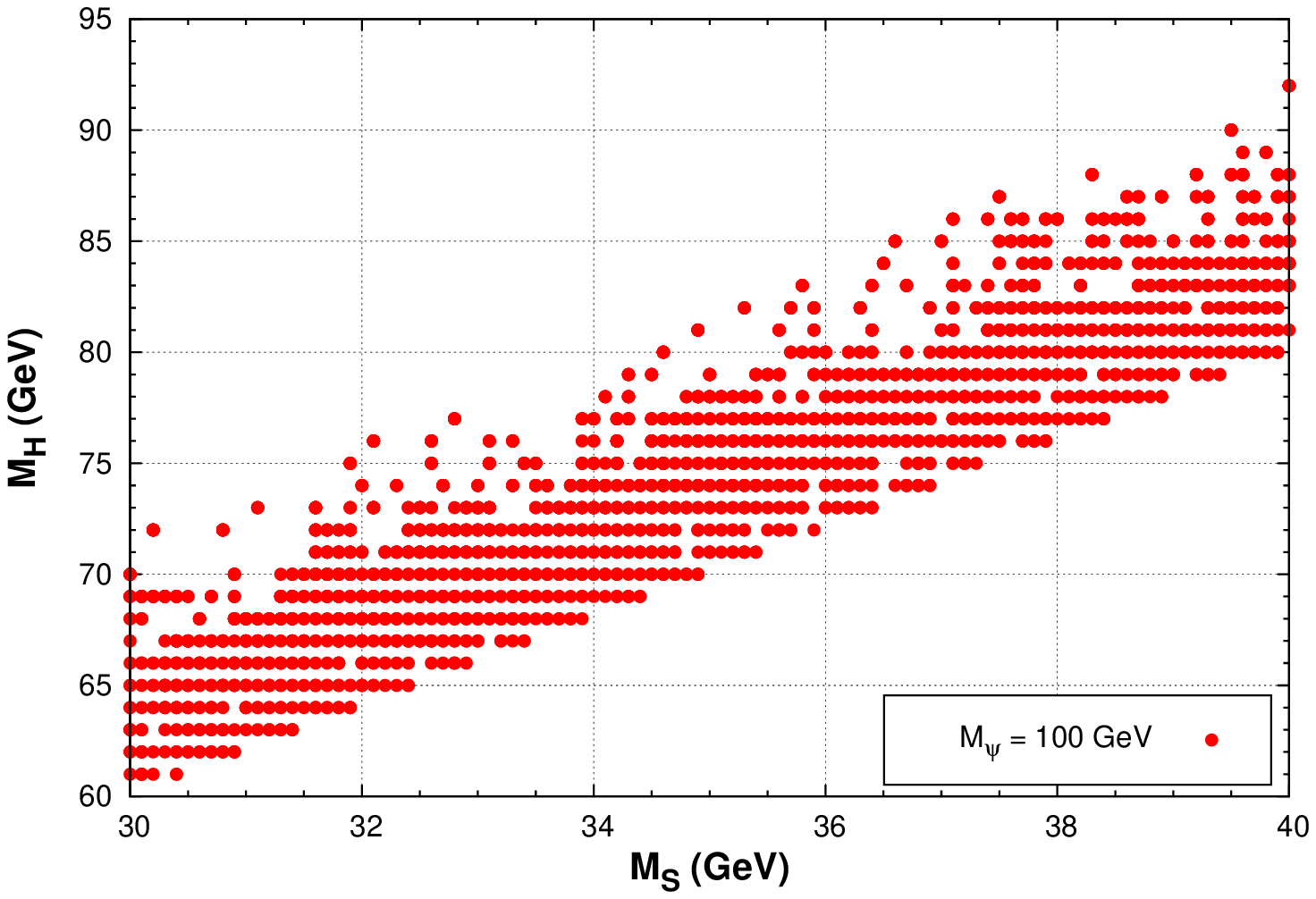}
\caption{Left panel : Variations of the fractional contributions of the
dark matter component $\psi$ with its mass $M_{\psi}$ for different values
of the gauge coupling $g_{\rm X}$ and the mass of $S$ ($M_{S}$).$~~~~~$
Right panel : Variations of the allowed ranges of $M_{H}$ with $M_S$ while
the mass of heavier dark matter is fixed at 100 GeV.}
\label{plot4}
\end{figure}
\paragraph{}
Left panel of Fig. \ref{plot4} describes the variations of the ratio
$\frac{\Omega_{\psi}h^2}{\Omega_{\rm T}h^2}$
with the heavier dark matter mass ($M_{\psi}$) for different values
of the gauge coupling $g_{\rm X}$ and the mass ($M_S$) of $S$. The red band represents
the variations for $g_{\rm X} = 0.25$, $M_{S}=35$ GeV while the green and blue coloured
bands are for two different values of $M_{S}$ namely, $30$ and $35$ GeV respectively with
the same gauge coupling $g_{\rm X}=0.30$. From this figure (left panel of Fig. \ref{plot4})
it appears that the contribution of the heavier dark matter component $\psi$
to the overall dark matter density increases with its mass $M_{\psi}$.
However, the ratio $\frac{\Omega_{\psi}h^2}{\Omega_{\rm T}h^2}$
decreases with $g_{\rm X}$ and 
remains nearly same for the variations of the
mass of the lighter dark matter component $S$ from 30 GeV to 35 GeV.
This increment of $\frac{\Omega_{\psi}h^2}{\Omega_{\rm T}h^2}$
is evident from Eqs. (\ref{psipsiz1z1}, \ref{sigma-psi}) which indicate that the
annihilation cross sections of $\psi$ for the channels $\psi \bar{\psi}
\rightarrow Z^{\prime} Z^{\prime}$, $S S^\dagger$ decrease
with $M_{\psi}$. On the other hand as the annihilation
cross sections of these two channels are proportional
to $g^4_{\rm X}$ therefore the fractional
contribution of $\psi$ to the total relic density
decreases with $g_{\rm X}$.
The right panel of Fig. \ref{plot4}
represents how the range of allowed values of $M_{H}$
vary with the mass of $S$ for $M_{\psi}=100$ GeV.
Needless to mention that, for not only each point
in the right panel of Fig. \ref{plot4} but for every other plots given in both panels of
Figs. \ref{plot1} - \ref{plot4}, all the possible constraints
discussed in the Section \ref{model} namely, the vacuum stability bounds,
the relic density constraints, the constraints obtain from the results of
dark matter direct detection experiments and LHC etc. are always satisfied.

{\bf Signal strength of the scalar $H$}-
We have also computed the signal strength of the
scalar boson $H$ for its various allowed decay channels
($H\rightarrow X \bar{X}$). The signal strength ($R^{\prime}_{X\bar{X}}$)
of $H$ for a particular decay mode is defined as
the ratio of production cross section times the branching
ratio of that decay channel of $H$ with the
SM expectation of these quantities. Thus the signal strength
of $H$ can be expressed as
\begin{eqnarray}
R^{\prime}_{X \bar{X}}~~ = \frac{\sigma_H}{\sigma_H^{\rm SM}}
\frac{{\rm BR}(H \rightarrow X \bar{X})}
{~~~~{\rm BR}(H \rightarrow X \bar{X})^{\rm SM}}\ ,
\label{signal-strenght1}
\end{eqnarray}
\begin{figure}[h!]
\includegraphics[height=9cm,width=7cm,angle=-90]{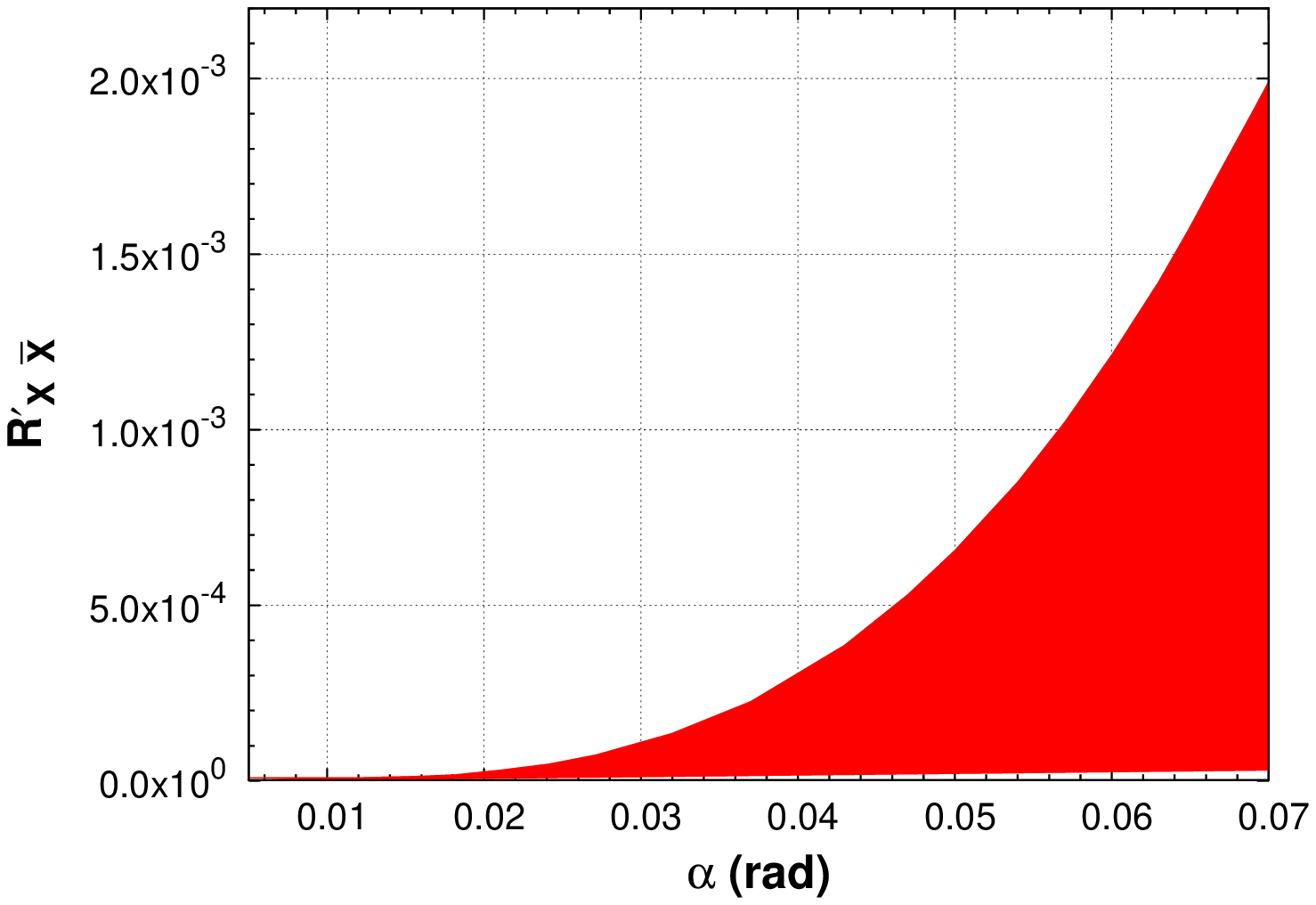}
\includegraphics[height=9cm,width=7.3cm,angle=-90]{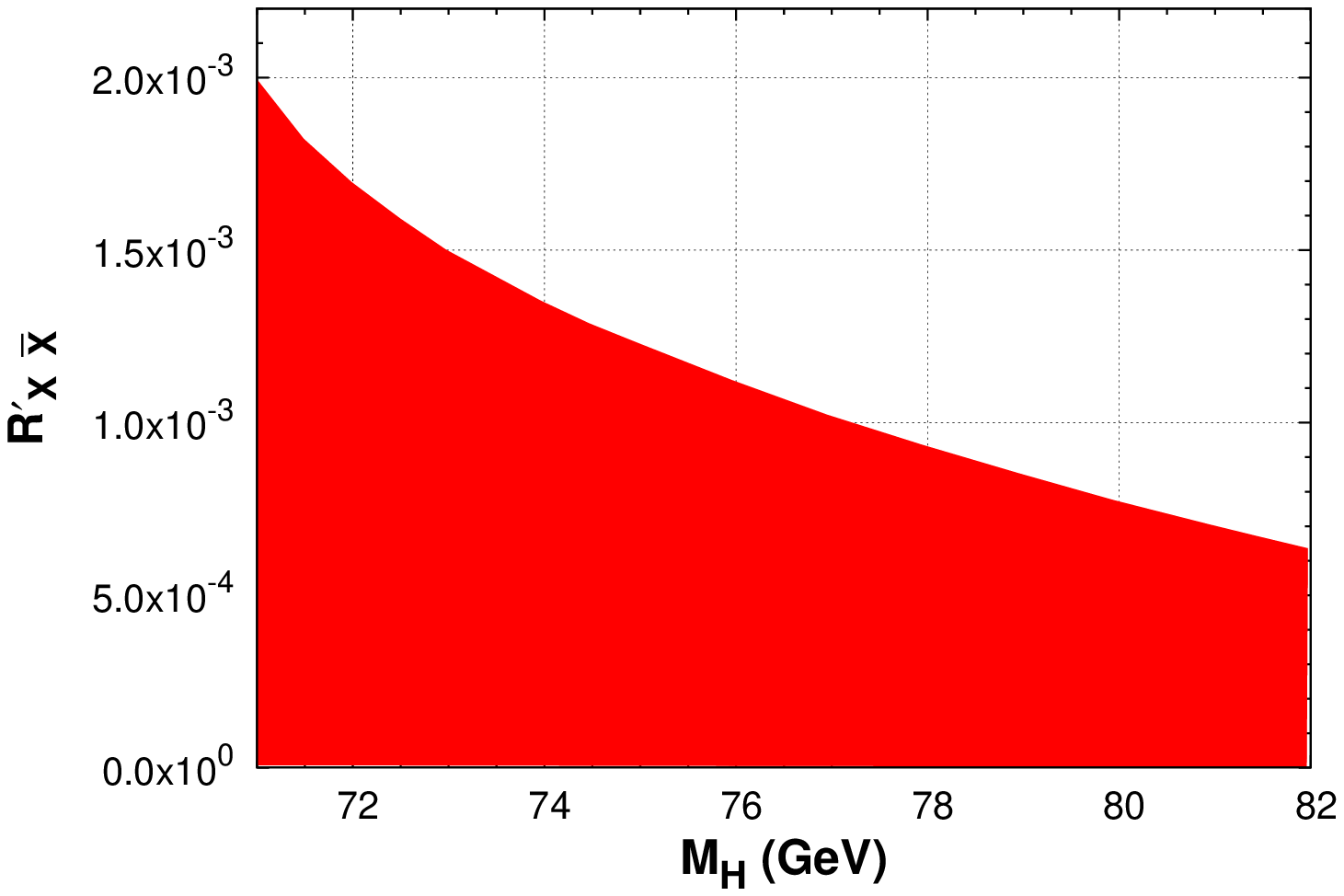}
\caption{Left (Right) panel- Variation of signal strength of the
scalar boson $H$ for a particular decay channel $H\rightarrow X \bar{X}$
with mixing angle $\alpha$ (mass $M_H$ of $H$). Here $X$ is any SM particle.}
\label{rpxx}
\end{figure}
In the left panel of Fig. \ref{rpxx} we show the variation of
signal strength ($R^{\prime}_{X\bar{X}}$) of the scalar boson $H$
for a particular decay channel $H\rightarrow X\bar{X}$ ($X$ is any
SM particle) with the allowed range of the mixing angle $\alpha$ between
$h$ and $H$ obtained from right panel of Fig. \ref{plot1}.
The variation of the same quantity ($R^{\prime}_{X\bar{X}}$)
with the mass\footnote{These plots are drawn for the mass of the
lighter dark matter particle $M_S$ = 35 GeV. We have
shown earlier (right panel of Fig. \ref{plot4})
that for $M_S=35$ GeV, both the relic density
as well as the direct detection criteria are
simultaneously satisfied only when $M_H$ lies
between $\sim 70$ GeV to $82$ GeV.} 
of $H$ is also shown in the right panel of Fig. \ref{rpxx}.
The increase of $R^{\prime}_{X\bar{X}}$ with $\alpha$ is
due to the fact that coupling strength of $H$ with
any SM particle $X$ is proportional to the mixing angle $\alpha$.
It is evident from both the plots of Fig. \ref{rpxx} that within the
allowed ranges of $\alpha$ and $M_{H}$, the signal strength
of $H$ is extremely weak and its value is $\la 2\times 10^{-3}$
in units of the Standard Model production cross section of $H$.   

\begin{figure}[h!]
\centering
\includegraphics[height=11cm,width=8cm,angle=-90]{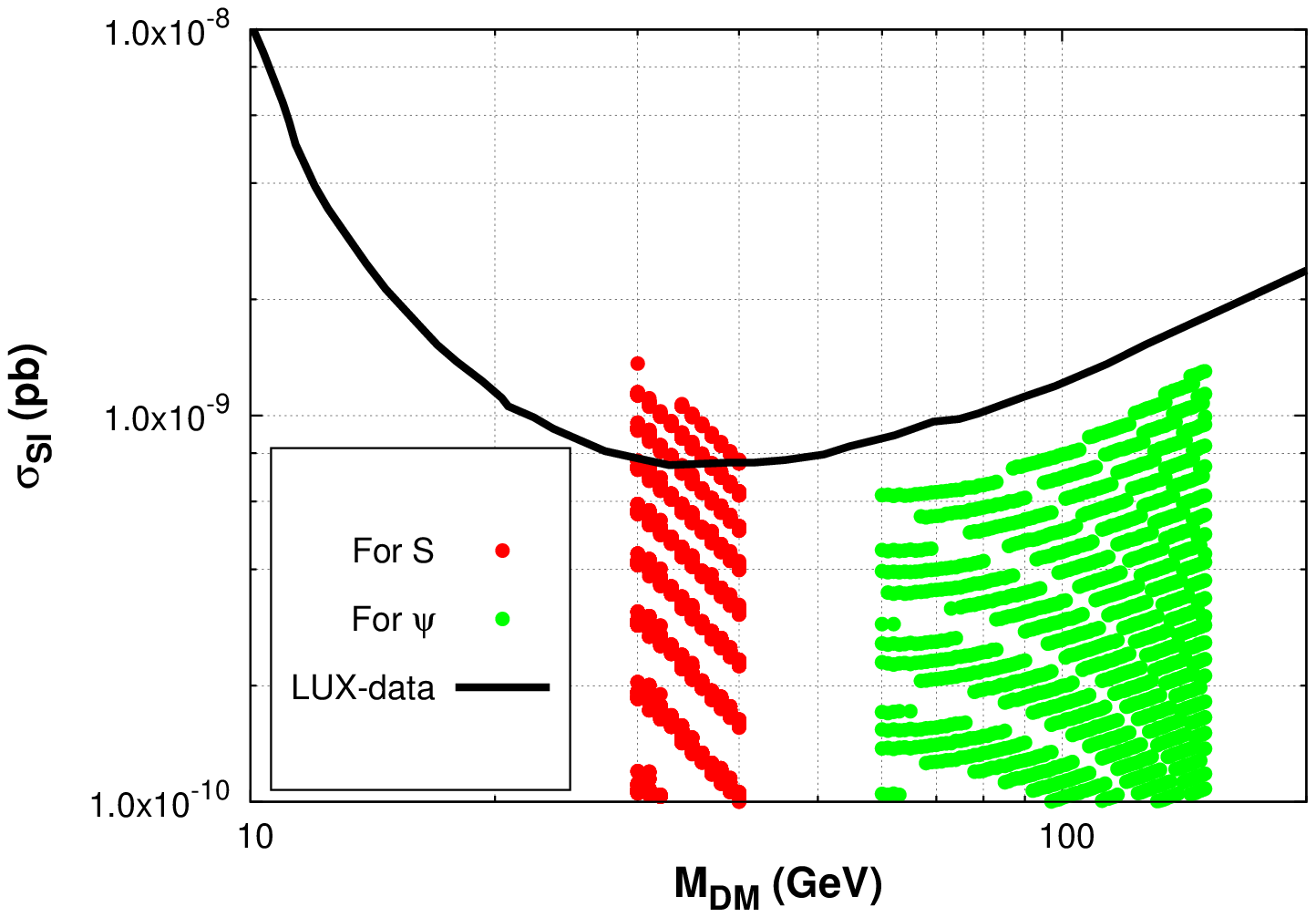}
\caption{Spin independent ``effective scattering cross sections" of the dark matter
component $S$ (red coloured region) and $\psi$ (green coloured region). Limits
obtained from LUX experiment are denoted by black solid line.}
\label{plot5}
\end{figure} 
\paragraph{}
We have already discussed in the Section \ref{model} that
in order to compare the direct detection results computed using
a multicomponent dark matter model, one should rescale the
spin independent scattering cross sections
for both the dark matter components by an
appropriate factor (see Eq. (\ref{effective-scatter}))
which manifests the existence of multicomponent dark matter. 
In view of this, the spin independent ``effective scattering cross sections"
between each of the dark matter candidate ($S$, $\psi$) and
the nucleon are plotted in Fig. \ref{plot5}. Here red coloured
zone represents ${{\sigma^\prime}^{\,i N\rightarrow i N }_{\rm SI}}$
for the candidate $S$ ($i=S$) however, for the heavier dark matter candidate
$\psi$ ($i=\psi$) it is indicated by the green coloured band. For the comparison
with the latest direct detection experimental results the data \cite{Akerib:2013tjd}
obtained from the LUX experiment are superimposed on the same figure (Fig. \ref{plot5}).
From the Fig. \ref{plot5} it is seen that although some portions,
only for the lighter DM component $S$, in the $\sigma_{\rm SI}-M_{\rm DM}$ plane 
are already excluded by the LUX experiment, still there exist
enough allowed regions in the $\sigma_{\rm SI}-M_{\rm DM}$ plane
for both the dark matter components $S$ and $\psi$ which can be
tested by more sensitive (``ton-scale") direct detection
experiments in near future. 
%%%%%%%%%%%%%%%%%%%%%%%%%%%%%%%%%%%%%%%%%%%%%%%%%%%%%%%%%%%%%%%
\paragraph{}
In Table \ref{tab2} we show the ranges of values of different observable
quantities, namely $\alpha$, $\theta_{\rm NB}$, $R_{X\bar{X}}$,
$R^{\prime}_{X\bar{X}}$, ${\rm BR}_{\rm inv}$, $\Gamma_h$, $\Gamma_{H}$,
$\frac{\Omega_S h^2}{\Omega_{\rm T} h^2}$, $\frac{\Omega_\psi h^2}
{\Omega_{\rm T}h^2}$, ${\sigma^\prime}^{\,S N\rightarrow S N }_{\rm SI}$
,${\sigma^\prime}^{\,\psi N\rightarrow \psi N }_{\rm SI}$, which are
allowed by various theoretical, observational and experimental constraints
listed in Section \ref{model}. These ranges are obtained for the
chosen values of the kinetic mixing parameter $\epsilon = 1.0\times 10^{-3}$
and the U(1)$_{\rm X}$ gauge coupling $g_{\rm X}=0.3$. The other relevant model
parameters namely $M_{S}$, $M_{\psi}$, $M_{Z^{\prime}}$, $M_{H}$, $\lambda_{2,\,3}$,
required for computing these allowed ranges of different
observables, are varied over their entire ranges mentioned in Eq. (\ref{para-ranges}). 
\begin{table}[h!]
\begin{center}
\vskip 0.5cm
\begin{tabular} {|c|c|c|c|c|c|}
\hline
\hline
$\alpha$ & $\theta_{\rm NB}$ & $R_{X\bar{X}}$ & $R^{\prime}_{X\bar{X}}$
&${\rm BR}_{\rm inv}$&$\Gamma_h$\\
(rad)&(rad)&&&of $h$&(GeV)\\
\hline
$\la 7.0\times$&$\sim (0.6-1.65)$&$\ga 0.80 $ & $\la 2.0\times 10^{-3}$&
$\la 20\%$&$\sim (4.1-5.1)$\\
$10^{-2}$& $\times 10^{-3}$&&&&$\times 10^{-3}$\\
\hline
\hline
\end{tabular}
\begin{tabular}{|c|c|c|c|c|}
$\Gamma_{H}$&$\frac{\Omega_S h^2}{\Omega_{\rm T} h^2}$&$\frac{\Omega_\psi h^2}
{\Omega_{\rm T}h^2}$&${\sigma^\prime}^{\,S N\rightarrow S N }_{\rm SI}$
&${\sigma^\prime}^{\,\psi N\rightarrow \psi N }_{\rm SI}$\\
(GeV)&&&(pb)&(pb)\\
\hline
$\sim(0.08-2.3)$&$\sim 0.7-0.94$&$\sim 0.06-0.3$&$\sim(0.1-0.8)$&$\sim(0.1-1.3)$\\
$\times 10^{-4}$&&&$\times 10^{-9}$&$\times 10^{-9}$\\
\hline
\hline
\end{tabular}
\end{center}
\caption{Allowed ranges of various observable
quantities computed using the present two component DM model
for $\epsilon = 1.0\times10^{-3}$, $g_{\rm X}=0.3$.
Other relevant model parameters, required for this computation,
are scanned over their entire considered ranges
mentioned in Eq. (\ref{para-ranges}).}
\label{tab2}
\end{table}
%%%%%%%%%%%%%%%%%%%%%%%%%%%%%%%%%%%%%%%%%%%%%%%%%%%%%%%%%%%%%%%%%%%%%%%%%%%%
\section{1-3 GeV $\gamma$ excess from Galactic Centre.}
\label{1-3gev-gamma}
%%%%%%%%%%%%%%%%%%%%%%%%%%%%%%%%%%%%%%%%%%%%%%%%%%%%%%%%%%%%%%%%%%%%%%%%%%%%%
In this section our endeavour is to explain
recently observed excess $\gamma$-rays 
at an energy range 1-3 GeV by Fermi-LAT from the Galactic Centre
region within the framework of the proposed
two component dark matter model.
In the present two component DM scenario, lighter dark matter
component namely, $S$ having mass in the range $30\sim40$
GeV annihilates with its own antiparticle $S^\dagger$
and thereby produces ${\rm b} \bar{\rm b}$ pair in the final state
with branching ratio $\sim 85\%-90\%$. Thereafter those b-quarks
hadronise to produce $\gamma$-rays that
can be detected by Fermi-LAT. The annihilation process
($SS^{\dagger}\rightarrow {\rm b}\bar{\rm b}$)
of the dark matter component $S$ proceeds mainly
through the exchange of SM like Higgs boson $h$ and $H$.
The Feynman diagram for this annihilation process is shown
in the left panel of Fig. \ref{anni-dia}.
The differential $\gamma$-ray flux due to
self annihilation of $S$ in the GC region is given by \cite{Cirelli:2010xx}
\footnote{In Ref. \cite{Cirelli:2010xx}, there is an extra half factor in the
expression of differential $\gamma$-ray flux (Eq. (\ref{gamma-flux}))
originating from the annihilation of dark matter particle which
is not its own antiparticle. In our case, as already mentioned before,
the lighter dark matter candidate is represented by a complex scalar
field $S$. Hence, $S$ is not self conjugate and the extra half
factor which should be present in Eq. (\ref{gamma-flux}) due to
the nature of dark matter particle is already taken into account when
we have define the expression of thermally averaged annihilation cross section
of $S$ (see Eq. (\ref{thermal-ave}) for the definition of
$\langle{\sigma {\rm v}}_{SS^\dagger\rightarrow {\rm b} {\bar{\rm b}}}\rangle$).},  
\begin{eqnarray}
\frac{d\Phi_{\gamma}}{d \Omega d E} &=& \frac{r_{\odot}}{8\pi}
\left(\frac{\rho_{\odot}}{M_{S}}\right)^2 ~\bar{J}~ 
{\langle{\sigma {\rm v}}_{SS^\dagger\rightarrow {\rm b} {\bar{\rm b}}}\rangle}^\prime
\frac{d N^{\rm b}_{\gamma}}{dE}\,\,,
\label{gamma-flux}
\end{eqnarray}
where $\frac{dN^{\rm b}_{\gamma}}{dE}$ is the energy spectrum of photons
produced from the hadronisation processes of ${\rm b}$ quarks\footnote{originate
from the self annihilation of $S$ through the process $S S^\dagger
\rightarrow {\rm b} \bar{\rm b}$.}. We have
used the numerical values of the photon spectrum %($\frac{dN^{\rm b}_{\gamma}}{dE}$)
for different values of photon energy $E_{\gamma}$, given in Ref. \cite{Cirelli:2010xx}.
In the above $\rho_{\odot} = 0.3$ GeV$/$cm$^3$ is the dark matter density at the
solar location which is nearly $r_{\odot}\simeq 8.5$ kpc away from the
Galactic Centre. The quantity $\bar{J}$ for the case of dark matter annihilation in the GC
can be expressed as,
\begin{eqnarray}
\bar {J} &=& \frac{4}{\Delta \Omega} \int \int db~dl \cos b~J( b, l) \,\,,
\label{jbar}
\end{eqnarray}
with
\begin{eqnarray}
J(l, b) &=& \int_{\rm l.o.s} \frac{d\mathfrak{s}}{r_\odot} 
\left(\frac{\rho(r)}{\rho_{\odot}}\right)^2 \,\, ,
\label{j}
\end{eqnarray}
and
\begin{eqnarray}
\Delta\Omega = 4 \int dl \int db \cos b \,\, ,
\label{solidangle} 
\end{eqnarray}
\begin{eqnarray}
r &=& \left(r_{\odot}^2 + {\mathfrak{s}}^2 - 2\,r_{\odot}\,{\mathfrak{s}}
\cos b \cos l\right)^{{1}/{2}}\,\,\, .
\label{radius}
\end{eqnarray}
In Eqs. (\ref{jbar}, \ref{solidangle}, \ref{radius}), $l$ and $b$
represent the galactic longitude and latitude respectively. While computing
the values of $\bar{J}$ we perform the integral over a region
which is situated within a radius of $5^o$ \cite{Daylan:2014rsa} around the GC.  
Integral over $\mathfrak{s}$ in Eq. (\ref{j}) is along the line of sight (l.o.s) distance.
The quantity ${\langle{\sigma {\rm v}}_{SS^\dagger
\rightarrow {\rm b} {\bar{\rm b}}}\rangle}^\prime$
in Eq. (\ref{gamma-flux}) is the ``effective annihilation cross section" which 
is in the product of annihilation cross section of the channel
$SS^\dagger\rightarrow {\rm b} \bar{\rm b}$ and square of the contribution of the component
$S$ to the total dark matter relic density
($\Omega_{\rm T} h^2$) {\it i.e.}
\begin{eqnarray}
{\langle{\sigma {\rm v}}_{SS^\dagger\rightarrow {\rm b} {\bar{\rm b}}}\rangle}^\prime
= \xi^{2}_{S}
{\langle{\sigma {\rm v}}_{SS^\dagger\rightarrow {\rm b} {\bar{\rm b}}}\rangle}  \,\, ,
\label{modi-anni}
\end{eqnarray}
where 
\begin{eqnarray}
\xi_{S} = \frac{\Omega_{S}}{\Omega_{\rm T}} \,\,
\label{xi_s}
\end{eqnarray}
is the fractional relic density of the component $S$.
The use of ${\langle{\sigma {\rm v}}_{SS^\dagger
\rightarrow {\rm b} {\bar{\rm b}}}\rangle}^\prime$
(``effective annihilation cross section") instead of actual annihilation cross section
for the channel $SS^\dagger\rightarrow {\rm b} \bar{\rm b}$
(${\langle{\sigma {\rm v}}_{SS^\dagger\rightarrow {\rm b} {\bar{\rm b}}}\rangle}$)
in Eq. (\ref{gamma-flux}) is needed since we are working in a framework with
more than one component of the dark matter. Note that if the
entire dark sector is composed of only one type of particle (say $S$)
then $\xi_{S} =1$, therefore
${\langle{\sigma {\rm v}}_{SS^\dagger\rightarrow
{\rm b} {\bar{\rm b}}}\rangle}^\prime$ and ${\langle{\sigma {\rm v}}
_{SS^\dagger\rightarrow {\rm b} {\bar{\rm b}}}\rangle}$ are identical.
The expression of ${\langle{\sigma {\rm v}}_{SS^\dagger\rightarrow
{\rm b} {\bar{\rm b}}}\rangle}$ is given in Eq. (\ref{sigmaff}) of Section \ref{boltz-eqn}.
Computation of $\gamma$-ray flux using Eq. (\ref{gamma-flux}) requires
the nature of the variation of the dark matter density
in the neighbourhood regions of the Galactic Centre
with the distance $r$. In short one needs to know the DM halo profile
$\rho(r)$ as a function of $r$. In the present work we use
%In order to compute the $\gamma$-ray flux using Eq. (\ref{gamma-flux})
%one has to know how the dark matter density in the neighbourhood regions of the GC
%varies with the distance $r$, {\it i.e.} in other word the functional dependence of $\rho(r)$
%with $r$. This functional relationship between $\rho(r)$ and $r$ is known as halo profile
%of DM. Since the actual functional form (halo profile) is still unavailable to us, 
%we use an approximate dark matter halo profile namely 
the NFW profile \cite{Navarro:1996gj} with $\gamma = 1.26$ \cite{Daylan:2014rsa}.
The general expression of NFW profile is given by, 
\begin{eqnarray}
\rho_{\rm NFW} = \rho_s \frac{\left(\frac{r}{r_s}\right)^{-\gamma}}
{\left(1+\frac{r}{r_s}\right)^{3-\gamma}} \,\,,
\end{eqnarray}
where scale radius $r_s =$ 20 Kpc. The normalisation constant $\rho_s$ (scale density)
is obtained by demanding that at the solar location ($r = r_{\odot}$)
the dark matter density should be 0.3 GeV$/$cm$^3$.
\begin{figure}[h!]
\centering
\subfigure[$M_{\psi} = 100$ GeV]
{\includegraphics[height=8cm,width=6cm,angle=-90]{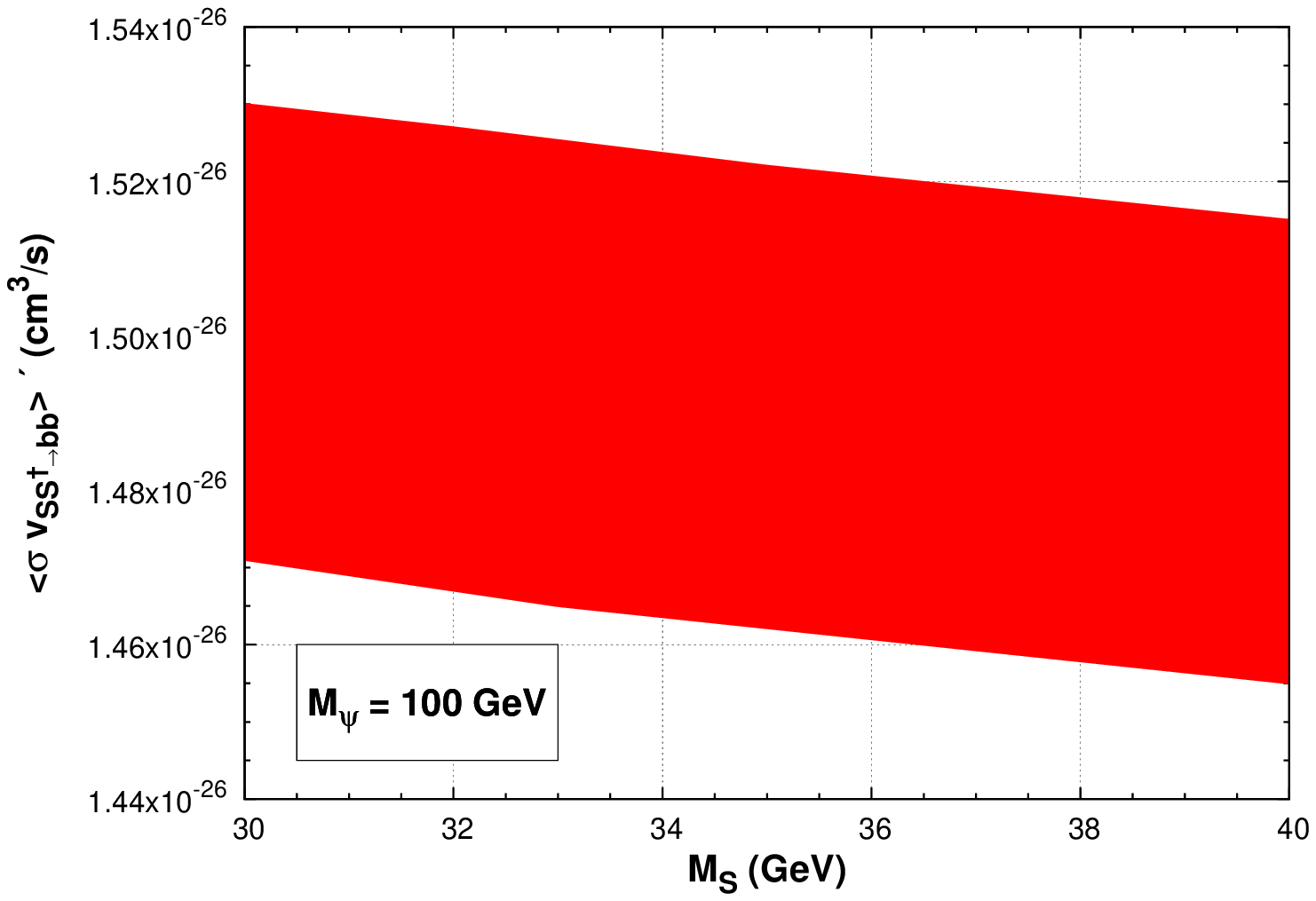}}
\subfigure[$M_{\psi} = 80$ GeV]
{\includegraphics[height=8cm,width=6cm,angle=-90]{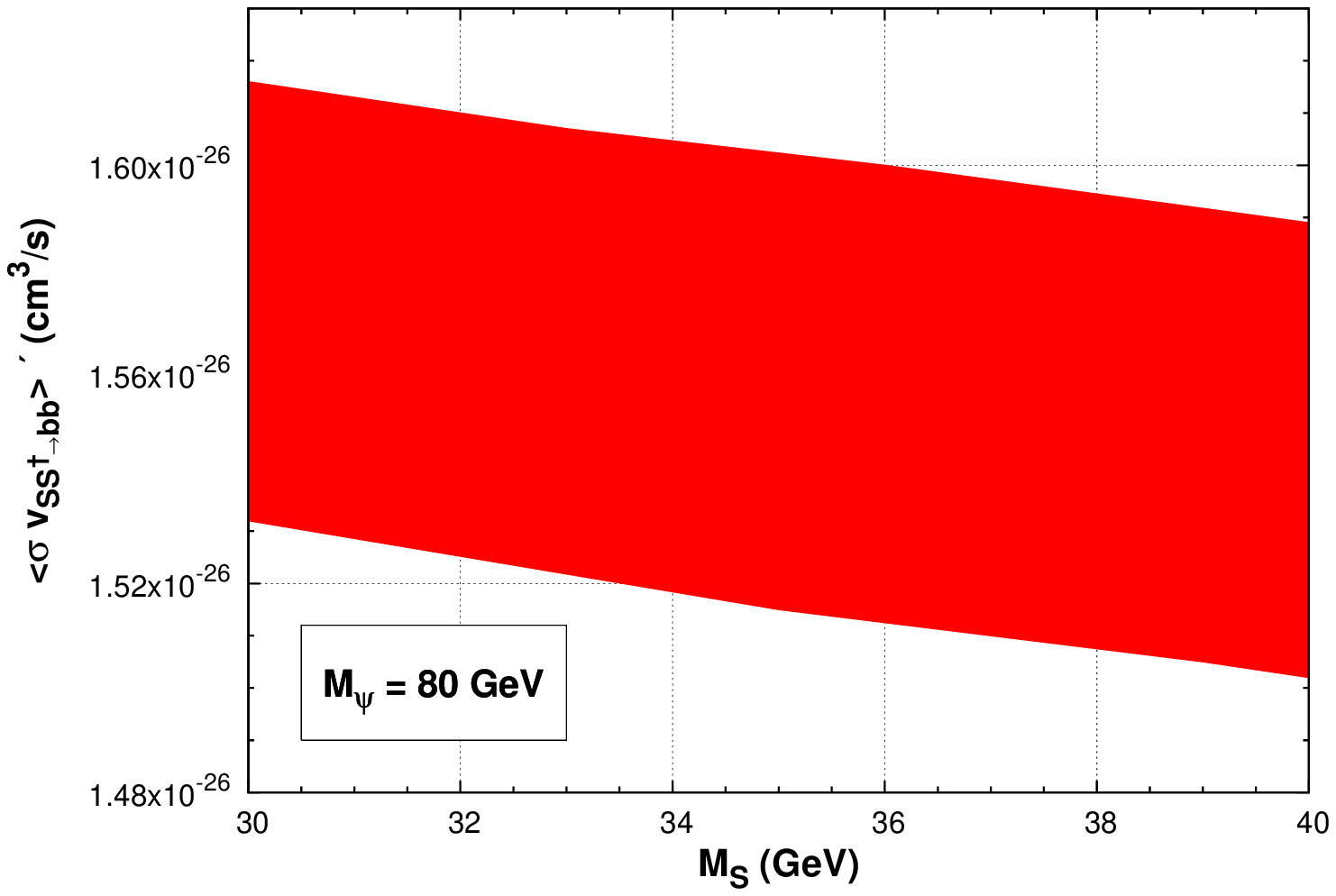}}\\
\subfigure[$M_{\psi} = 60$ GeV]
{\includegraphics[height=9cm,width=6cm,angle=-90]{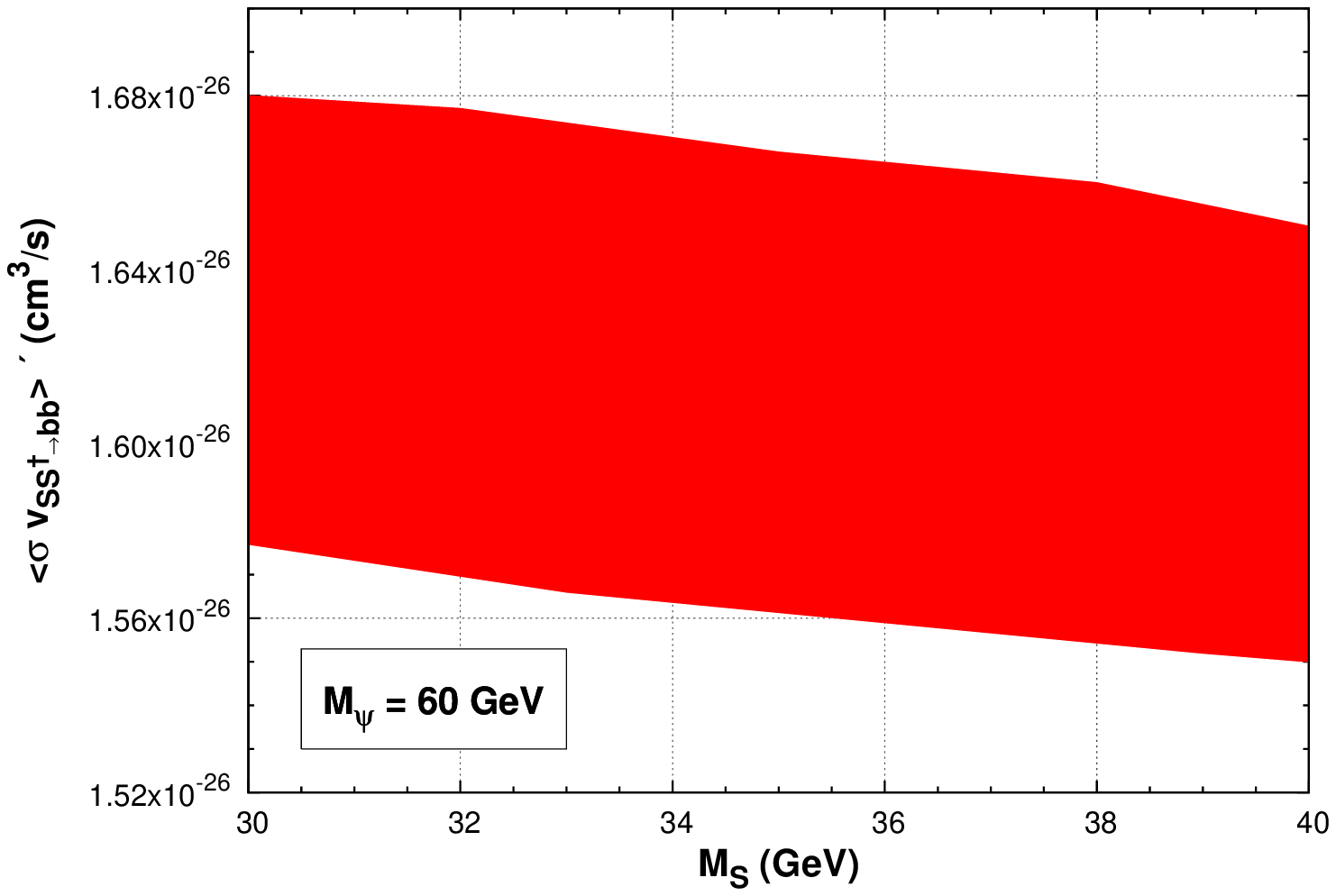}}
\caption{Left-Panel : Variations of 
${\langle{\sigma {\rm v}}_{SS^\dagger\rightarrow {\rm b} {\bar{\rm b}}}\rangle}^\prime$
with the mass of the dark matter component $S$ for different
values of the mass of heavier dark matter component $\psi$.}
\label{flux-plot1}
\end{figure}

\begin{figure}[h!]
\centering
\subfigure[$M_{\psi} = 100$ GeV]{
\includegraphics[height=8cm,width=7cm,angle=-90]{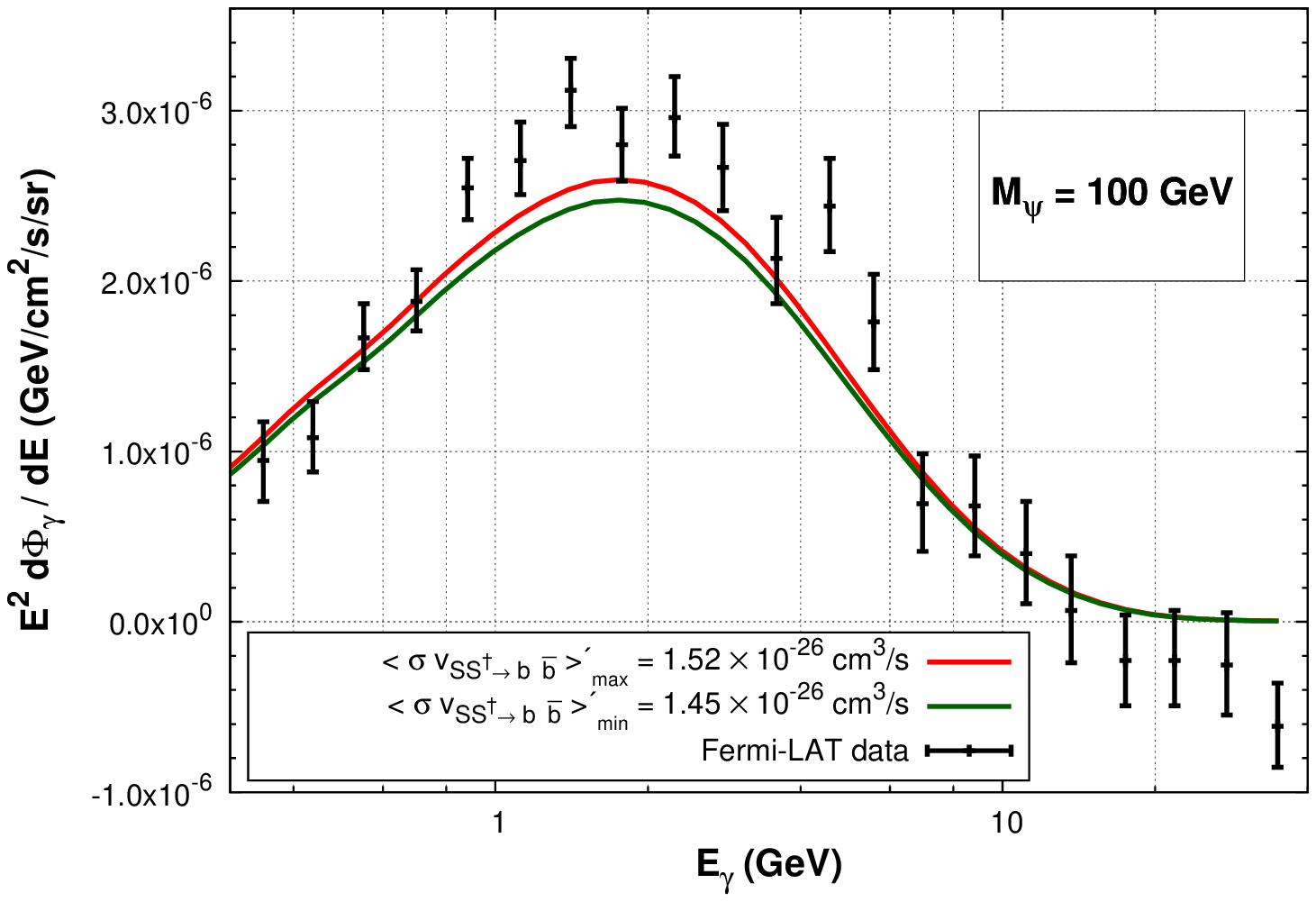}}
\subfigure[$M_{\psi} = 80$ GeV]
{\includegraphics[height=8cm,width=7cm,angle=-90]{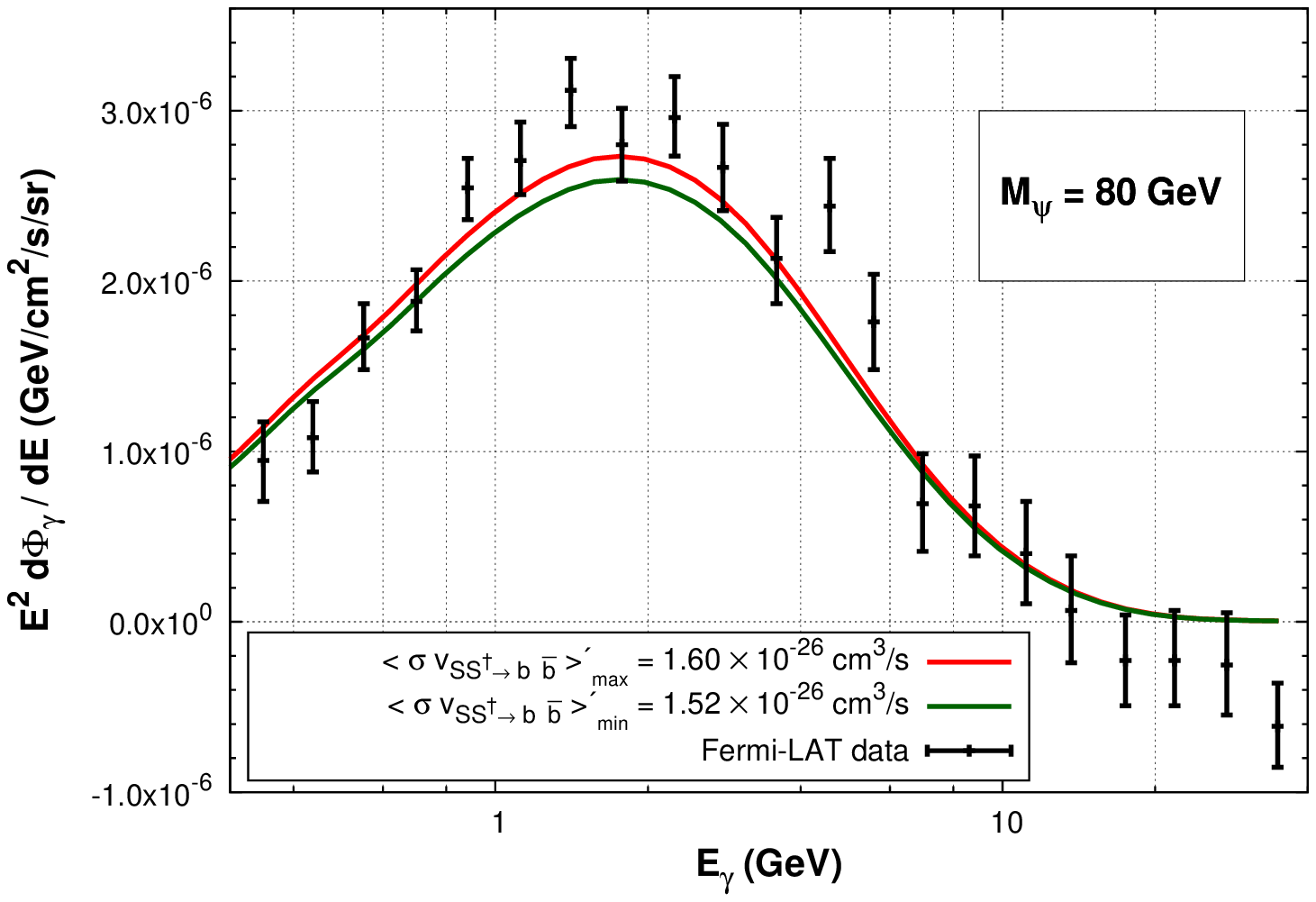}}\\
\subfigure[$M_{\psi} = 60$ GeV]
{\includegraphics[height=9cm,width=8cm,angle=-90]{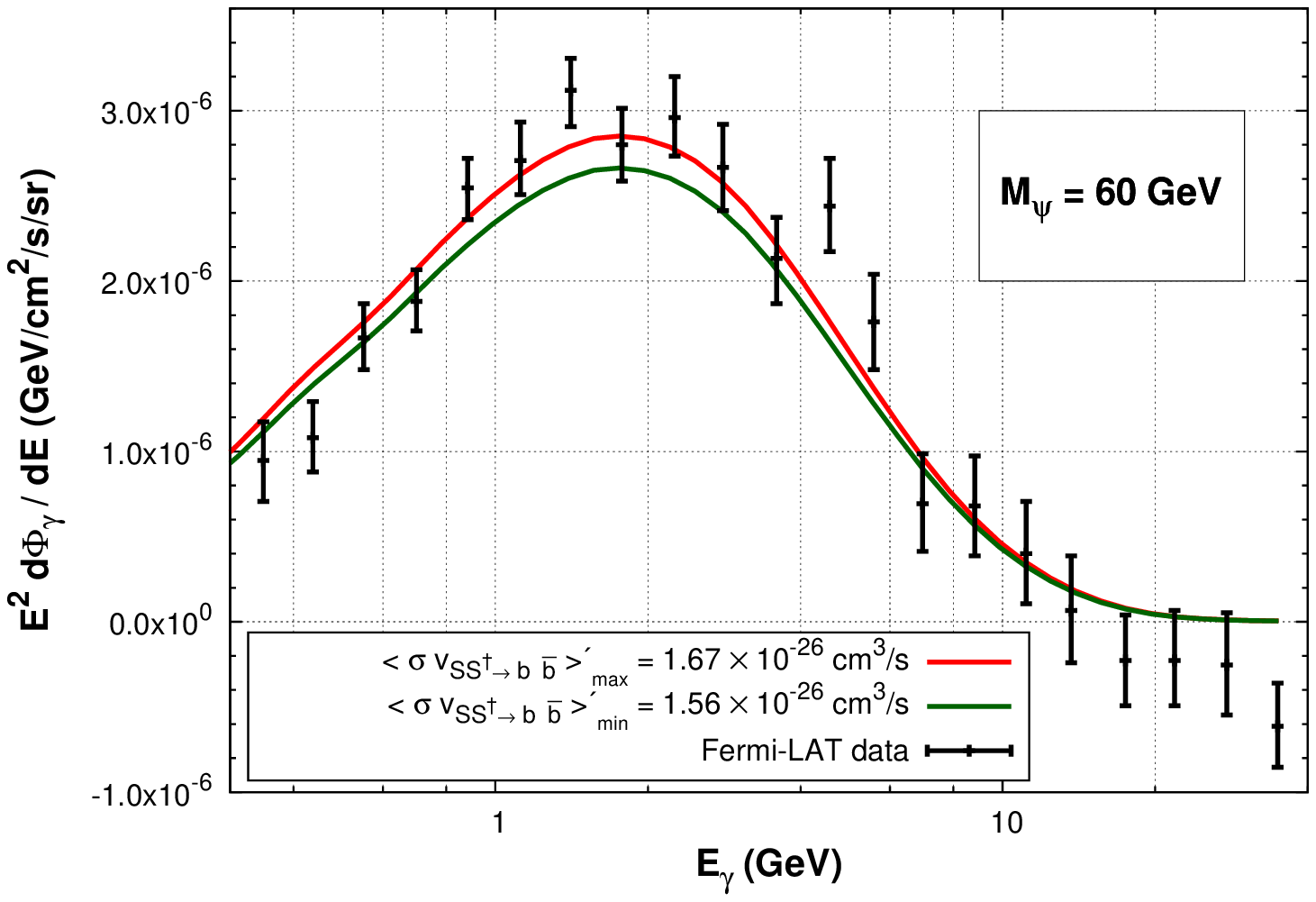}}
\caption{Variations of $\gamma$-ray flux obtain from the annihilation
of $S$ in the Galactic Centre as a function of the photon energy (${\rm E}_{\gamma}$)
for three different values of $M_{\psi}$.}
\label{flux-plot2}
\end{figure}
\paragraph{}
In Fig. \ref{flux-plot1}(a-c) we show the variations of the ``effective
annihilation cross section" 
${\langle{\sigma {\rm v}}_{SS^\dagger\rightarrow {\rm b} {\bar{\rm b}}}\rangle}^\prime$
(defined in Eq. (\ref{modi-anni})) for annihilation channel
$SS^\dagger\rightarrow {\rm b} \bar{\rm b}$ with the mass of the
the dark matter component $S$ in the range of 30 GeV - 40 GeV for three different
values of $M_{\psi}$ namely, 60, 80 and 100 GeV.
Since the quantity 
${\langle{\sigma {\rm v}}_{SS^\dagger\rightarrow {\rm b} {\bar{\rm b}}}\rangle}^\prime$
is the product of annihilation cross section of $S$ for the channel
$SS^\dagger\rightarrow {\rm b} \bar{\rm b}$
and square of the fractional contribution of $S$ to the total
relic density ($\xi^2_S$) (see Eq. (\ref{modi-anni})), therefore
in all the plots (a-c) of Fig. \ref{flux-plot1}
we have taken only those values of both the quantities
${\langle{\sigma {\rm v}}_{SS^\dagger\rightarrow {\rm b} {\bar{\rm b}}}\rangle}$
and $\xi_{S}$ for which the total dark matter density lies within the range predicted by the
PLANCK experiment (Eq. (\ref{planck-limit})). All the plots in Fig. \ref{flux-plot1}
show that the ``effective annihilation cross section" of the channel
$SS^\dagger\rightarrow {\rm b} {\bar{\rm b}}$
decreases as the mass ($M_{\psi}$) of the heavier
dark matter component increases. From Eq. (\ref{omega_tot}) and
Eq. (\ref{xi_s}) we know that the fractional contributions of
both the dark matter components namely, $\psi$ and $S$
are related through the relation
\begin{eqnarray}
\xi_{\psi} + \xi_{S} = 1 \,\,.
\label{frac-densities-relation}
\end{eqnarray}  
Now, $\xi_{\psi}$ increases with $M_{\psi}$ (see left panel of Fig. \ref{plot4})
therefore in order to satisfy the above relation given in Eq. (\ref{frac-densities-relation})
the fraction contribution ($\xi_S$) of the lighter dark
matter component to the overall density
decreases with the increase of heavier dark matter mass. This results in an
enhancement in the pair annihilation rate
(${\langle{\sigma {\rm v}}_{SS^\dagger\rightarrow {f} {\bar{f}}}\rangle}$, $f$ is any
SM fermion except top quark) of the dark matter component $S$.
Since
${\langle{\sigma {\rm v}}_{SS^\dagger\rightarrow {\rm b} {\bar{\rm b}}}\rangle}^\prime$
is proportional to $\xi^2_S$ (see Eq. (\ref{modi-anni}))
hence, the ``effective annihilation cross section" of
$S$ for the channel $SS^\dagger\rightarrow {\rm b}\bar{\rm b}$ decreases with $M_{\psi}$.
Similarly, one can understand the variations of
${\langle{\sigma {\rm v}}_{SS^\dagger\rightarrow {\rm b} {\bar{\rm b}}}\rangle}^\prime$
with the mass of $S$ using left panel of Fig. \ref{plot2}. From all the
plots (a-c) of Fig. \ref{flux-plot1} it appears the ``effective annihilation
cross section" of the dark matter candidate $S$ for the annihilation channel
$SS^{\dagger} \rightarrow {\rm b} \bar{\rm b}$ lies within the range
$(1.45\sim 1.68)\times 10^{-26}{\rm cm^3}/{\rm s}$. This
allowed range of   
${\langle{\sigma {\rm v}}_{SS^\dagger\rightarrow {\rm b} {\bar{\rm b}}}\rangle}^\prime$
falls well below the current bounds on the dark matter annihilation
cross section into ${\rm b} \bar{\rm b}$ final state from
collider results and other various measurements like
antiproton flux, positron flux, diffuse radio emission, CMB etc.
(see left panel of Fig. 1 of Ref. \cite{Kong:2014haa}). 
\paragraph{}
The plots (a-c) in Fig. \ref{flux-plot2} show the variations of the $\gamma$-ray flux
obtained from the regions surrounding the Galactic Centre due to the self annihilation of
the dark matter candidate $S$ of mass 35 GeV into ${\rm b} \bar{\rm b}$ final state for
three different values of $M_{\psi} = 100, 80$ and 60 GeV respectively. In each of the plots
(a-c) of Fig. \ref{flux-plot2} the red and green solid lines represent the $\gamma$-ray fluxes
computed by using maximum and minimum allowed values of the ``effective annihilation
cross section"
(${\langle{\sigma {\rm v}}_{SS^\dagger\rightarrow {\rm b}
{\bar{\rm b}}}\rangle}^\prime$)
for a particular value of the mass of heavier dark matter component $\psi$.
These quantities are represented by
${\langle{\sigma {\rm v}}_{SS^\dagger\rightarrow {\rm b}
{\bar{\rm b}}}\rangle}^\prime_{\rm max}$
and
${\langle{\sigma {\rm v}}_{SS^\dagger\rightarrow {\rm b}
{\bar{\rm b}}}\rangle}^\prime_{\rm min}$
respectively. The range of allowed values of
${\langle{\sigma {\rm v}}_{SS^\dagger\rightarrow {\rm b} {\bar{\rm b}}}\rangle}^\prime$
with the mass of $S$ for three different values of $M_{\psi}$ ($M_{\psi}=$ 60, 80 and 100 GeV)
are given in plots (a-c) of Fig. \ref{flux-plot1}. The black vertical lines in each of the plots
of Fig. \ref{flux-plot2} represent the Fermi-LAT data and corresponding error bars.
Since the allowed ranges of the ``effective annihilation cross section"
of the dark matter particle $S$ 
for the channel $S S^\dagger \rightarrow
{\rm b} \bar{\rm b}$ decrease with the increase of $M_{\psi}$
(see plots (a-c) of Fig. \ref{flux-plot1}
and the related discussions), therefore the corresponding
$\gamma$-ray fluxes (Eq. (\ref{gamma-flux}))
shown in Fig. \ref{flux-plot2} also decrease with $M_{\psi}$.
Therefore, comparing the gamma-ray fluxes computed for three different
values of $M_{\psi}$ with the Fermi-LAT data, we find that the
$\gamma$-ray fluxes obtained for $M_{\psi}=60$ and 80 GeV
agree well with the experimental data. Moreover, the $\gamma$-ray
fluxes computed for $M_{\psi}\geq 100$ GeV are incompatible
with the available Fermi-LAT data.
 
\begin{figure}[h!]
\centering
\includegraphics[height=10cm,width=8cm,angle=-90]{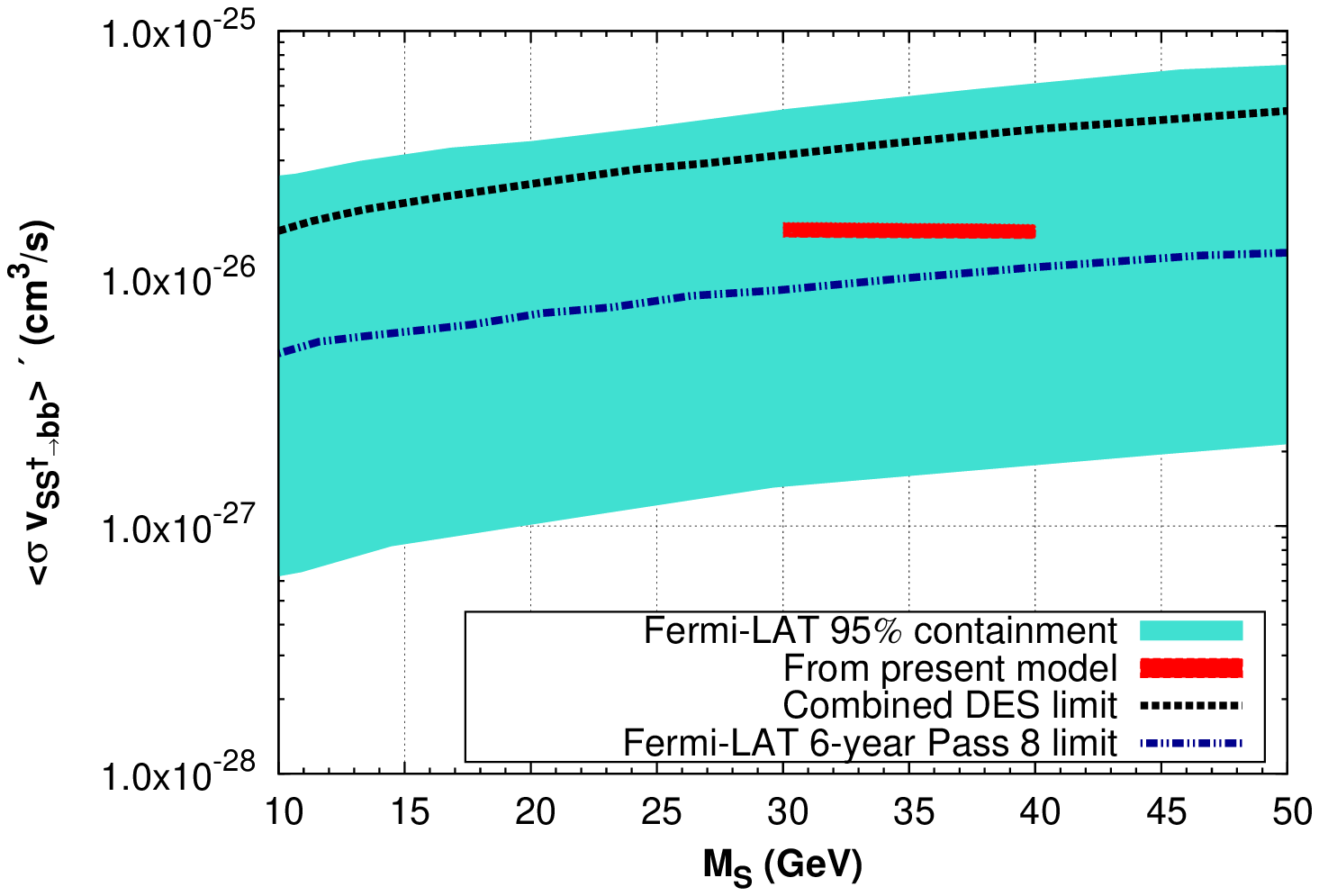}
\caption{Comparison between the range of allowed values of ``effective
annihilation cross section" of DM candidate
$S$ (for ${\rm b} \bar{\rm b}$ annihilation channel)
obtained from the present model and the 95\% C.L. upper limits as well as the
corresponding error bar reported by both Fermi-LAT and DES collaborations
after their analyses of data from dwarf spheroidal galaxies.}  
\label{compare-sigmavbb}
\end{figure} 
\paragraph{}
We have also compared the range of allowed values of the
``effective annihilation cross section"
($\langle\sigma {\rm v}_{SS^{\dagger}\rightarrow {\rm b} \bar{\rm b}}\rangle$)
of the dark matter candidate $S$ for the channel
$SS^\dagger\rightarrow {\rm b} \bar{\rm b}$ with the limits obtained from
the analyses of both Fermi-LAT collaboration \cite{Ackermann:2015zua} and
DES collaboration \cite{Drlica-Wagner:2015xua}. In Fig. \ref{compare-sigmavbb}
the dashed blue line represents the upper limits on dark matter
annihilation cross section for ${\rm b} \bar{\rm b}$ annihilation
channel. This limit is obtained after a combined analysis
of six years of Pass 8 Reprocessed data from 15 dwarf spheroidal
galaxies (dSphs) by the Fermi-LAT collaboration. The corresponding
95\% C.L. error band is shown by the turquoise coloured contour. The
red coloured patch is the range of allowed values of
$\langle\sigma {\rm v}_{SS^{\dagger}\rightarrow {\rm b} \bar{\rm b}}\rangle^\prime$
that is derived from the present model for different values of
relevant model parameters (see Fig. \ref{flux-plot1}) while
the 95\% C.L. combined upper limits on $\langle\sigma {\rm v}_{{\rm b} \bar{\rm b}}\rangle$,
obtained from the analysis of 8 new dSphs by the DES collaboration,
is described by the black dashed line. Although it appears from
Fig. \ref{compare-sigmavbb} that the allowed values of
$\langle\sigma {\rm v}_{SS^{\dagger}\rightarrow {\rm b} \bar{\rm b}}\rangle^\prime$
obtained from the present two component DM model which produce the
observed flux lie above the upper limits given by
Fermi-LAT from their analysis of 15 dSphs's data (blue dashed line),
the allowed values of
$\langle\sigma {\rm v}_{SS^{\dagger}\rightarrow {\rm b} \bar{\rm b}}\rangle^\prime$
still fall within the 95\% C.L. error band reported by Fermi-LAT (turquoise
coloured band). Needless to mention here that the range of
allowed values of $\langle\sigma {\rm v}_{SS^{\dagger}
\rightarrow {\rm b} \bar{\rm b}}\rangle^\prime$
satisfy the limits reported by the DES collaboration (black dashed line).
We have also checked the gamma-ray flux from the GC using other dark matter
density profile available in literature, namely the Einasto profile \cite{einasto, einasto1}.
We have found that in order to explain the Fermi-LAT observed gamma-ray flux
from the self annihilation of a 35 GeV $S$ particle using Einasto dark
matter profile, one needs $\langle \sigma {\rm v}_{SS^{\dagger} \rightarrow
{\rm b} {\rm \bar{b}}}\rangle^{\prime} \sim 4.8\times 10^{-26}$ cm$^3/$s which violates
both the upper limits reported by Fermi-LAT (blue dashed line in Fig. \ref{compare-sigmavbb})
and DES (black dashed line in Fig. \ref{compare-sigmavbb}) collaborations.      
%%%%%%%%%%%%%%%%%%%%%%%%%%%%%%%%%%%%%%%%%%%%%%%%%%%%%%%%%%%%%%%%%%%%%%%
\section{Summary}
\label{summary}
%%%%%%%%%%%%%%%%%%%%%%%%%%%%%%%%%%%%%%%%%%%%%%%%%%%%%%%%%%%%%%%%%%%%%%%
In the present work, we propose a dark matter model which is an extension
of the Standard Model of particles physics in all three sectors namely gauge,
fermionic as well as scalar and contains two different types of dark matter
candidates. Therefore, in this two component dark matter model the role of
two dark matter candidates are played by a complex scalar field $S$ and a
Dirac fermion $\psi$ respectively. Although, both of these dark sector
particles ($S$, $\psi$) are singlet under SM gauge group (${\rm SU}(3)_{\rm c}\times
{\rm SU}(2)_{\rm L} \times {\rm U}(1)_{\rm Y}$)
but possess a non zero ${\rm U(1)}_{\rm X}$ charge
which ensures their stability. Thus, in addition to SM gauge group
%(${\rm SU}(2)_{\rm L} \times {\rm U}(1)_{\rm Y}$)
we have an additional local ${\rm U(1)}_{\rm X}$ gauge
symmetry under which all the SM particles
(including the Higgs boson) behave like singlet. Besides the dark matter
component $S$, the scalar sector of the present model is composed of
another complex singlet $\Phi_s$ and a ${\rm SU}(2)_{\rm L}$
doublet ($\Phi$) (the usual Higgs doublet). Both $\Phi$ and
$\Phi_s$ possess non zero VEVs namely, $v$ and $v_s$
respectively which spontaneously break the ${\rm SU}(2)_{\rm L}
\times {\rm U}(1)_{\rm Y}\times {\rm U(1)}_{\rm X}$
symmetry to a remnant U(1)$_{\rm em}\times\mathbb{Z}_2$
symmetry. This residual $\mathbb{Z}_2$ symmetry stabilize
the fields $S$ and $\psi$. 
As a result of this symmetry breaking, the neutral components
of $\Phi$ and $\Phi_s$ mix with each other and we get two
physical scalars, namely $h$ and $H$ with a nonzero mixing between
angle $\alpha$ between them. The symmetry breaking phenomenon is
manifested by the presence of five gauge bosons in the model
such as $W^\pm$, $Z$, $Z^\prime$ and $A$. Among these five
gauge bosons only $W^\pm$ has non zero electrical charge
and $A$ remains massless which is identified as the ``photon"
(mediator of the electromagnetic interaction). 
We have taken into account all the relevant
constraints which get affected due to the presence of nonzero mixing angles
between both $Z-Z^{\prime}$ and $h-H$. These include electroweak precision observables,
electroweak oblique parameters, $\rho$ parameter, bounds from the LHC
results on the signal strength and the invisible decay width of the
SM Higgs boson etc. Considering $S$ and $\psi$ as the two possible
candidates for the dark matter particles
in the Universe, their viability is examined by computing the
total relic abundance at the present epoch and the scattering cross sections
off the detector nuclei. In order to find the total relic abundance which is
the sum of the individual relic abundances of both the
dark matter components, we have solved  
two coupled Boltzmann equations for $\psi$ and $S$ at the present epoch.
We find that for a wide range of values of the model parameters
the total relic density of the two dark matter candidates
falls within the range specified by the PLANCK experiment.
We have compared the spin independent ``effective scattering cross sections" for
both the dark matter candidates off the detector nuclei with the latest
results of LUX experiment. We find that although, some portions of only
the lighter dark matter component $S$ of present two component dark matter model
have already been excluded by the results of LUX experiment but still
their exist enough regions in the $\sigma_{\rm SI}-M_{\rm DM}$
plane which can be tested by the ``ton-scale" direct detection
experiments in near future.
Finally, we have computed the $\gamma$-ray flux originated
from the self annihilation of the dark matter candidate $S$ into
${\rm b} \bar{\rm{b}}$ final state at the Galactic Centre region. We
find that our two component dark matter model also
shows an excess in the $\gamma$-ray spectrum obtained from the GC region
at an energy range $1\sim 3$ GeV from the annihilation of
dark matter candidate $S$, having mass in the range $30\sim40$ GeV.
The resulting $\gamma$-ray flux becomes lower as the mass splitting
between the two dark matter components increases.
In the end, we conclude that the $\gamma$-ray fluxes
computed for $M_{\psi}=60$ GeV and 80 GeV with
$M_S=35$ GeV and $\langle \sigma {\rm v}_{SS^{\dagger} \rightarrow
{\rm b} {\rm \bar{b}}}\rangle^\prime \sim (1.52-1.67)\times 10^{-26}$ cm$^3/$s
agree well with the Fermi-LAT data. Moreover the $\gamma$-ray fluxes
for $M_{\psi}\geq 100$ GeV are incompatible with the experimental data.
Finally, we have compared the range
of allowed values of annihilation cross section with the
limits reported by both the Fermi-LAT and DES collaborations. We have
found that the range of annihilation cross section for the ${\rm b} \bar{\rm b}$
annihilation channel predicted from this present model is in right ballpark
with reported limits on $\langle {\sigma {\rm v}_{{\rm b} \bar{\rm b}}}\rangle$
by the Fermi-LAT and DES collaborations.   
%%%%%%%%%%%%%%%%%%%%%%%%%%%%%%%%%%%%%%%%%%%%%%%%%%%%%%%%%%%%%%%%%%%%%%%%%%%      
\section{Acknowledgements}
%%%%%%%%%%%%%%%%%%%%%%%%%%%%%%%%%%%%%%%%%%%%%%%%%%%%%%%%%%%%%%%%%%%%%%%%%%%
Author would like to thank D. Adak, A. Dutta Banik and D. Majumdar for many
useful suggestions and discussions. Author would also like to acknowledge
Department of Atomic Energy (DAE), Govt. of India for their financial assistance. 
%\bibliography{gamma.bib}{} 
%\bibliographystyle{jhep}
\providecommand{\href}[2]{#2}\begingroup\raggedright\endgroup
\end{document}